\documentstyle[12pt,aps,eqsecnum,amssymb]{revtex}
\tighten
\newcommand{\be}{\begin{equation}}
\newcommand{\ee}{\end{equation}}
\newcommand{\ba}{\begin{eqnarray}}
\newcommand{\ea}{\end{eqnarray}}
\newcommand{\uz}{{\underline z}{\vphantom{z}}}
\newcommand{\bz}{{\bf z}}
\newcommand{\zz}{{z,\cc\uz}}
\newcommand{\upartial}{{\underline\partial}{\vphantom{\partial}}}
\newcommand*{\cc}[1]{ \rlap{$\stackrel{*}{\phantom{#1}}$}#1 }
\newcommand*{\ccc}[1]{\stackrel{*}{#1}\!\!{\vphantom{#1}}}
\newcommand{\uGamma}{\hat{\underline\Gamma}{\vphantom{\Gamma}}}
\newcommand{\barM}{\hat{\bar M}{\vphantom{M}}}
\newcommand{\barMM}{\hat{\bar{\bf M}}{\vphantom{M}}^2}
\newcommand{\qq}{{\frac 12}}
\renewcommand{\>}{{\rangle}}

\newcommand{\RS}{{\mathbb{R}}}
\newcommand{\CS}{{\mathbb{C}}}
\newcommand{\diag}{\mathop{{\rm diag}}}
\newcommand{\sign}{\mathop{{\rm sign}}}
\newcommand{\da}{{\dot\alpha}}
\newcommand{\db}{{\dot\beta}}

\begin{document}
\title{FIELDS ON THE POINCAR\'{E} GROUP: \\
Arbitrary Spin Description and Relativistic Wave Equations}
\author{D M Gitman${}^a$, A L Shelepin${}^{a,b}$ \thanks{%
E-mail: gitman@fma.if.usp.br, alex@shelepin.msk.ru}}
\address{${}^a$Instituto de F\'{\i}sica, Universidade de S\~ao Paulo,\\
Caixa Postal 66318, 05315-970--S\~ao Paulo, SP, Brazil \\
${}^b${Moscow Institute of Radio Engenering, Electronics and Automation,}\\
Prospect Vernadskogo, 78, 117454, Moscow, Russia }
\maketitle
\bigskip
\begin{abstract}
In this paper, starting from pure group-theoretical point of view, we develop
a regular approach to describing particles with different spins in the
framework of a theory of scalar fields on the Poincar\'e group. Such fields
can be considered as generating functions for conventional spin-tensor fields.
The cases of 2, 3, and 4 dimensions are elaborated in detail. Discrete
transformations $C,P,T$ are defined for the scalar fields as automorphisms of
the Poincar\'e group. Doing a classification of the scalar functions, we
obtain relativistic wave equations for particles with definite spin and
mass. There exist two different types of scalar functions (which describe the
same mass and spin), one related to a finite-dimensional nonunitary
representation and another one related to an infinite-dimensional unitary
representation of the Lorentz subgroup. This allows us to derive both usual
finite-component wave equations for spin-tensor fields and positive energy
infinite-component wave equations.
\end{abstract}

\section{Introduction}

Traditionally in field theory particles with different spins are described
by multicomponent spin-tensor fields on Minkowski space. However, it is
possible to use for this purpose scalar functions as well, which depend on
both Minkowski space coordinates and on some continuous bosonic variables
corresponding to spin degrees of freedom. For the first time, such fields
were introduced in \cite{GinTa47,BarWi48,Yukaw50,Shiro51} in connection with
the problem of constructing relativistic wave equations (RWE). Fields of
this type may be treated as ones on homogeneous spaces of the Poincar\'e
group. A systematic development of such point of view was given by
Finkelstein \cite{Finke55}.
He also gave a classification and explicit constructions of homogeneous
spaces of the Poincar\'e group, which contain Minkowski space.
The next logical step was done by Lur{\c{c}}at
\cite{Lurca64} who suggested to construct quantum field theory on the
Poincar\'e group. One of the motivations was to give a dynamical role to the
spin. Some development of these ideas was given in
\cite{BacKi69,Kihlb70,BoyFl74,Arodz76,Tolle78,Tolle96,Drech97}.
For example, different homogeneous spaces were described, as well as
possibilities to introduce interactions in spin phase space, and to
construct Lagrangian formulations were studied. The authors of \cite{BacKi69}
arrived at the conclusion that eight is the lowest dimension of a homogeneous
space suitable for a
description of both half-integer and integer spins. However, no convinced
physical motivation for the choice of homogeneous spaces was presented, and
the interpretation of additional degrees of freedom and of corresponding
quantum numbers remained an open problem.

In this paper, starting from pure group-theoretical point of view, we develop
a regular approach to describing particles with different spins in the
framework of a theory of scalar fields on the Poincar\'e group. Such fields
can be considered as generating functions for conventional spin-tensor
fields. In this language the problem of constructing RWE of different types
is formulated from a unique position.

In our consideration, we use scalar fields on the proper Poincar\'e group,
i.e., fields on the ten-dimensional manifold; this manifold is a direct
product of Minkowski space and of the manifold of the Lorentz subgroup.
These fields arise in our constructions in course of the study of a
generalized regular representation (GRR).
That provides a possibility to analyze then all the representations of the
Poincare group. The study of GRR implies a wide use of harmonic
analysis method \cite{Vilen68t,VilKl91,BarRa77,ZhelSc83}. In a sense, this
method is an alternative to one of induced representations suggested by
Wigner \cite{Wig39}\ (see \cite{BarRa77,Macke68,KimNo86,Ohnuk88}).
It turns out that the fields on the Poincar\'{e} group can be considered as
generating functions for usual spin-tensor fields on Minkowski space, thus we
naturally obtain all results for the latter fields. However, sometimes it is
more convenient to formulate properties and equations for spin-tensor fields
in terms of the generating functions.
Moreover, the problem of constructing RWE looks very natural in the language
of the scalar fields on the group. We show that this problem can be formulated
as a problem of a classification of different scalar fields. For this purpose,
in accordance with the general theory of harmonic analysis, we consider
various sets of commuting operators and identify constructing RWE with
eigenvalue problems for this operators. We succeeded to define discrete
transformations for the scalar fields using some automorphisms of the proper
Poincare group.
The space of scalar fields on the group turns out to be closed with respect
to the discrete transformations. One ought to say that the latter
transformations are of fundamental importance for constructing RWE and for
their analysis. Consideration of the discrete transformations helps us to
give right physical interpretation for quantum numbers which appear in
course of the classification of the scalar fields.

The paper is organized as follows.

In Sect. 2 we introduce the basic objects of our study, namely, scalar fields
$f(x,\bz)$. The scalar fields depend on $x$, which are coordinates on
Minkowski space, and on $\bz$, which are coordinates on the Lorentz subgroup.
The complex coordinates $\bz$ describe spin degrees of freedom. It is
shown that these fields are generating functions for usual spins-tensor
fields. Classifying the scalar fields with the help of various sets
of commuting operators on the group, we get description of irreps of the
group. We formulate a general scheme of constructing RWE in this language in
any dimensions. We introduce discrete transformations in the space of the
scalar functions and we relate these transformations to automorphisms of the
proper Poincar\'e group.

In Sect. 3 we apply the above general scheme to detailed study of scalar
fields on two-dimensional Poincar\'e and Euclidean groups. In particular,
we construct RWE and analyze their solutions.

Three-dimensional Poincar\'e and Euclidean group case is considered in
Sect. 4. Besides finite-component equations, we also construct positive
energy RWE assotiated with unitary infinite-dimensional irreps of 2+1 Lorentz
group. These equations, in particular, describe particles with fractional
spins.

In Sect. 5 we study scalar fields on the $3+1$ proper Poincar\'{e} group. A
connection of the present consideration with other approaches to RWE theory
is considered in detail. In particular, we pay significant attention to
equations with subsidiary conditions. General first-order Gel'fand--Yaglom
equations (including Bhabha equations), Dirac--Fierz--Pauli equations, and
Rarita--Schwinger equations arise in the present consideration as well. This
give a regular base for comparison of properties of various RWE.

Doing the classification of scalar functions in 2, 3, and 4 dimensions, we
obtain equations describing fields with fixed mass and spin. In Sect. 6 we
consider the general features of these equations.

One ought to say that the construction of RWE is elaborated in detail only
for the massive case. We plane to discuss the massless case in a later
article.

\section{Fields on the proper Poincar{\'{e}} group and spin description}

\subsection{Parametrization of the Poincar\'{e} group}

Consider Poincar\'e group transformations
\begin{equation}  \label{vector}
x^{\prime\nu}=\Lambda^\nu_{\;\;\mu} x^\mu + a^\nu
\end{equation}
of coordinates $x=(x^\mu,\;\mu=0,\dots,D)$ in $d=D+1$-dimensional
Minkowski space, 
$ds^2= \eta_{\mu\nu}dx^\mu dx^\nu$,
$\eta_{\mu\nu}=\mathop{{\rm diag}} (1,-1,\dots,-1)$.
The matrices $\Lambda$ define rotations in Minkovski space and belong to the
vector representation of $O(D,1)$ group. We are also going to
consider $D$-dimensional Euclidean case in which $ds^2= \eta_{ik}dx^i dx^k$,
and $\eta_{ik}=\mathop{{\rm diag}} (1,1,\dots,1)$, $i,k=1,\dots,D$. Here
the matrices $\Lambda$ belong to the vector representation of $O(D)$ group.

The transformations (\ref{vector}) which can be obtained continuously from
the identity form the proper Poincar\'{e} group $M_0(D,1)$ with the elements
$g=(a,\Lambda)$. Corresponding homogeneous transformations ($a=0$) form the
proper Lorentz group $SO_0(D,1)$. In the Euclidean case we deal with $M_0(D)$
and $SO(D)$ respectively. The composition law and the inverse element of
these groups have the form
\begin{equation}  \label{vector2}
(a_2,\Lambda_2)(a_1,\Lambda_1)=(a_2+\Lambda_2 a_1,\Lambda_2\Lambda_1), \quad
g^{-1}=(-\Lambda^{-1}a, \Lambda^{-1}).
\end{equation}
Thus, the groups $M_0(D,1)$ and $M_0(D)$ are semidirect products
\[
M_0(D,1)=T(d)\times )SO_0(D,1), \qquad M_0(D)=T(D)\times )SO_(D),
\]
where T(d) is $d$-dimensional translation group.

There exists one-to-one correspondence between the vectors $x$ and
$2\times 2$ Hermitian matrices $X$ in pseudo-Euclidean spaces of 2,3 and 4
dimensions,%
\footnote{ We use two sets of $2\times 2$ matrices
 $\sigma_\mu=(\sigma_0,\sigma_k)$ and $\bar\sigma_\mu=(\sigma_0,-\sigma_k)$,
 \[
 \sigma_0=\left(\begin{array}{cc} 1 & 0 \\ 0 & 1 \end{array} \right) ,\quad
 \sigma_1=\left(\begin{array}{cc} 0 & 1 \\ 1 & 0 \end{array} \right) ,\quad
 \sigma_2=\left(\begin{array}{cc} 0 & -i \\ i & 0 \end{array}\right) ,\quad
 \sigma_3=\left(\begin{array}{cc} 1 & 0 \\ 0 & -1 \end{array}\right) .
 \]
 }
\be
x\leftrightarrow X,\quad X=x^\mu\sigma_\mu.
\ee
Namely:
\begin{eqnarray}
&&d=3+1:\quad X=\left(\begin{array}{cc}
x^0+x^3 & x^1-ix^2 \\ x^1+ix^2 & x^0-x^3 \end{array} \right),  \label{XM31}
\\
&&d=2+1:\quad X=\left(\begin{array}{cc}
x^0 & x^1-ix^2 \\ x^1+ix^2 & x^0 \end{array} \right),  \label{XM21}
\\
&&d=1+1:\quad X=\left(\begin{array}{cc}
x^0 & x^1 \\ x^1 & x^0 \end{array} \right).  \label{XM11}
\end{eqnarray}
In all the above cases
\be
\det X=\eta _{\mu \nu }x^{\mu }x^{\nu },\quad x^{\mu }=\frac{1}{2}%
\mathop{\rm Tr}(X\bar\sigma^\mu).
\ee
In Euclidean spaces of 2 and 3 dimensions a similar correspondence has the
form
\begin{eqnarray}
&&D=3:\qquad X=\left(\begin{array}{cc}
x^3 & x^1-ix^2 \\ x^1+ix^2 & -x^3 \end{array} \right),  \label{XM3} \\
&&D=2:\qquad X=\left(\begin{array}{rr}
x^2 & x^1 \\ x^1 & -x^2 \end{array} \right).  \label{XM2}
\end{eqnarray}

If $x$ is subjected to a transformation (\ref{vector}), then $X$
transforms as follows (see, for example, \cite{Vilen68t}:
\begin{equation}  \label{spinor0}
X^{\prime}=UXU^\dagger +A,
\end{equation}
where $A=a^\mu\sigma_\mu$ and $U$ are some $2\times 2$ complex matrices
obeying the conditions
\begin{equation}  \label{ULam}
\sigma_\nu\Lambda^\nu_{\;\;\mu}=U\sigma_\mu U^\dagger.
\end{equation}
Eq. (\ref{ULam}) relates the matrices $\Lambda$ and $U$. There are many
$U$ which correspond to the same $\Lambda$. We may fix this arbitrariness
imposing the condition
\be  \label{detU}
\det U=1,
\ee
which does not contradict to the relation $\det U=e^{i\phi}$, which follows
from (\ref{ULam}). However, even after that, there is no one-to-one
correspondence between $\Lambda$ and $U$, namely two matrices ($U$,$-U$)
correspond to one $\Lambda$. Considering both $U$ and $-U$ as representatives
for $\Lambda$, we in fact go over from $SO_0(D,1)$ to its double covering
group ${\rm Spin}(D,1)$, or, in the Euclidean case, from $SO(D)$ to its
double covering group ${\rm Spin}(D)$. In the dimensions under consideration
the groups ${\rm Spin}(D,1)$ and ${\rm Spin}(D)$ are isomorphic
to the following ones:%
\footnote{We denote the complex conjugation by $*$ above the quantities.}
\begin{eqnarray}
&&d=3+1: \quad U\in SL(2,C), \quad U=\left(
\begin{array}{cc}
u^1_1 & u^1_2 \\
u^2_1 & u^2_2
\end{array}
\right) ,\quad u_1^1 u_2^2-u_1^2 u_2^1=1,  \label{XZM31} \\
&&d=2+1: \quad U\in SU(1,1), \quad U=\left(
\begin{array}{cc}
u_1 & u_2 \\
\cc u_2 & \cc u_1
\end{array}
\right), \;\; |u_1|^2-|u_2|^2=1,  \label{XZM21} \\
&&D=3: \qquad U\in SU(2), \quad U=\left(
\begin{array}{rr}
u_1 & u_2 \\
-\cc u_2 & \cc u_1
\end{array}
\right), \;\; |u_1|^2+|u_2|^2=1,  \label{XZM3} \\
&&d=1+1: \quad U\in SO(1,1), \quad U=\left(
\begin{array}{cc}
\cosh \frac\phi 2 & \sinh \frac\phi 2 \\
\sinh \frac\phi 2 & \cosh \frac\phi 2
\end{array}
\right), \qquad  \label{XZM11} \\
&&D=2: \qquad U\in SO(2), \quad U=\left(
\begin{array}{rr}
\cos \frac\phi 2 & \sin \frac\phi 2 \\
-\sin \frac\phi 2 & \cos \frac\phi 2
\end{array}
\right).  \label{XZM2}
\end{eqnarray}
Considering nonhomogeneous transformations and retaining both elements $U$
and $-U$ in the consideration, we go over from the groups $M_0(D,1)$ and
$M_0(D)$ to the groups
\[
M(D,1)=T(d)\times ){\rm Spin}(D,1),\qquad M(D)=T(D)\times ){\rm Spin}(D)
\]
respectively. As it is known, that allows one to avoid double-valued
representations for half integer spins. Thus, there exists one-to-one
correspondence between the elements $g$ of the groups $M(D,1)$, $M(D)$ and
two $2\times 2$ matrices, $g\leftrightarrow (A,U)$. The first one $A$
corresponds to translations and the second one $U$ corresponds to rotations.
Eq. (\ref{spinor0}) describes the action of $M(D,1)$ on Minkowski space
(the latter is coset space $M(D,1)/{\rm Spin}(D,1)$).
As a consequence of (\ref{spinor0}), one can obtain the composition law and
the inverse element of the groups $M(D,1)$, $M(D)$:
\begin{equation}
(A_2,U_2)(A_1,U_1)=(U_2 A_1 U_{2}^{\dagger}+A_2, \; U_{2}U_{1})\;, \quad
g^{-1}=(-U^{-1}A(U^{-1})^{\dagger}, \; U^{-1})\;.  \label{comp1}
\end{equation}
The matrices $U$ in the dimensions under consideration satisfy the following
identities:
\begin{eqnarray}
&&U\in SL(2,C):\quad \sigma_2U\sigma_2=(U^T)^{-1};  \label{auto31} \\
&&U\in SU(1,1):\quad \sigma_1U\sigma_1=\cc U, \quad
\sigma_2U\sigma_2=(U^T)^{-1}, \quad \sigma_3U\sigma_3=(U^\dagger)^{-1},
\label{auto21} \\
&&U\in SU(2):\qquad \sigma_2U\sigma_2=(U^T)^{-1}=\cc U.  \label{auto3}
\end{eqnarray}

An equivalent picture arise in terms of the matrices $\overline X%
=x^\mu\bar\sigma_\mu$. Using the relation $\overline X=\sigma_2X^T\sigma_2$,
the transformation law for $X$ (\ref{spinor0}), and the identity
(\ref{auto31}), one can get
\begin{equation}  \label{xbar}
\overline X^{\prime}=(U^\dagger )^{-1}\overline XU^{-1} +\overline A.
\end{equation}
Thus, $\overline X$ are transformed by means of the elements
$(\overline A,(U^\dagger )^{-1})$. The relation
$(A,U)\to (\overline A,(U^\dagger )^{-1})$ defines an automorphism of the
Poincar\'{e} group $M(D,1)$. In Euclidean case the matrices $U$ are unitary,
and the latter relation is reduced to $(A,U)\to (-A,U)$.

The representation of the Poincar\'e transformations in the form
(\ref{spinor0}) is closely related to a representation of finite rotations in
$\RS^d$ in terms of the Clifford algebra. In higher dimensions the
transformation law has the same form, where $A$ is a vector element and $U$
corresponds to an invertible element (spinor element) of the Clifford algebra
\cite{BenTu88}.
Besides, the representation of the finite transformations in the form
(\ref{spinor0}) can be useful for spin description by means of Grassmannian
variables $\xi$, since $\xi$ and $\partial \xi$ give a realization of the
Clifford algebra \cite{Berez66}.

\subsection{Regular representation and scalar functions on the group}

It is well known \cite{ZhelSc83,Vilen68t,VilKl91} that any irrep of a group
$G$ is contained (up to the equivalence) in a decomposition of a GRR. Thus,
the study of GRR is an effective method for the analysis of irreps of the
group. Consider, first, the left GRR $T_L(g)$, which is defined in the space
of functions $f(g_0)$, $g_0\in G$, on the group,
as
\begin{equation}
T_L(g)f(g_0)=f^{\prime }(g_0)=f(g^{-1}g_0),\;g\in G.  \label{Lreg0}
\end{equation}
As a consequence of the relation (\ref{Lreg0}), we can write
\begin{equation}
f'(g'_0)=f(g_0),\quad g'_0=gg_0.  \label{3.2}
\end{equation}
Let $G$ be the group $M(3,1)$, and we use the parametrization of its
elements by two $2\times 2$ matrices (one hermitian and another one from $%
SL(2,C)$), which was described in the previous Sect. At the same time, using
such a parametrization, we choose the following notations:
\begin{equation}
g\leftrightarrow (A,U)\,,\;g_0\leftrightarrow (X,Z)\,,  \label{3.3}
\end{equation}
where $A,X$ are $2\times 2$ hermitian matrices and $U,Z\in SL(2,C).$ The map
$\;g_0\leftrightarrow (X,Z)$ creates the correspondence
\begin{eqnarray}
&&g_0\leftrightarrow (x,z,\uz),\quad \text{where} \quad x=(x^\mu), \;
z=(z_\alpha), \; \uz=(\uz_\alpha),
\nonumber \\ \label{3.4}
&&\mu =0,1,2,3, \; \alpha =1,2, \; z_1\uz_2-z_2\uz_1=1,
\end{eqnarray}
by virtue of the relations
\be
X=x^\mu\sigma_\mu, \quad
Z=\left(\begin{array}{cc} z_1 & \uz_{1} \\ z_2 & \uz_2 \end{array} \right)
\in SL(2,C)\,.  \label{Z4}
\ee
On the other hand, we have the correspondence
$g'_0 \leftrightarrow (x',z',\uz')$,
\begin{eqnarray} \nonumber
&&g'_0=gg_0 \leftrightarrow
(X',Z')=(A,U)(X,Z)=(UXU^{+}+A,UZ) \leftrightarrow (x',z',\uz'),
\\  \label{3.6}
&&x^{\prime\mu}\sigma_\mu=X'=UXU^{+}+A \quad \Longrightarrow \quad
 x^{\prime\mu}=(\Lambda_0)_{\;\,\nu }^\mu x^\nu +a^\mu, \quad
 \Lambda\leftarrow U\in SL(2,C),
\\  \label{3.7}
&&\left(\begin{array}{cc} z'_1 & \uz'_1 \\ z'_2 & \uz'_2 \end{array} \right)
 = Z'= UZ       \; \Longrightarrow \;
 z'_{\alpha}=U_{\alpha}^{\;\;\beta}z_{\beta}, \;
 \uz'_{\alpha}=U_{\alpha}^{\;\;\beta}\uz_{\beta}, \;
 U=(U_\alpha^{\;\;\beta}),\; z'_1\uz'_2-z'_2\uz'_1=1.
\end{eqnarray}
Then the relation (\ref{3.2}) takes the form
\ba
&&f'(x',z',\uz')=f(x,z,\uz),  \label{3.9}
\\ \label{3.10}
&&x^{\prime\mu}=(\Lambda_0)_{\;\,\nu}^\mu x^\nu+a^\mu, \quad
 \Lambda\leftarrow U\in SL(2,C),
\\ \label{lawZ}
&& z'_{\alpha}=U_{\alpha}^{\;\;\beta}z_{\beta}, \quad
 \uz'_{\alpha}=U_{\alpha}^{\;\;\beta}\uz_{\beta}, \quad
 z_1\uz_2-z_2\uz_1 = z'_1\uz'_2-z'_2\uz'_1=1.
\ea

The relations (\ref{3.9})-(\ref{lawZ}) admit a remarkable interpretation.
We may treat $x$ and $x'$ in these relations as position coordinates in
Minkowski space (in different Lorentz refrence frames)
related by proper Poincare transformations, and the sets $(z,\uz)$ and
$(z,\uz')$ may be treated as spin coordinates in these Lorentz frames. They
are transformed according to the formulas (\ref{lawZ}). Carrying
two-dimensional spinor representation of the Lorentz group, the variables $z$
and $\uz$ are invariant under translations as one can expect for spin degrees
of freedom.
Thus, we may treat sets $(x,z,\uz)$ as points in a position-spin space with
the transformation law (\ref{3.10}), (\ref{lawZ}) under the change from one
Lorentz reference frame to another. In this case equations
(\ref{3.9})-(\ref{lawZ}) present the transformation law for scalar
functions on the position-spin space.

On the other hand, as we have seen, the sets $(x,z,\uz)$ are in one-to-one
correspondence to the group $M(3,1)$ elements. Thus, the functions
$f(x,z,\uz)$ are still functions on this group. That is why we often call
them scalar functions on the group as well, remembering that the term
''scalar'' came from the above interpretation.

Remember now that different functions of such type correspond to different
representations of the group $M(3,1)$. Thus, the problem of classification
of all irreps of this group is reduced to the problem of a classification
of all scalar functions on position-spin space.
However, for the purposes of such classification, it is natural to restrict
ourselves by the scalar functions which are analytic both in $z,\uz$ and in
$\cc z,\cc\uz$ (or, simply speaking, which are differentiable with respect
to these arguments). Further such functions are denoted by
$f(x,z,\uz,\cc z,\cc\uz)=f(x,\bz)$, $\bz=(z,\uz,\cc z,\cc\uz)$.

Consider now the right GRR $T_R(g)$. This representation is defined in the
space of functions $f(g_0)$, $g_0\in G$ as
\begin{equation}
T_R(g)f(g_0)=f'(g_0)=f(g_0g),\quad g\in G,  \label{Rreg0}
\end{equation}
As a consequence of the relation (\ref{Rreg0}), we can write
\begin{equation}
f'(g'_0)=f(g_0),\quad g'_0=g_0g^{-1}.
\end{equation}
In the case of the proper Poincare group, the right transformations act on
$g_0\leftrightarrow (X,Z)$ according to the formula
\begin{equation}
g'_0=g_0g^{-1}\leftrightarrow (X',Z')=(X+Z^{-1}A(Z^{-1})^{\dagger },ZU^{-1}),  \label{Rarg2}
\end{equation}
hence $x^{\prime\mu} =x^\mu +L_{\;\;\nu }^\mu a^\nu $, where the matrix $L$
depends on $\bz$,
$\sigma_\nu L_{\;\;\mu }^\nu = Z^{-1} \sigma_\mu (Z^{-1})^\dagger$.
The transformations for $x,\bz$ do not admit similar to the left GRR case
interpretation. In particular, the transformation low for $x$ does not look
as a Lorentz transformation. On the other hand, the study of the right GRR is
useful for the purposes of the classification of the Poincare group irreps,
since the generators of the right GRR are used to construct complete sets of
commuting operators on the group.


\subsection{Generators of generalized regular representations}

Generators of the left GRR correspond to translations and rotations. They
can be written as
\be  \label{Lgen}
\hat{p}_{\mu }=-i\partial /\partial x^{\mu }, \quad
 \hat{J}_{\mu\nu }= \hat{L}_{\mu\nu }+ \hat{S}_{\mu \nu},
\ee
where $\hat{L}_{\mu\nu }=i(x_\mu\partial _\nu -x_\nu\partial _\mu )$
are angular momentum operators, and $\hat{S}_{\mu\nu }$ are spin
operators depending on $\bz$ and $\partial/\partial \bz$. An explicit form
of spin operators is given in the Appendix.

The algebra of the generators (\ref{Lgen}) has the form
\ba
&&[\hat{p}_{\mu },\hat{p}_{\nu }]=0,\quad  [\hat{J}_{\mu\nu},\hat{p}_{\rho}]
 =i(\eta_{\nu \rho}\hat{p}_{\mu} -\eta_{\mu \rho}\hat{p}_{\nu}), \quad
\nonumber \\
&&[\hat J_{\mu \nu },\hat J_{\rho \sigma }] =i\eta _{\nu \rho }\hat J_{\mu\sigma }
 -i\eta _{\mu \rho }\hat J_{\nu \sigma }-i\eta _{\nu \sigma }\hat J_{\mu \rho}
 +i\eta _{\mu \sigma }\hat J_{\nu \rho }\;. \label{komm0}
\ea

In the space of Fourier transforms
\begin{equation}
\varphi (p,\bz) = (2\pi )^{-d/2}\int f(x,\bz)e^{ipx}dx
\end{equation}
the left GRR acts as (one has to use (\ref{Lreg0})):
\be  \label{Lregp}
T_{L}(g)\varphi (p,\bz) = e^{iap^{\prime}} \varphi (p^{\prime},g^{-1}\bz),
\quad p^{\prime}=g^{-1}p \leftrightarrow P^{\prime}=U^{-1}P(U^{-1})^\dagger,
\quad P=p_\mu \sigma^\mu .
\ee
One can see that $\det Z$ and $\det P=p^2$ are invariant under the
transformations%
\footnote{ $p^2=\eta^{\mu\nu}p_\mu p_\nu$. Since we do not use $p$ with
 upper indices, this does not lead to a misunderstanding.}
(\ref{Lregp}) and that $p^2$ is an eigenvalue of the Casimir operator
$\hat p^2$.

For the groups $M(D)$ there are two types of representations depending
on $p^2$: 1) $p^2\ne 0$; 2) $p^2=0$; then all $p_i=0$, and irreps are
labelled by eigenvalues of Casimir operators of the rotation subgroup.

For the groups $M(D,1)$ there are four types of representations
depending on the eigenvalues $m^2$ of the Casimir operator $\hat p^2$:
1) $m^2>0$; 2) $m^2<0$ (tachyon); 3) $m^2=0$, $p_0\ne 0$ (massless particle);
4) $m^2=p_0=0$, irreps are labelled by eigenvalues of the Casimir operators
of the Lorentz subgroup, and the corresponding functions do not depend on $x$.

For decomposing the left GRR we are going to construct a complete
set of commuting operators in the space of functions on the group.
Together with the Casimir operators some functions of right generators%
\footnote {The physical meaning of the right generators is not so
 transparent. However, one can remember that the right generators of $SO(3)$
 in the nonrelativistic rotator theory are interpreted as operators of
 angular momentum in a rotating body-fixed reference frame
 \cite{Wigne59,LanLi3,BieLo81}.}
may be included in such a set. Therefore it is necessary to know the explicit
form of right generators. As a consequence of the formulas
\ba
&&T_{R}(g)f(x,\bz) = f(xg,\;\bz g),\quad xg \leftrightarrow X + ZAZ^{\dagger},
\quad \bz g \leftrightarrow ZU,        \label{Rreg1}
\\
&&T_{R}(g)\varphi (p,\bz) = e^{-ia'p} \varphi (p, \bz g), \quad
a'\leftrightarrow A'=ZAZ^{\dagger}   \label{Rregp}
\ea
one can obtain
\be  \label{Rgen}
\hat{p}_{\mu }^R=-(L^{-1}(\bz))^{\;\;\nu}_\mu p_\nu, \quad
\hat{J}_{\mu\nu }^R= \hat{S}_{\mu \nu}^R,
\ee
where $L\in SO(D,1)$ (or $L\in SO(D,1)$ in the Euclidean case).
The operators of right translations can also be written in the form
${\hat P}^R=-Z^{-1}\hat{P}(Z^{-1})^\dagger$; operators $\hat{S}_{\mu\nu}$ and
$\hat{S}^R_{\mu\nu }$ are left and right generators of ${\rm Spin}(D,1)$
(or ${\rm Spin}(D)$) and depend on $\bz$ only. All the right generators
(\ref{Rgen}) commute with all the left generators (\ref{Lgen}) and obey the
same commutation relations (\ref{komm0}).

In accordance with theory of harmonic analysis on Lie groups
\cite{ZhelSc83,BarRa77} there exists a complete set of commuting operators,
which includes Casimir operators, a set of the left generators and a set
of right generators (both sets contain the same number of the generators).
The total number of commuting operators is equal to the number of
parameters of the group. In a decomposition of the left GRR the
nonequivalent representations are distinguished by eigenvalues of the
Casimir operators, equivalent representations are distinguished by
eigenvalues of the right generators, and the states inside the irrep are
distinguished by eigenvalues of the left generators.

In particular, Casimir operators of spin Lorentz subgroup are functions of
$\hat{S}^R_{\mu\nu }$ (or $\hat{S}_{\mu\nu }$) and commute with all the left
generators (with left translations and rotations), but do not
commute with generators of the right translations. These operators
distinguish equivalent representations in the decomposition of the left GRR.
Notice that some aspects of the theory of harmonic analysis on the 3+1 and
2+1 Poincar\'e groups were considered in
\cite{Ridea66,Hai69,Hai71} and \cite{GitSh97} respectively.

If GRR acts in the space of all functions on the group $G$, then a regular
representation acts in the space of functions $L^2(G,\mu)$, such that the
norm
\be \label{L2}
\int \cc f(g)f(g)d\mu(g)
\ee
is finite \cite{VilKl91,ZhelSc83}, where $d\mu(g)$ is an invariant measure
on the group. The regular representation is unitary, as it follows from
(\ref{L2}) and from the invariance of the measure. However we will
also use nonunitary representations (in particular, finite-dimensional
representations of the Lorentz group).
Therefore we consider the GRR as a more useful concept.

\subsection {Fields on the Poincar\'{e} group}

As we have shown, the relations associated with the left GRR (\ref{Lreg0})
define the transformation law for coordinates ($x,\bz$) on the position-spin
space under the change from one Lorentz reference frame to another. The
equations
\ba
&&f'(x',\bz')=f(x,\bz),  \label{scal_xz}
\\
&&x'=gx=\Lambda x+a \leftrightarrow UXU^\dagger + A, \quad
 \bz'=g\bz \leftrightarrow UZ. \label{scal_xz1}
\ea
define a scalar field on this space (i.e. a scalar field on the Poincar\'{e}
group). In contrast to scalar field on Minkowski space, this field is
reducible with respect to both mass and spin.

Consider the transformation laws of $x$ and $\bz$ in various dimensions
more detail.

In two-dimensional case matrices $Z$ depend on only one parameter
(angle or hyperbolic angle, see (\ref{XZM11}),(\ref{XZM2})).
The functions on the group depend on $x=(x^\mu)$ and $z=e^{\alpha}$
(or $x=(x^k)$ and $z=e^{i\alpha}$ in Euclidean case); it is appropriate
to consider these functions as functions of real parameter $\alpha$ directly.

In three-dimensional case according to (\ref{XZM21}),(\ref{XZM3})
\be  \label{Z3}
D=3: \quad Z=\left(\begin{array}{rr} z_1 & -\cc z_{\dot 2} \\ z_2 &
 \cc z_{\dot 1} \end{array}\right) ,\qquad
d=2+1: \quad
 Z=\left(\begin{array}{cc} z_1 & \cc z_{\dot 2} \\ z_2 & \cc z_{\dot 1}
 \end{array}\right), \qquad \det Z=1.
\ee
Functions $f(x,\bz)$ depend on $x=(x^\mu)$ (in Euclidean case $x=(x^k)$) and
$\bz=(z, \cc z)$, where $z$ are the elements of the first column of matrix
(\ref{Z3}). Let us write the relation (\ref{scal_xz1}) for $d=2+1$ in
component-wise form
\ba  \label{transf21}
&&x^{\prime\nu} \sigma_{\nu\alpha\da }=
  U_\alpha^{\;\;\beta} x^\mu\sigma_{\mu\beta\db }
  \cc U_{\;\;\dot\alpha}^{\dot\beta} + a^\mu\sigma_{\mu\alpha\da } ,
\\
&&z'_\alpha = U_\alpha^{\;\;\beta} z_\beta , \quad
 \cc z'_{\dot\alpha} =\cc U_{\dot\alpha}^{\;\;\dot\beta} \cc z_{\dot\beta}, \quad
 z^{\prime\alpha} = (U^{-1})^{\alpha}_{\;\;\beta} z^\beta , \quad
 \cc z^{\prime\dot\alpha} =(\cc U^{-1})^{\dot\alpha}_{\;\;\dot\beta} \cc z^{\dot\beta}.
\ea
Undotted and dotted indices correspond to spinors transforming
by means of matrix $U$ and complex conjugate matrix $\cc U$.
Invariant tensor $\sigma_{\nu\alpha\da }$ has one vector index and two
spinor indices of distinct types.

For the group $M(3,1)$ matrices $Z$, $\det Z=1$, has the form (\ref{Z4});
the elements $z^\alpha$ and $\uz^\alpha$ of first and second columns of
matrix (\ref{Z4}) are subjected to the same transformation law.
The functions $f(x,\bz)$ depend on $x=(x^\mu)$ and $\bz=(z,\cc z,\uz,\cc\uz)$.
The main reason to consider not real parameters (for example, real and
imaginary parts of $z,\uz$), but of $z,\uz$ and $\cc z,\cc\uz$, is the fact
that the complex variables are subjected to simple transformation rule.
Besides, the use of spaces of analytic and antianalytic functions is suitable
for the problem of decomposition of GRR.

According to (\ref{scal_xz1}) and (\ref{xbar}) one may write the
transformation law of $x^\mu$, $z_\alpha$, $\cc z_{\dot\alpha}$ in
component-wise form
\ba  \label{transf}
&&x^{\prime\nu} \sigma_{\nu\alpha\da }=
  U_\alpha^{\;\;\beta} x^\mu\sigma_{\mu\beta\db }
  \cc U_{\;\;\dot\alpha}^{\dot\beta} + a^\mu\sigma_{\mu\alpha\da},  \qquad
  x^{\prime\nu} \bar\sigma_\nu^{\;\;\da\alpha }=
  (\cc U^{-1})^{\dot\alpha}_{\;\;\dot\beta}x^\mu\bar\sigma_\mu^{\;\;\db\beta}
  (U^{-1})^{\;\;\alpha}_{\beta} + a^\mu\bar\sigma_\mu^{\;\;\da\alpha},
\\
&&z'_\alpha = U_\alpha^{\;\;\beta} z_\beta , \quad
 \cc z'_{\da} =\cc U_{\da}^{\;\;\db} \cc z_{\db}, \quad
 z^{\prime\alpha} = (U^{-1})^{\alpha}_{\;\;\beta} z^\beta , \quad
 \cc z^{\prime\da} =(\cc U^{-1})^{\da}_{\;\;\db} \cc z^{\db}.
\ea
It is easy to see from (\ref{transf}) that the tensors
\be
\sigma_{\mu\alpha\da}= (\sigma_\mu)_{\alpha\da},\quad
\bar\sigma_\mu^{\;\;\da\alpha} = (\bar\sigma_\mu)^{\da\alpha}
\ee
are invariant. These tensors
are usually used to convert vector indices into spinor ones and vice versa
or to construct vector from two spinors of different types:
\be
x^\mu = \frac 12 \bar\sigma^{\mu\da\alpha} x_{\da\alpha}, \quad
x_{\alpha\da} = \sigma_{\mu\alpha\da} x^\mu, \quad
q^\mu = \frac 12 \bar\sigma^{\mu\da\alpha} z_\alpha \cc z_{\dot\alpha}.
\ee

In consequence of the unimodularity of $2\times 2$ matrices $U$ there exist
invariant antisymmetric tensors
$\varepsilon^{\alpha\beta}=-\varepsilon^{\beta\alpha}$,
$\varepsilon^{\da\db}=-\varepsilon^{\db\da}$,
$\varepsilon^{12}=\varepsilon^{{\dot 1}{\dot 2}}=1$,
$\varepsilon_{12}=\varepsilon_{{\dot 1}{\dot 2}}=-1$.
Now spinor indices are lowered and raised according to the rules
\be
z_\alpha=\varepsilon_{\alpha\beta}z^\beta, \quad
z^\alpha=\varepsilon^{\alpha\beta}z_\beta,
\ee
and in particular one can get $\sigma_{\mu\alpha\da}=\bar\sigma_{\mu\da\alpha}$.
Below we will also use the notations
$\partial_\alpha={\partial}/{\partial z^\alpha}$,
$\partial^\da   ={\partial}/{\partial \cc z_\da}$, and correspondingly
$\partial^\alpha=-{\partial}/{\partial z_\alpha}$,
$\partial_\da   =-{\partial}/{\partial \cc z^\da}$.

In the framework of theory of the scalar functions on the Poincar\'{e} group
{\it a standard spin description in terms of multicomponent functions
arises under the separation of space and spin variables}.

Since $\bz$ is invariant under translations, any function $\phi(\bz)$
carry a representation of the Lorentz group. Let a function $f(h)=f(x,\bz)$
allows the representation
\be  \label{lor1}
f(x,\bz)= \phi^n(\bz)\psi_n(x),
\ee
where $\phi^n(\bz)$ form a basis in the representation space of the
Lorentz group. The latter means that one may decompose the functions
$\phi^n(\bz')$ of transformed argument $\bz' =g\bz$ in terms of the
functions $\phi^n(z)$:
\be  \label{z_bas}
\phi^n (\bz') = \phi^l (\bz)L_l^{\;\;n}(U).
\ee
An action of the Poincar\'{e} group on a line $\phi^n(\bz)\phi^n(\bz)$
is reduced to a multiplication by matrix $L(U)$, where $U\in {\rm Spin}(D,1)$,
$\phi (\bz')=\phi (\bz)L(U)$.

Comparing the decompositions of the function $f'(x',\bz')=f(x,\bz)$ over
the transformed basis $\phi (\bz')$ and over the initial basis $\phi (\bz)$,
$$
f'(x',\bz')=\phi (\bz')\psi'(x')
= \phi (\bz)L(U) \psi'(x')=\phi (\bz)\psi (x),
$$
where $\psi (x)$ is a column with components $\psi_n(x)$, one may obtain
\be  \label{nfield}
\psi' (x')=L(U^{-1})\psi (x),
\ee
i.e. the transformation law of a tensor field on Minkowski space.
This law correspond to the representation of the Poincar\'e group acting in
a linear space of tensor fields as follows
$T(g)\psi (x)=L(U^{-1})\psi (\Lambda^{-1}(x-a))$.
According to (\ref{z_bas}) and (\ref{nfield}), the functions $\phi (z)$
and $\psi (x)$ are transformed under contragradient representations of the
Lorentz group.

For example, let us consider scalar functions on the Poincar\'e group
$f_1(x,\bz)=\psi_\alpha (x)z^\alpha$ and
$f_2(x,\bz)=\bar\psi_\alpha (x)\cc z^\alpha$,
which correspond to spinor representations of Lorentz group.
According to (\ref{lor1}) and (\ref{nfield})
\be
\psi'_\alpha (x')= U_\alpha^{\;\;\beta}\psi_\beta (x), \quad
\bar\psi'_{\dot\alpha} (x')=
\cc U_{\dot\alpha}^{\;\;\dot\beta}\bar\psi_{\dot\beta} (x).
\ee
The product $\psi_\alpha (x) \bar\psi^{*\alpha}(x)$ is Poincar\'{e} invariant.

Thus tensor fields of all spins are contained in the decomposition of the
field (\ref{scal_xz}) on the Poincar\'e group, and the problems of their
classification and construction of explicit realizations
are reduced to problem of the decomposition of the left GRR.

Notice that above we reject the phase transformations, which correspond to
$U=e^{i\phi}$. This transformations of $U(1)$ group do not change space-time
coordinates $x$, but change the phase of $\bz$. According to (\ref{z_bas})
and (\ref{nfield}) that leads to the transformation of phase of tensor field
components $\psi_n(x)$. Taking account of this transformations means the
consideration of the functions on the group
$T(d)\times ){\rm Spin}(D,1)\times U(1)$.


\subsection {Automorphisms of the Poincar\'e group and
             discrete transformations: P,C,T}

Let us consider elements $g\leftrightarrow (A,U)$, $g_0\leftrightarrow (X,Z)$
of the Poincar\'e group $M(D,1)$. It is easy to see that transformations
\ba
&&(A,U)\to(\overline A,(U^\dagger )^{-1}),
\quad (X,Z)\to(\overline X,(Z^\dagger )^{-1}),        \label{auto00}\\
&&(A,U)\to(\cc A,\cc U),\quad (X,Z)\to(\cc X,\cc Z),  \label{auto1}  \\
&&(A,U)\to(-A,U),\quad (X,Z)\to(-X,Z)                 \label{auto2}
\ea
are outer involutory automorphisms of the group and generate finite group
consisting of eight elements.

The automorphisms (\ref{auto00})-(\ref{auto2}) define discrete
transformations of space-time and spin coordinates $x,\bz$. The substitution
of transformed coordinates into the functions $f(x,\bz)$ (or into the
generators (\ref{Lgen})) leads to change signs of some physical variables.
(Notice that the substitution both into the functions and into the generators
leaves signs unaltered.)

The space reflection (or parity transformation $P$) is defined by the
relations $x^0\to x^0$, $x^k\to -x^k$, or $X\to \overline X$. If $X$ is
transformed by means of the group element $(A,U)$, then $\overline X$ is
transformed by means of the group element $(\overline A,(U^\dagger )^{-1})$,
see (\ref{xbar}). Therefore the space reflection represents a
realization of the automorphism (\ref{auto00}) of the Poincar{\'e} group
\be  \label{P}
(X,Z)\stackrel{P}{\to}(\overline X,(Z^\dagger )^{-1}).
\ee
Thus, under the space reflection $x$ and $\bz$ have to be changed in all the
constructions according to (\ref{P}). In particular, for the momentum
$P=p_\mu\sigma^\mu$ we obtain $P\to \overline P$,
where $\overline P=p_\mu\bar\sigma^\mu$. The generators of the rotations are
not changed and the generators of the boosts change their signs only.

The time reflection transformation $T'$ is defined by the relation
$x^\mu\to (-1)^{\delta_{0\mu}}x^\mu$, or $X\to -\overline X$, and corresponds
to the composition of automorphisms (\ref{auto00}) and (\ref{auto2}):
\be  \label{T}
(X,Z)\stackrel{T'}{\to}(-\overline X,(Z^\dagger )^{-1}).
\ee
Inversion $PT'$, $(X,Z)\stackrel{PT'}{\to}(-X,Z)$, corresponds to the
automorphism (\ref{auto2}).

Automorphism of complex conjugation (\ref{auto1}) means substitution
$i\to -i$,
\be \label{C}
f(x,\bz)\stackrel{C}{\to} \cc f(x,\bz).
\ee
One can show that in the framework of the characteristics related to the
Poincar\'{e} group this transformation corresponds to the charge
conjugation. Both the transformation (\ref{C}) and charge conjugation
change signs of all the generators, $\hat p_\mu\to -\hat p_\mu$,
$\hat L_{\mu\nu}\to -\hat L_{\mu\nu}$, $\hat S_{\mu\nu}\to -\hat S_{\mu\nu}$.
Below, considering RWE, we will see that
transformation (\ref{C}) change also the sign of current vector $j^\mu$.

The time reversal $T$ is defined by the relation $X\to -\overline X$
(the time reflection transformation $T'$), with the supplementary condition
of energy sign conservation that means $P\to \overline P$. Therefore,
the conditions
$\hat p_\mu\to -(-1)^{\delta_{0\mu}}\hat p_\mu$,
$\hat L_{\mu\nu}\to -(-1)^{\delta_{0\mu}+\delta_{0\nu}}\hat L_{\mu\nu}$,
$\hat S_{\mu\nu}\to -(-1)^{\delta_{0\mu}+\delta_{0\nu}}\hat S_{\mu\nu}$
take place. The transformation $CT'$ obeys these conditions.

However, it is known \cite{UmeKaT54,KemPoP59} that it is possible to give two
distinct definitions of time reversal transformation obeying conditions
mentioned above. Wigner time reversal $T_w$ leaves the total charge (and
correspondingly $j^0$) unaltered, and reverses the direction of current $j^k$.
Schwinger time reversal $T_{sch}$ \cite{Schwi51} leaves the current
$j^k$ invariant and reverses the charge.

The transformation $CT'$ changes the sign of $j^0$ and therefore
can be identify with Schwinger time reversal, $T_{sch}=CT'$.
$CPT_{sch}$-transformation corresponds to the inversion $(X,Z){\to}(-X,Z)$.
Wigner time reversal $T_w$ and $CPT_w$-transformation can be defined
considering both outer and inner automorphisms of the proper Poincar\'e
group \cite{BucGiS00}. Namely, $CPT_w=I_xI_z$, where $I_z$ is defined as
\be
(X,Z)\stackrel{I_z}{\to}(X, Z(-i\sigma_2))
\ee
and is a composition of the inner automorphism
$(X,Z)\to (\overline X^T,(Z^T)^{-1})$ and of the rotation by the angle $\pi$.
Wigner time reversal is the composition of above considered
transformations, $T_w=I_zCT=I_zT_{sch}$.

The improper Poincar\'e group is defined as a group, which includes
continuous transformations of the proper Poincar\'e group $g\in M(D,1)$
and the space reflection $P$.

In the Euclidean case the space reflection is reduced to the substitution
$(X,Z)\stackrel{P}{\to}(-X,Z)$. The charge conjugation inverts the
momentum and spin orientation.


\subsection {Equivalent representations}

In the decomposition of scalar field (\ref{scal_xz}) on the Poincar\'e group
(or, that is the same, of the left GRR) there are equivalent representations
distinguished by the right generators.

Remember that representations $T_1(g)$ and $T_2(g)$ acting in linear spaces
$L_1$ and $L_2$ respectively are equivalent if there exists
an invertible linear operator $A:L_1\rightarrow L_2$ such that
\be
AT_1(g)=T_2(g)A.
\ee
In particular, the left and the right GRR of a Lee group G are equivalent.
The operator $(Af)(g)=f(g^{-1})$ realizes the equivalence
\cite{ZhelSc83,Vilen68t}.

Let us consider functions $f(x,\bz)$ belonging to two equivalent
representations in the decomposition of the left GRR of the group $M(D,1)$
(or $M(D)$). If the representations $T_1(g)$ and $T_2(g)$ acting in the
different subspaces $L_1$ and $L_2$ of the space of functions on the group
are equivalent, then
$$
AT_1(g)f_1(x,\bz)=T_2(g)Af_1(x,\bz), \quad f_2(x,\bz)=Af_1(x,\bz),
$$
where $f_1(x,\bz)\in L_1$ and $f_2(x,\bz)\in L_2$.
In particular, if operator $A:L_1\rightarrow L_2$ is a function of the
right translations generators $\hat p^R_\mu$, then one can't map the
function $f_1(x,\bz)$ to the function $f_2(x,\bz)$ by the group
transformation, which leaves the interval square invariant.
Therefore the physical equivalence of the states, that correspond to
equivalent irreps in the decomposition of the scalar field $f(x,\bz)$,
is not evident at least.

Below we will consider a number of examples in various dimensions. In
particular, in the framework of the representation theory of three-dimensional
Euclidean group $M(3)$ irreps characterized by different spins (but with the
same spin projection on the direction of propagation) are equivalent.
There are no contradictions in the fact that in this case different
particles are described by equivalent irreps since it is not possible to map
corresponding wave functions one into another by the rotations or
translations of the frame of references.

In some cases more general consideration may be based on the representation
theory of an extended group. In the framework of the latter there are
two possibilities: either irreps labelled by different eigenvalues of right
generators of initial group are nonequivalent or some equivalent irreps
of initial group are combined into one irrep.
For example, in nonrelativistic theory spin becomes the characteristic of
nonequivalent irreps after the extantion of $M(3)$ up to Galilei group.
In 3+1 dimensions for $m>0$ the proper Poincar\'e group representations
characterized by different chiralities are equivalent.
If we going from the Lorentz group to the group $SO(3,2)$, then all
characterized by spin $s$ states with different chiralities $\lambda$,
$\lambda=-s,-s+1,\dots,s$ are combined into one irrep.

The space of functions $f(x,\bz)$ contains functions transforming under
equivalent representations of the proper Poincar\'e group and is sufficiently
wide to define discrete transformations, including space reflection, time
reflection, and charge conjugation.
These discrete transformations associated with automorphisms of the group
also combine equivalent irreps of proper Poincar\'e group into one
representation of the extended group. For example, in 3+1 dimensions space
reflection combines two equivalent irreps of the proper group labelled by
$\lambda$ and $-\lambda$ into one irrep of the improper group.

Besides, as we will see below, the different types of RWE (finite-component
and infinite-component equations) are also associated with equivalent
representations in the decomposition of the left GRR.

Thus initially it is appropriate to consider all representations in the
decomposition of the scalar field on the Poincar\'e group, including
equivalent ones. In this sense we note the close analogy with the theory of
nonrelativistic three-dimensional rotator \cite{Wigne59,LanLi3,BieLo81}.
In the latter theory one considers functions on the rotation group $SU(2)$
and two sets of operators: angular momentum operators in an inertial
laboratory (space-fixed) frame (left generators $\hat J_i^L$) and angular
momentum operators in a rotating (body-fixed) frame (right generators
$\hat J_i^R$). The classification of the rotator states is based on the use
of the complete set of commuting operators which, apart from $\hat {\bf J}^2$ and
$\hat J_3^L$, includes also $\hat J_3^R$. Operator $\hat J_3^R$ distinguishes
equivalent representations in the decomposition of the left GGR of the
rotation group and corresponds to the quantum number which does not depend on
the choice of the laboratory frame. This quantum number plays a significant
role in the theory of molecular spectra.
In 3+1-dimensional case there exist two analogs of $\hat J_3^R$, namely
$\hat B_3^R=\hat S_{03}^R$ and $\hat S_3^R=\hat S_{12}^R$, which act in the
space of functions on the Poincar\'e group.
As we will see below, the first may be interpreted as a chirality operator,
and the second allows to distinguish particles and antiparticles.

\subsection {Quasiregular representations and spin description}

The consideration of GRR of the Poincar\'e group ensures the possibility
of consistent description of particles with arbitrary spin by means of
scalar functions on $\RS^d\times {\rm Spin}(D,1)$.
At the same time, for description of spinning particles it is possible to
use the spaces $\RS^d\times M$, where $M$ is some homogeneous space of the
Lorentz group (one or two-sheeted hyperboloid, cone, complex disk,
projective space and so on); see, for example,
\cite{BacKi69,Kihlb70,BoyFl74,Wig63,KimWi87,BieBrT88,HasSi92,KuzLy95,LyaSeS96,DerGi99,Drech97}
and \cite{JacNa91,Plyus91,Plyus92,CorPl96} for 3+1 and 2+1-dimensional cases
respectively. In some papers fields on homogeneous spaces are considered;
in other papers such spaces are treated as phase spaces of some classic
mechanics, and the latter are treated as models of spinning relativistic
particles.

These spaces appear in the framework of the next group-theoretical scheme.
Let us consider the left quasiregular representation of the Poincar\'e group
\be
T(g)f(g_0H)=f(g^{-1}g_0H), \quad H\subset {\rm Spin}(D,1).
\label{kvazi1}
\ee
$H$ is a subgroup of ${\rm Spin}(D,1)$, and since $x$ is invariant under
right rotations (see (\ref{Rreg1}))
$$
g_0\leftrightarrow (X,Z), \quad g_0H\leftrightarrow (X,ZH).
$$
Therefore the relation (\ref{kvazi1}) defines the representation of the
Poincar\'e group in the space of functions $f(x,zH)$ on
\be
\RS^d\times ({\rm Spin}(D,1)/H).
\ee
In the decomposition of the representation in the space of functions on
${\rm Spin}(D,1)/H$ (or $\RS^d\times ({\rm Spin}(D,1)/H)$) there is,
generally speaking, only part of irreps of the Lorentz (or Poincar\'e) group.
In particular, the case $H\sim {\rm Spin}(D,1)$ corresponds to scalar field
on Minkowski space. The classification and description of homogeneous spaces
of 3+1 Poincar\'e and Lorentz groups one can find in
\cite{Finke55,BacKi69,GelGrV66}.

Thus the consideration of quasiregular representations allows one to
construct a number of spin models classified by subgroups
$H\subset {\rm Spin}(D,1)$. But the existence of nontrivial subgroup $H$
leads to rejection of the part of equivalent (with different characteristics
with respect to the Lorentz subgroup) or, possibly, nonequivalent irreps of
the Poincar\'e group.

\subsection {Relativistic wave equations}

The problem of RWE construction for particles with arbitrary spin in various
dimensions is far from its completion and continues to attract significant
attention. To describe massive particles of spin $s$ in four dimensions one
usually employs the equations connected with the representations
$(\frac{s}{2}\,\frac{s}{2})$ and $(\frac{2s\pm 1}{4}\,\frac{2s\mp 1}{4})$ of
the Lorentz group (see, for example, \cite{Ohnuk88,BucKu95}). These equations
admit Lagrangian formulations \cite{FiePa39,SinHa74,SinHa74a}, but for $s>1$
minimal electromagnetic coupling leads to a noncasual propagation \cite
{Wight78,Zwanz78}. On the other hand, all known equations with casual
solutions either have a redundant number of independent components (as
equations \cite{Hurle71,Krugl99} for representations $(s\,0)$ and $(0\,s)$
have) or describe many masses and spins simultaneously, as Bhabha equations
\cite{Luban42,Bhabh45,KraNi77} do. Besides the problem of interaction of
higher spin fields, one may mention attempts to construct RWE with a
completely positive energy spectrum
\cite{Major32,GelMiS63,StoTo68,Dir71,Dir72a}
and RWE for fractional spin fields
\cite{JacNa91,Plyus91,Plyus92,GitSh97}.

With respect to mathematical methods used, it is possible divide all
approaches to RWE construction in three groups.

The first approach, which follows Refs. \cite{Dir36,FiePa39,RarSc41,BarWi48},
deals with equations for symmetric spin tensors. It allows one to describe
fields with fixed mass and spin and also to construct RWE which admit
Lagrangian formulation; however, as was mentioned above, for $s>1$ we face
the problem of noncasual propagation.

The second approach, which follows Refs. \cite
{Kemme39,Luban42,Bhabh45,Haris47,GelYa48,GelMiS63}, is devoted to studying
RWE of the form $(\alpha^\mu \hat p_\mu-\varkappa)\psi(x)=0$, and is based on
the use of algebraic properties of $\alpha$-matrices. These equations admit
Lagrangian formulation, however, for $s>1$ they describe a
nonphysical spectrum of particles: a decreasing mass with increasing spin.

The third approach is connected to the use of some supplementary variables
to describe spin degrees of freedom and initially was suggested for RWE
with a mass spectrum (see \cite{GinTa47,Ginzb56}). It was used for
constructing positive energy wave equations \cite{StoTo68,Dir71,Dir72a},
equations describing gauge fields \cite{Vasil92,Vasil96}, and for anyon
equations \cite{JacNa91,Plyus91,Plyus92,GitSh97}.

From the point of view of the approach which we developed above, the
problem of constructing RWE looks like a selection of invariant subspaces
in the space of functions on the group.

The classification of the scalar functions can be based on the use of the
operators $\hat C_k$ commuting with $T_L(g)$ (and correspondingly with all
the left generators). For these operators, as a consequence of a relation
$\hat C f(x,\bz)=cf(x,\bz)$, one can obtain that
$\hat C f'(x,\bz)=cf'(x,\bz)$, where $f'(x,\bz)=T_L(g)f(x,\bz)$. Therefore,
different eigenvalues $c$ correspond to subspaces, which are invariant with
respect to action of $T_L(g)$. The invariant subspaces correspond to
subrepresentations of the left GRR.

In addition to the Casimir operators, for the classification one may use
the right generators since all the right generators commute with all the left
generators. The right generators, as was mentioned, distinguish equivalent
representations in the decomposition of the left GRR.

There is some freedom to choose the commuting operators which are functions
of the right generators of the Poincar\'e group. We will use only functions
of the generators of the right rotations (\ref{Rgen}), in particular, for the
coordination with standard formulation of the theory.

Following the general scheme of harmonic analysis, for $M(D,1)$ one may
consider the system consisting of $d$ equations
\be \label{rveCaz}
\hat C_kf(x,\bz)=c_kf(x,\bz),
\ee
where $\hat C_k$ are the Casimir operators of the Poincar\'e group and of the
spin Lorentz subgroup. These operators constitute a subset of the complete set
of commuting operators on the Poincare group.
Just the system we will use for $d=2+1$ below.

On the other hand, there exist some additional requirements associated with
the physical interpretation. In the first place, in massive case the system
must be invariant under space reflection in order to describe states with
definite parity. Secondly, it is often supposed that the system contains an
equation of the first order in $\partial/\partial t$
(approach based on the first order equations advocated mainly in
\cite{Bhabh45,KraNi76,Birit79}).%
\footnote{
 As a consequence of relativistic invariance, a linear in
 $\partial/\partial t$ equation can be either first order or infinite order
 in space derivatives (square-root Klein-Gordon equation
 \cite{Suc63,Samar84,BriEnS91,Smith93}). The latter type of equations are
 naturally obtained in the theory of Markov processes for probability
 amplitudes \cite{She97}.
 }

The Casimir operators of the Poincar\'e group are the functions of the
generators $\hat p_\mu$ and $\hat J_{\mu\nu}$. In odd dimensions there exists
linear in $\hat p_\mu$ Casimir operator, since the invariant tensor
$\varepsilon^{\mu\dots\nu}$ has also odd number of indices. As we will see
below, in 2+1 dimensions the system (\ref{rveCaz}) is invariant under space
reflections.

In even dimensions the invariant tensor $\varepsilon^{\mu\dots\nu}$ also has
even number of indices, and therefore linear in $\hat p_\mu$ Casimir operator
does not exist. Besides, in even dimensions under space reflection irrep of
proper Poincar\'e group is mapped onto equivalent irrep labelled by another
eigenvalues of the Casimir operators of spin Lorentz subgroup. The linear
combinations of basis elements of these two irreps form the bases of two
labelled by intrinsic parity $\eta=\pm 1$ irreps of improper Poincar\'e group
including space reflection.

Nevertheless, in even dimensions there exists operator
$\hat C'= \hat p_\mu \hat\Gamma^\mu$, where $\hat \Gamma^\mu=
\hat \Gamma^\mu(\bz,\partial/\partial \bz)$, commuting with all left generators
and connecting the states which are interchanged under space reflections.
In contrast to the Casimir operators this operator is not a function of
generators of Poincar\'e group and does not commute with some right
generators. A first order equation
\be \label{rveDir}
\hat p_\mu \hat\Gamma^\mu f(x,\bz)=\varkappa f(x,\bz)
\ee
interlocks, at least, two irreps of the group $M(D,1)$ characterized by
different eigenvalues of the Casimir operator of spin Lorentz subgroup.
Equations (\ref{rveCaz}) and (\ref{rveDir}) have the same form; namely,
invariant operator acts on the scalar function $f(x,\bz)$ on the group $M(D,1)$.
The addition of the operators $\hat \Gamma^\mu$ means in fact the
extension of the Lorentz group up to more wide group
(in particular, in four dimensions to the 3+2 de Sitter group $SO(3,2)$).
Equation (\ref{rveDir}) replaces the equations of the system
(\ref{rveCaz}), which are not invariant under space reflection.

In the approach under consideration equations have the same form for all
spins. The separation of the components with fixed spin and mass is realized
by fixing eigenvalues of the Casimir operators of the Poincar\'e group
(or the operator $\hat p_\mu \hat\Gamma^\mu$).
Fixing the representation of the Lorentz group, under which $\phi(\bz)$
transforms in the decomposition
$$
f(x,\bz)=\phi^n(\bz)\psi_n(x),
$$
one can obtain RWE in standard multicomponent
form. This fixation is realized by the Casimir operator of spin Lorentz
subgroup.

There are two types of equations to describe one and the same spin,
one on functions $f(x,\bz)$, where $\phi^n(\bz)$ transforms under
finite-dimensional nonunitary irrep of the Lorentz group, and another on
functions $f(x,\bz)$, where $\phi^n(\bz)$ transforms under infinite-dimensional
unitary irrep of the Lorentz group.
In matrix representation these equations are written in the form of
finite-component or infinite-component equations respectively.
The latter type of equations (for example, Majorana equations
\cite{Major32,GelMiS63,Fradk66,StoTo68}) is interesting
because it gives the possibility to combine the relativistic invariance with
probability interpretation. Desirability of this combination was emphasized
in \cite{Dir72}.

\medskip

Let us briefly consider the possibility of existence of particles with
fractional spin. The restrictions on the spin value arise in the
representation theory of $M(D)$ and $M(D,1)$ if one restricts the
consideration by (1) unitary, (2) finite-dimensional (with respect to the
number of spin components) or (3) single-valued representations. (The latter
means that the representation acts in the space of single-valued functions.)
The restriction by single-valued functions (often supposed in mathematical
papers related to a classification of representations) is omitted in some
physical problems that allows to consider particles with fractional spin
(anyons). Thus, we will also consider multi-valued representations of $M(D)$
and $M(D,1)$ in the space of the functions $f(x,\bz)$ on the group.
These representations correspond to single-valued representations of the
universal covering group.

\section {Two-dimensional case}
\subsection {Field on the group $M(2)$}

In two-dimensional case the general formulas become simpler.
Matrices $U$ (\ref{XZM2}) of $SO(2)$ subgroup depend on only one parameter,
namely an angle $\phi$, $0\le\phi\le 4\pi$. Using the correspondence
$g_0\leftrightarrow(X,Z(\theta /2))$, $g\leftrightarrow(A,U(\phi /2))$
one may write the action of GRR:
\ba
&&T_L(g)f(x,\theta /2)=f(x',\theta /2 -\phi /2),
\\  \nonumber
&&\quad x_1'=(x_1-a_1)\cos \phi+(x_2-a_2)\sin \phi, \quad
x_2'=(x_2-a_2)\cos \phi-(x_1-a_1)\sin \phi ,
\\
&&T_R(g)f(x,\theta /2)=f(x'',\theta /2+\phi /2),
\\  \nonumber
&&\quad x_1''=x_1+a_1\cos \theta-a_2\sin \theta,
 \quad x_2''=x_2+a_2\cos \theta+a_1\sin \theta.
\ea
Left and right generators, which correspond to parameters $\theta$ and
$\phi$, are given by
\ba
&&\hat p_k=-i\partial _k,\quad \hat J=\hat L+\hat S,
\\
&&\hat p_k^R=i\Lambda _k^{\;\;i} \partial _i,\quad \hat J^R=-\hat S,
\ea
where
$$
\hat L=i(x_1\partial _2 -x_2\partial _1 )=-i\frac{\partial}{\partial \varphi },
\quad \hat S={-i} \frac{\partial}{\partial \theta} ,\quad \Lambda =
\left(\begin{array}{cc} \cos \theta  & \sin \theta  \\
                       -\sin \theta  & \cos \theta  \end{array} \right).
$$
The functions on the group are ones on $\RS^2\times S^1$, and
invariant measure on the group is
$$
d\mu (x,\theta)=(4\pi)^{-1}dx_1dx_2d\theta, \quad
-\infty <x<+\infty ,\quad 0\le\theta <4\pi.
$$
We will consider two complete sets of commuting operators:
$\hat p_1$, $\hat p_2$, $\hat S$ and $\hat { p}^2$, $\hat J$, $\hat S$.
The eigenfunctions of these operators are
\ba
&&\<x_1 x_2 \theta|p_1\, p_2\, s\> =
(2\pi)^{-1} \exp(ip_1x_1+ip_2x_2+is\theta ), \label{m2bas0}
\\
&&\<r\varphi \theta|p\, j\, s\> =
(2\pi)^{-1/2} i^lJ_l(pr)\exp(il\varphi)\exp(is\theta ), \label{m2bas1}
\ea
where $l=j-s$ is orbital momentum, $J_l(pr)$ is the Bessel function.
Irreps are labelled by eigenvalues $p^2$ of the Casimir operator $\hat p^2$.
For $p\ne 0$ the representation is irreducible, for $p=0$ it decomposes into
one-dimensional irreps of spin subgroup $U(1)$, which are labelled by
eigenvalues $s$ of the spin projection operator (or, simply speaking, spin
operator) $\hat S$.

At $p\ne 0$ the representations characterized by the spin $s$ and $s'=s+n$,
where $n$ is integer number, are equivalent. Really, operator $\hat S$
commutes with all left generators, but does not commute with the
generators of right translations, which mix spin and space coordinates.
Operators $\hat p^R_+=p_1^R-ip_2^R$ and $\hat p^R_-=p_1^R+ip_2^R$ are
raising and lowering operators with respect to spin $s$
\be
\hat p^R_\pm  |p_1\, p_2\, s\>=(ip_1 \pm p_2) |p_1\, p_2\, s\pm 1\>.
\ee
Right translations do not conserve both interval (distance) and spin $s$.

The functions (\ref{m2bas1}) satisfy the relations of orthogonality and
completeness
\ba
&&\int \<p\, j\, s|r\varphi \theta\>\<r\varphi \theta|p\, j\, s\>
 rdrd\varphi d\theta = \frac{\delta (p-p')}{p} \delta_{jj'}
 \delta_{ss'},
\\
&&\int \sum_{l,s}\<r\varphi \theta|p\, j\, s\>
\<p\, j\, s|r\varphi \theta\> d{p}
 =\frac {\delta (r-r')}r \delta (\varphi-\varphi ')\delta (\theta-\theta').
\ea
It means that we have obtained the decomposition of left regular
representation. Spin operator $\hat S$ distinguishes equivalent irreps
(except the case $p=0$, when irreps are labelled by its eigenvalues). The
decomposition of the functions of $\theta$ on the eigenfunctions of $\hat S$
corresponds to the Fourier series expansion of functions on a circle.

Thus the representations of $M(2)$ are single-valued for integer and
half-integer $s$. The fractional values of $s$ correspond to multi-valued
representations. Irreps are equivalent if are labelled by the same $p\ne 0$
and the difference $s-s'=n$ is an integer number.
For fixed $p\ne 0$ there are only two nonequivalent single-valued
representations, which correspond to integer and half-integer spin.
Nonequivalent multi-valued representations for fixed $p\ne 0$ are labelled
by $\tilde s \in [0,1)$, $\tilde s =s-[s]$.

\subsection {Field on the group $M(1,1)$}

Matrices $U$ (\ref{XZM11}) of $SO(1,1)$ subgroup, which is isomorphic to
an additive group of real numbers, depend on a hyperbolic angle $\phi$.
Using the correspondence
$g_0\leftrightarrow(X,Z(\theta /2))$, $g\leftrightarrow(A,U(\phi /2))$,
one may write the action of GRR:
\ba
&&T_L(g)f(x,\theta /2)=f(x',\theta /2-\phi /2),
\\ \nonumber
&&\quad x^{\prime 0}=(x^0-a^0)\cosh \phi+(x^1-a^1)\sinh \phi,\quad
 x^{\prime 1}=(x^1-a^1)\cosh \phi+(x^0-a^0)\sinh \phi,
\label{LM11}
\\
&&T_R(g)f(x,\theta /2)=f(x'',\theta /2+\phi /2),
\\ \nonumber
&&\quad x^{\prime\prime 0}=x^0+ a^0\cosh \theta - a^1\sinh \theta,\quad
 x^{\prime\prime 1}=x^1+a^1\cosh \theta - a^0\sinh \theta.
\nonumber
\ea
The functions on the group are ones on $\RS^2\times \RS$, and
invariant measure on the group can be written as
$$
d\mu (x,\theta)=dx^0dx^1d\theta, \quad
-\infty <x,\theta <+\infty.
$$
As above, we will consider two complete sets of commuting operators,
$\hat p_1$, $\hat p_2$, $\hat S$ and $\hat { p}^2$, $\hat J$, $\hat S$,
where $\hat J=\hat L+\hat S$, $\hat L=i(x^0\partial^0+x^1\partial^1)$,
$\hat S=-i\partial/\partial\theta$. The eigenfunctions of the first set are
\be
\<x^0 x^1\theta |p_1\, p_2\, \lambda\>=
(2\pi )^{-3/2} \exp(ip_\mu x^\mu +i\lambda\theta ), \label{m11bas0}
\ee
where $\lambda$ is an eigenvalue of the spin projection (chirality) operator
$\hat S$. The form of eigenfunctions of the second set depends on the type
of irrep. There are four types of unitary irreps labelled by eigenvalue
$m^2$ of operator $\hat p^2$ \cite{LShSh94}.

1. $m^2>0$. Representations correspond to the particles of nonzero mass, the
eigenfunctions of operators $\hat p^2$, $\hat J$, $\hat S$ are
\be
\<r\varphi \theta|m\, j\, \lambda\> = (4\pi)^{-1}
i\exp(\pi l/2) H_{il}^{(2)}(\pm mr)\exp(il\varphi)\exp(i\lambda\theta ),
\label{M11_1}
\ee
where $H_{il}^{(2)}(mr)$ is Hankel function, $r^2=(x^0)^2-(x^1)^2$, and
$\pm$ corresponds to the sign of energy $p_0$.

2. $m^2<0$. Representations correspond to tachyons, which in $d=1+1$ are more
similar to usual particles because of symmetry between space and time
variables. The form of $\<r\varphi \theta|m\, j\, \lambda\>$ coincides with
(\ref{M11_1}), but $m$ is imaginary.

3. $m=0$, $p_1=\pm p_0$. Representations correspond to the massless particles.
According to (\ref{Lregp}), for the action of finite transformations $T_0(g)$
on the functions $f(p, \pm p,\theta/2)$ one may obtain
$$
T_0(g)f(p,\pm p,\theta/2) = e^{iap'}f(p',\,\pm p',\, \theta /2-\phi /2),
\quad p'=e^{\mp \phi}p.
$$
Therefore the representation $T_0(g)$ is reducible and splits into four
irreps differed by the signs of $p_0$ and $p_1=\pm p_0$, and reducible
representation, which corresponds to $m=p_0=0$.

4. $m=p_0=0$. This representation decomposes into sum of one-dimensional
irreps of the group $SO(1,1)$, which are labelled by eigenvalues of $\hat S$.

There are no integer value restrictions for the spectrum of $\hat S$, and
chirality can be fractional, $-\infty <\lambda<+\infty $.
The decomposition of the functions $f(x,\theta)$ in therms of the
eigenfunctions of $\hat S$ corresponds to the Fourier integral expansion of
functions on a line. The equivalence of the representations characterized by
different $\lambda$ is related to the fact that, like in Euclidean case,
operator $\hat S$ does not commute with right translations.

One can convert vector indices into spinor indises and vice versa with the
help of the formula (\ref{spinor0}). In the case under consideration matrices
$U$ are real and symmeric, $X'=UXU$, or in component-wise form
$x^{\prime\nu} \sigma_{\nu\alpha\alpha'}=
U_\alpha^{\;\;\beta} \sigma_{\mu\beta\beta' } x^\mu U_{\;\;\alpha'}^{\beta'}$,
and there exists one type of spinor indeces only. Denoting elements of
the first column of matrix $Z$ transforming under spinor representation of
$SO(1,1)$ by $z_\alpha$, $z_1=\cosh (\theta/2)$, $z_2=\sinh (\theta/2)$,
we obtain for components of vector and antisymmetric tensor
\be  \label{11spvect}
q^\mu = \sigma^{\mu\alpha\beta }z_\alpha z_\beta, \quad
q^0=\cosh \theta, \quad q^1=\sinh \theta, \quad
q^{01}=\sigma^{01\alpha\beta}z_\alpha z_\beta =i.
\ee
There exist two invariant tensors $\eta^{\mu\nu}$ and $\varepsilon^{\mu\nu}$,
which can be used for raising of indices. This is related to the fact that
vectors $(x^0\, x^1)$ and $(x^1\, x^0)$ have the same transformation rule,
and one can construct invariant from two vectors by two different ways:
$\eta^{\mu\nu}q_\mu q'_\nu=\cosh(\theta-\theta')$,
$\varepsilon^{\mu\nu}q_\mu q'_\nu=\sinh(\theta-\theta')$.

\subsection {Relativistic wave equations in 1+1 dimensions}

An irrep of the group $M(1,1)$ can be extract from GRR by
fixing the sign of $p_0$ and eigenvalues of operators $\hat p^2$, $\hat S$,
\ba
&&\hat p^2 f(x,\theta) = m^2 f(x,\theta), \label{11m}
\\
&&\hat S   f(x,\theta) = \lambda  f(x,\theta), \label{11s}
\ea
where chirality $\lambda$ distinguishes equivalent irreps labelled by
identical eigenvalues $m^2$ of the Casimir operator $\hat p^2$. Solutions of
this system have the form $f(x,\theta)=\psi(x)e^{i\lambda\theta}$, where
$\hat p^2\psi(x) = m^2\psi(x)$.

According to (\ref{P}), space reflection converts $e^{i\lambda\theta}$ to
$e^{-i\lambda\theta}$. Irreps of the improper Poincar\'e group are labelled
by mass $m$, $\sign p_0$, intrinsic parity $\eta=\pm 1$, and spin
$s=|\lambda|$ (as above, $s$ distinguishes equivalent irreps). In the rest
frame it is easy to write down functions with mentioned characteristics:
\be \label{11par}
e^{\pm imx^0}(e^{i\lambda\theta}\pm e^{-i\lambda\theta}).
\ee
States with arbitrary momentum can be obtained from (\ref{11par}) by
hyperbolic rotations and form the basis of unitary irrep of improper group.
On the other hand, the problem arise to construct equations that unlike
the system (\ref{11m})-(\ref{11s}) are invariant under improper Poincar\'e
group and have solutions with definite parity. These equations should
combine states with chiralities $\pm \lambda$.

The general form of the linear in $\hat p^\mu$ equations is
\be  \label{eqs11}
\hat p_\mu \hat\Gamma^\mu f(x,\theta)=\varkappa f(x,\theta),
\ee
where $\hat\Gamma^\mu=\hat\Gamma^\mu (\theta,\partial /\partial\theta)$.
For invariance of (\ref{eqs11}) under space reflection $P$ and hyperbolic
rotations the operator $\hat p_\mu \hat\Gamma^\mu$ must commute
with $P$ and $\hat J=\hat L+\hat S$, whence
\be  \label{G11comm}
\hat\Gamma^\mu \stackrel{P}{\to}(-1)^{\delta_{1\mu}} \hat\Gamma^\mu, \qquad
[\hat\Gamma^0,\hat S]=i\hat\Gamma^1, \qquad
[\hat\Gamma^1,\hat S]=i\hat\Gamma^0.
\ee
The operators
\be \label{G11}
\hat\Gamma^0= s\cosh\theta- \sinh\theta \,\frac{\partial}{\partial\theta}, \quad
\hat\Gamma^1=s\sinh\theta-\cosh\theta \,\frac{\partial}{\partial\theta}, \quad
[\hat\Gamma^0,\hat\Gamma^1]=-i\hat S
\ee
obey these relations. One may construct the operators, which raise and lower
chirality $\lambda$ by 1,
\be
\hat\Gamma_+=\hat\Gamma^0-\hat\Gamma^1=
e^{-\theta} (s+{\partial}/{\partial\theta}), \quad
\hat\Gamma_-=\hat\Gamma^0+\hat\Gamma^1=
e^{\theta}(s-{\partial}/{\partial\theta}).
\ee
Operators $\hat \Gamma^0$, $\hat \Gamma^1$, and
$\hat\Gamma^2=-i\hat S=-\partial/\partial\theta$ obey the commutation
relations of the generators of the $SO(2,1)\sim SU(1,1)$ group:
$$
[\hat \Gamma^a,\hat\Gamma^b]=\epsilon^{abc}\hat\Gamma_c, \quad
\hat\Gamma_a=\eta_{ab}\hat\Gamma^b, \quad \eta_{00}=\eta_{22}=-\eta_{11}=1,
\quad \hat \Gamma_a\hat\Gamma^a=s(s+1).
$$

Thus, if the symmetry with respect to the space reflection takes place,
the condition of mass irreducibility (\ref{11m}) can be supplemented
by equation (\ref{eqs11}) instead of (\ref{11s}). This means the passage
to the new set of commuting operators, namely from $\hat p_\mu$,
$\hat S$ to $\hat p_\mu$, $\hat p_\mu\hat\Gamma^\mu$. Let us consider the system
\ba  \label{11m1}
&&\hat p^2 f(x,\theta)=m^2 f(x,\theta),
\\   \label{11s1}
&&\hat p_\mu \hat\Gamma^\mu f(x,\theta)=ms f(x,\theta).
\ea
The operator $\hat S$ does not commute with $\hat p_\mu \hat\Gamma^\mu$,
and the particle with nonzero mass described by equation (\ref{11s1})
can't be characterized by certain chirality. In the rest frame $p_0=\pm m$,
and the functions $f(x,\theta)=e^{\pm imx^0}\phi(\theta)$ should be
eigenfunctions of operator $\hat \Gamma^0$ with eigenvalues $\pm s$.
The equation
$$
\hat\Gamma^0\phi(\theta)=
(s\cosh \theta -(\sinh \theta )\partial/\partial\theta)\phi(\theta)
=\varkappa \phi(\theta)
$$
for $\varkappa=\pm s$ has solutions $[\cosh(\theta/2)]^{2s}$ and
$[\sinh(\theta/2)]^{2s}$ respectively. Below we will consider two cases.

1. The solutions of the system (\ref{11m1})-(\ref{11s1}) are sought in the
space of polynomials of $e^{-\theta/2}$ and $e^{\theta/2}$ that correspond
to finite-dimensional nonunitary representations of $SU(1,1)$.
Corresponding representations of $SO(1,1)$ subgroup are also nonunitary.
For these representations generator $\hat S$ is anti-Hermitian,
and it is convenient to redefine chirality operator as $i\hat S$.
In the rest frame a general solution of the system (\ref{11m1})-(\ref{11s1}) is
\be  \label{plain11}
f(x,\theta)= C_1 e^{imx^0}[\cosh(\theta/2)]^{2s}
           + C_2 e^{-imx^0}[\sinh(\theta/2)]^{2s},
\ee
where $2s$ is integer positive number.
Therefore for an unique spin $s$ there are only two independent components
(with positive and negative frequency).
The space inversion takes $\theta$ to $-\theta$, and in the rest frame
solutions with different sign of $p_0$ and half-integer $s$ are characterized
by opposite parity $\eta$. For integer $s$ all solutions are characterized
by $\eta=1$. Plane wave solutions, which correspond to moving particle, can
be obtained from (\ref{plain11}) by a hyperbolic rotation by the angle
$2\phi$:
$$
f_{m,s}(x,\theta)=
C_1 e^{ik_0x^0+ik_1x^1} \left( \cosh[(\theta+\phi)/2] \right)^{2s} +
C_2 e^{-ik_0x^0-ik_1x^1} \left(\sinh[(\theta+\phi)/2] \right)^{2s},
$$
where $k_0=m\cosh 2\phi$, $k_1=m\sinh 2\phi$.

In the ultrarelativistic limit $\phi\to\pm\infty$ we have two states with
chirality $\lambda=\pm s$ respectively.
Thus, if in the rest frame one may distinguish two components with positive
and negative frequency, then in massless limit one may distinguish two
components with positive and negative chirality.

Matrix form of the system (\ref{11m1})-(\ref{11s1}) can be obtained by the
decomposition of $f(x,\theta)$ over the basis $e^{\lambda\theta/2}$,
$\lambda=-s,-s+1,\dots,s$. There are $2s+1$ components $\psi(x)$ in this
form, but only two of them are independent.
Notice that representations of $SO(1,1)$ of the form $e^{\lambda\theta}$,
are nonunitary for real $\lambda$ and integral over $\theta$ is divergent.
One can redifine the norm of a state with the help of scalar product
in the space of multicomponent functions $\psi(x)$, but this product
is not positive definite.

For $s=1/2$, substituting the function
$f(x,\theta)=\psi_1(x) e^{\theta/2}+\psi_2(x) e^{-\theta/2}$ into equation
(\ref{11s1}), we obtain two-dimensional Dirac equation \cite{AbdAbR91}
\be
\hat p_\mu\gamma^\mu \Psi(x) = m \Psi(x), \qquad
\gamma^0=\sigma_1, \quad \gamma^1=-i\sigma_2, \quad 2\hat S=\gamma^3=\sigma_3.
\ee
where $\Psi(x)=(\psi_1(x)\;\psi_2(x))^T$.
Matrix $\gamma^3=\gamma^0\gamma^1$ corresponds to chirality operator and
satisfies the condition $[\gamma^3,\gamma^\mu ]_+=0$. On the other hand,
this matrix corresponds to hyperbolic rotation, and similarly to 3+1 case
one can write $\gamma^\mu\gamma^\nu=\eta^{\mu\nu}-i\sigma^{\mu\nu}$, where
$\sigma^{01}=i\gamma^3$. Invariant scalar product has the form
$|\psi_1(x)|^2-|\psi_2(x)|^2$.

For $s=1$, substituting the function
$f(x,\theta)=\psi_{11}(x) e^{\theta}+\psi_{12}(x)+\psi_{22}(x) e^{-\theta}$
into equation (\ref{11s1}), we obtain
\ba  \nonumber
&&(\hat p_\mu\Gamma^\mu -m)\Psi(x)=0, \quad
\\
&&\Gamma^0 = \frac 1{\sqrt{2}} \left( \begin{array}{ccc}
 0 & 1 & 0 \\ 1 & 0 & 1 \\ 0 & 1 & 0 \end{array} \right), \quad
\Gamma^1 = \frac 1{\sqrt{2}} \left( \begin{array}{ccc}
 0 & -1 & 0 \\ 1 & 0 & -1 \\ 0 & 1 & 0 \end{array} \right), \quad
\hat S =   \left( \begin{array}{ccc} 1 & 0 & 0 \\ 0 & 0 & 0 \\ 0 & 0 & -1
\end{array} \right), \label{M11spin1}
\ea
where $\Psi(x)=(\psi_{11}(x)\;\psi_{12}(x)/\sqrt{2}\;\psi_{22}(x))^T$.
Using (\ref{11spvect}) to convert spinor indices to vector ones:
${\cal F}_0=\psi_{22}-\psi_{11}$, ${\cal F}_1=\psi_{22}+\psi_{11}$,
$F_{01}=-F_{10}=-i\psi_{12}$, we obtain
$p_0{\cal F}_1 - p_1{\cal F}_0 = -im F_{01}$, $ip_0 F_{10}=m {\cal F}_1$,
$ip_1 F_{10}=m {\cal F}_0$.
Thus one can rewrite 1+1 "Duffin--Kemmer" equation (\ref{M11spin1}) in the
form, which is similar to Proca equations for in 3+1 dimensions
(see (\ref{31fDK}),(\ref{DK})),
\be  \label{M11spin1a}
\partial_\mu{\cal F}_\nu - \partial_\nu{\cal F}_\mu = mF_{\mu\nu}, \qquad
\partial^\nu F_{\mu\nu} = m{\cal F}_\mu.
\ee
As a consequence of (\ref{M11spin1a}) we obtain
$\partial_\mu{\cal F}^\mu=0$, $(\hat p^2 -m^2){\cal F}^\mu=0$.
But 1+1-dimensional case is distinctly different from 3+1-dimensional one
because the component $F_{01}=-F_{10}$ is characterized by zero chirality and
thus the roles of $F_{\mu\nu}$ and ${\cal F}_\nu$ are interchanged.

In the massless case the system (\ref{M11spin1a}) splits into two independent
equations for the components ${\cal F}_\mu$ and $F_{\mu\nu}$ respectively,
\ba \label{11max}
&&\partial_\mu{\cal F}_\nu - \partial_\nu{\cal F}_\mu=0, \qquad
\\ \label{11maxw}
&&\partial^\mu F_{\mu\nu}=0.
\ea
First equation has propagating solutions
$$
{\cal F}_0 = C_1e^{ip(x^0+x^1)} + C_2e^{ip(x^0-x^1)} ,\qquad
{\cal F}_1 = C_1e^{ip(x^0+x^1)} - C_2e^{ip(x^0-x^1)}
$$
obeying transversality condition  $\partial_\mu{\cal F}^\mu=0$.
Second equation (free two-dimensional Maxwell equation \cite{AbdAbR91})
corresponds to the components with zero chirality and has trivial solution
$F_{\mu\nu}=const$ only.
Notice that for real field $f^*(x,\theta)=f(x,\theta)$ components
${\cal F}_\mu$ and $F_{\mu\nu}$ also are real, and propagating solutions
do not exist for $m=0$.

If for $s=1/2$ and for $s=1$ the first equation of the system
(\ref{11m1})-(\ref{11s1}) is the consequence of the second equation,
then for $s>1$ there are the solutions of equation (\ref{11s1}) with mass
spectrum, $m_i |s_i|=ms$, $s_i=s,\,s-1,\dots,-s$.
For the extraction of characterized by certain mass $m$ and spin $s$
representations of improper Poincar\'e group it is necessary to use both
equations of the system.

Notice that the chirality $\lambda$ of a particle described by
(\ref{11m})-(\ref{11s}) can be fractional, but the spin $s$ of a particle
described by (\ref{11m1})-(\ref{11s1}) can be only integer or half-integer
for $m\ne 0$ and finite number of components $\psi(x)$.

Really, if $2s$ is not integer, then acting by the raising operator on the
state with label $\lambda=-s$, we not get into the state labelled by
$\lambda=s$ and connected with initial state by the space reflection;
moreover, the spectrum of $\lambda$ is not bounded above.
On the other hand, it is possible to develop an alternative approach
(in particular, for the massive particles with fractional spin) based on the
using of infinite-dimensional unitary irreps of $SO(2,1)$.

2. Let us consider now the solutions of (\ref{11m1})-(\ref{11s1}) in the
space of the square-integrable functions of $\theta$. In the rest frame, as
we have seen above, there are two types of the solutions.
The solutions $[\sinh(\theta/2)]^{2s}$ are not square-integrable for any $s$
since corresponding integral is divergent either at zero or at infinity.
The solutions $[\cosh(\theta/2)]^{2s}$ for $s<0$ are square-integrable:
$$
\int_{-\infty}^{+\infty}[\cosh(\theta/2)]^{4s}d\theta = 2B(1/2,\, 2s).
$$
Therefore in the space of square-integrable functions equation
(\ref{11s1}) has only positive energy solutions.
Solutions with $p_0<0$ correspond to the equation
$\hat p_\mu \hat\Gamma^\mu f(x,\theta)=-ms f(x,\theta)$.
Normalized positive energy solutions of the system (\ref{11m1})-(\ref{11s1})
for the particle with spin $|s|$ and momenta $p_0=m\cosh \phi$,
$p_1=m\sinh \phi$ are
\be  \label{11Major}
f(x,\theta)=(2\pi)^{-1}\left( 2 B(1/2,\, 2s)\right)^{-1/2}
e^{ip_0x^0+ip_1x^1} \left( \cosh[(\theta+\phi)/2] \right)^{-2|s|}.
\ee
In contrast to the case $d>2$ solutions with distinct $s$ are nonorthogonal.
The decomposition of the solutions (\ref{11Major}) over the functions
$e^{i\lambda\theta}$ (i.e. over $SO(1,1)$ unitary irreps) corresponds to the
Fourier integral expansion.
We will consider properties of the positive energy equations more detail
in 2+1-dimensional case below.

\section{Three-dimensional case}
\subsection {Field on the group $M(3)$}

The case of $M(3)$ group is characterized by many-dimensional spin space.
On the other hand, the constructions allow the simple physical interpretation.

Using the operators
$\hat J^{i} =\hat L^{i}+\hat S^{i} =(1/2)\epsilon^{ijk}\hat J_{jk}$,
it is possible to rewrite the commutation relations (\ref{komm0}) in the more
compact form
\be
[\hat{p}_{i},\hat{p}_{k}]=0,\quad [\hat{p}^{i},\hat{J}^{j}]
=i\epsilon ^{ijk}\hat{p}_{k}, \quad [\hat{J}^{i},
\hat{J}^{j}]=i\epsilon ^{ijk}\hat{J}_{k}\;.
\label{M3comm}
\ee
The invariant measure on the group is given by the formulas
\ba \label{M3mera}
d\mu(x,\bz) = C d^3x \delta (|z_1|^2+|z_2|^2-1) d^2z_1 d^2z_2=
\frac 1{16\pi^2}d^3x \sin \theta d\theta d\phi d\psi.
\\ \nonumber
-\infty <x<+\infty ,\quad 0<\theta <\pi ,\quad 0<\phi <2\pi ,\quad
-2\pi <\psi <2\pi ,
\ea
where $z_1=\cos \frac\theta 2 e^{i(\psi -\phi)/2}$,
$z_2=i\sin \frac\theta 2 e^{i(\psi +\phi)/2}$ are the elements of the
first column of matrix (\ref{Z3}), $z^2=-z_1$, $z^1=z_2$, and
$\theta ,\phi ,\psi$ are the Euler angles.
The spin projection operators acting in the space of the functions on the
group $f(x,\bz)$ have the form
\ba
&&\hat S_k=\frac 12(z\sigma_k\partial_z -
 \cc z\cc\sigma_k\partial_{\ccc z}\,),
 \quad z=(z^1\; z^2),\;
 \partial_z=(\partial/\partial{z^1} \;\partial/\partial{z^2})^T, \quad
 \nonumber
\\
&&\hat S^R_k = -\frac 12
 (\chi \cc\sigma_k\partial_\chi -\cc \chi
 \sigma_k\partial_{\ccc \chi}\,),  \quad \chi =(z^1\; -\!\cc z^{\dot 2}), \;
 \partial _\chi =(\partial/\partial{z^1}\;
 -\partial/\partial{\cc z^{\dot 2}})^T.
\label{M3S}
\ea
In terms of Euler angles one can obtain
\be
\hat S_3=-i\partial /\partial \phi, \quad \hat S_3^R=i\partial /\partial \psi.
\label{M3S3}
\ee

The operator $\hat {\bf p}^2$ and the operator of the spin projection on the
direction of propagation $\hat W={\bf \hat p\hat J}={\bf \hat p\hat S}$ are
the Casimir operators. The eigenvalues $S(S+1)$ of the Casimir operator of
rotation subgroup in $z$-space $\hat {\bf S}^2=\hat {\bf S}_R^2$ define
spin $S$.
complete sets of the commuting operators
$\{\hat p_k, \hat W, \hat {\bf S}^2, \hat S_R^3\}$,
$\{\hat {\bf p}^2,\hat W,\hat {\bf J}^2,\hat S_3,\hat {\bf S}^2, \hat S^R_3\}$
consist of six operators (two Casimir operators, two left generators, and two
right generators). The Casimir operator
$\hat W$ does not commute with $\hat L_k$ and $\hat S_k$ separately but
only with the generators $\hat J_k=\hat L_k+\hat S_k$, therefore there are
sets, which do not include $\hat W$, for example,
$\{\hat {\bf p}^2,\hat p_3,\hat L_3, \hat S_3,\hat {\bf S}^2, \hat S^R_3\}$
and $\{\hat p_\mu, \hat S_3, \hat {\bf S}^2, \hat S^R_3 \}$.

We will consider the first set since in this case eigenfunctions have the
most simple form. This set includes two Casimir operators, the operator of
spin square $\hat {\bf S}^2$ and the generator $\hat S^R_3$.
The latter two generators commute with all left generators but do not
commute with right generators and label equivalent representations in the
decomposition of the left GRR.

According to (\ref{M3S3}), the eigenfunctions of $\hat S^R_3$,
$\hat S^R_3|\dots n\>=n|\dots n\>$, have the form
$|\dots n\>=F(x,\theta,\phi)\exp(-in\psi )$ and are differed only
by a phase factor. As a consequence of the commutation relations of
generators $\hat S_k^R$ the operators
$\hat S^R_\pm =\hat S^R_1\pm i\hat S^R_2$ are the raising and lowering
operators for the eigenfunctions of $\hat S^R_3$
\be
\hat S^R_\pm |\dots n\> = C(S,n)|\dots n\pm 1 \>.
\label{sr_mix}
\ee
The intertwining operators $\hat S^R_\pm$ consist of the generators of right
rotations, which conserve the interval square according to (\ref{Rarg2}).
Moreover, the right rotations do not act on $x$. But there are no
transformations (rotations and translations) of the reference frame,
which connect the representations with different $n$.
Notice that the states labelled by $n$ and $-n$ are interchanged under
charge conjugation (\ref{C}).

The operator $\hat {\bf S}^2$ also labels equivalent representations of
$M(3)$ group. This operator commutes with all generators except right
translations, and therefore an intertwining operator is a function of the
latter generators. Right translations change both the interval and spin.
Therefore it is naturally to characterize free particle in three-dimensional
Euclidean space not only by momentum and spin projection on the direction of
propagation, but also by spin $S$.

There are two standard realizations of the representation spaces, which
correspond to eigenvalues $n=\pm 2S$ and $n=0$ of the operator $\hat S^R_3$.

The first realization is the space of analytic ($n=-2S$) functions $f(x,z)$
or antianalytic ($n=2S$) functions $f(x,\cc z)$ of two complex variables
$z^1, z^2$, $|z^1|^2+|z^2|^2=1$, i.e. the space of functions of two-component
spinors. In particular, according to (\ref{M3S}), for the space of analytic
functions we have
\be
\hat S_k = \frac 12 z\sigma_k \partial_z,
\label{M3SS}
\ee
$\hat S^R_3=-(z^1\partial/\partial z^1 + z^2\partial/\partial z^2$),
and $\hat {\bf S}^2= \hat S^R_3 (\hat S^R_3 - 1)$. The eigenfunctions of
the operator of spin square are polynomials of the power $2S$ in $z^1$, $z^2$.
The charge conjugation transformation connects equivalent irreps labelled by
$n=\pm 2S$ and the spaces of analytic and antianalytic function.
This transformation reverses the direction of momentum and spin.

The second realization is the space of functions, which do not depend on
the angle $\psi$, and corresponds to $n=0$.
It is the space of functions of five real variables on the manifold
$$
\RS^3\times S^2, \qquad d\mu = (4\pi)^{-1}d^3x\sin\theta d\theta d\phi .
$$
The point in the spin space (i.e. on the sphere $S^2\sim \CS P^1\sim SU(2)/U(1)$)
can be define by the spherical coordinates $\theta, \phi$, or by two complex
variables $z_1=\cos \frac\theta 2 e^{-i\phi /2},
z_2=\sin \frac\theta 2 e^{i\phi /2}$ (in this case one may use (\ref{M3SS})
for the spin projection operators), or by one complex number $z=z_1/z_2$
(this case corresponds to the realization in terms of projective space
$\CS P^1$).
In terms of variables $\theta, \phi$ the eigenfunctions of operators
$\hat S$, $\hat S_3$ are $P^s_S(\cos\theta )e^{is\phi}$, where
$P^s_S(\cos\theta )$ are associated Legendre functions \cite{Vilen68t}.

Let us consider eigenfunctions of the set of the operators
$\{\hat p_\mu, \hat W, \hat{\bf S}^2 \}$ in the space of analytic functions
of $z^1$, $z^2$:
\be   \label{systM3}
 \hat p_\mu f(x,z) = p_\mu f(x,z), \quad
 \hat {\bf S}^2 f(x,z) = S(S+1)f(x,z), \quad
 {\bf \hat p\hat S} f(x,z) = ps f(x,z).
\ee
The eigenfunctions of $\hat {\bf S}^2$ are polynomials of the power $2S$ in
$z$ (the unitary irreps of $SU(2)$ are finite-dimensional, therefore spin $S$
and spin projection on the direction of propagation $s$ are integer or
half-integer). Let $p_\mu =(0,0,p)$, then the normalized solutions of
the system (\ref{systM3}) are
$$
|0\,0\,p\,S\,s\>=(2\pi)^{-3/2}\left( \frac{(2S)!}{(S+s)!(S-s)!} \right)^{1/2}
(z^1)^{S+s} (z^2)^{S-s}e^{ix_3p}.
$$
The states with arbitrary direction of vector $p$ may be obtain by the
rotation $P=UP_0U^\dagger$,
$Z=UZ_0$, $P_0=p\sigma_3$, $Z_0=(z_1 \; z_2)^T$,
\be
|p_1\,p_2\,p_3\, S\, s\>=
(2\pi )^{-3/2}\left( \frac{(2S)!}{(S+s)!(S-s)!}\right)^{1/2}
(z^1 \cc u_1+z^2 \cc u_2)^{S+s} (-z^1 u_2+z^2 u_1)^{S-s}e^{ipx},
\ee
where $u_1$, $u_2$ are the elements of the first line of matrix $U$.
Notice that it is sufficient to use only two angles for the parametrization
of matrix $U$ since the initial state has a stationary subgroup $U(1)$.

For the rest particle $\hat {\bf p}^2=\hat {\bf p}\hat {\bf S} =0$ and only
in this case $M(3)$ irreps labelled by different $S$ are nonequivalent.

In general case functions corresponding to the particle of spin $S$
have the form
\ba \label{M3psi}
&&f_S(x,z)=\sum_{n=0}^{2S}\phi^n(z)\psi_n(x),
\quad \phi^n(z)=(C^n_{2S})^{1/2}(z^1)^{S-n}(z^2)^n,
\\
&&\int \cc f_S(x,z) f'_{S'}(x,z)d\mu(x,z)=
  \delta_{SS'}\int\sum_{n=0}^{2S}\cc\psi_n(x)\psi'_n(x) d^{\,3}x ,
   \label{M3int}
\ea
where $C_n^{2S}$ is the binomial coefficient, and $d\mu(x,z)$ is the invariant
measure (\ref{M3mera}). The relation (\ref{M3psi}) gives the connection
between the description by the functions $f(x,z)$ and the standard
description by the multicomponent functions $\psi_n(x)$.
It is easy to see that the action of the operators
$\hat S_k = \frac 12 z\sigma_k \partial_z$ on the function (\ref{M3psi})
reduces to the multiplication of the column $\psi(x)$ by $(2S+1)\times (2S+1)$
matrices $S_k$ of $SU(2)$ generators in the representation $T_S$,
$\hat S_k f(x,z)=\phi(z)S_k\psi(x)$.
Matrices $S_k$ obey the commutation relations of spin projection operators,
namely $[S^i,S^j]=i\epsilon ^{ijk} S_k$.

In particular, the linear function of $z^1$,$z^2$ corresponds to spin $S=1/2$,
and the action of the operators $\hat S_k$ on $\psi(x)$ is reduced to
the multiplication by $\sigma$-matrices,
$\hat S_k f(x,z)=\phi(z)\sigma_k\psi(x)$.

As was mentioned above, the operator $\hat {\bf S}^2$ is not a Casimir
operator of $M(3)$ and labels equivalent representations of the group.
This operator is the direct analog of the Lorentz spin operator in
pseudoeuclidean case, and we will consider its properties in details.

1. Operator $\hat {\bf S}^2$ is composed of right generators commuting with
all left generators and therefore is not changed under the
coordinate transformation (left transformations of the Euclidean group).
The right transformations do not change the spin projection $s$ on the
direction of propagation but change both spin $S$ and interval (distance).

2. Operator $\hat {\bf S}^2$ does not depend on $x$ and commutes with
operators $x_k,\hat p_k,\hat S_k$, therefore in the presence of interactions
is conserved for any Hamiltonian $\hat H=\hat H(x_k,\hat p_k,\hat S_k)$.

3. The eigenvalues of $\hat {\bf S}^2$ label irreps of the rotation subgroup
in the spin space and define the possible values of the spin projection $s$,
which can arise under the interactions.

Notice that in the representation theory of Galilei group (symmetry group of
nonrelativistic mechanics, which includes $M(3)$ and ensures more general
description) irreps labelled by different eigenvalues of $\hat {\bf S}^2$
are not equivalent. The classification of irreps of Galilei group can be
based on the use of two invariant equations. The Schr\"{o}dinger equation
fixes the mass $m$, and the second equation fixes the eigenvalue of spin
operator $\hat {\bf S}^2$ \cite{Hamer60,Levy63}.

\subsection {Field on the group $M(2,1)$ and fractional spin}

Using the operators
$\hat J^{\rho}=\hat L^{\rho}+\hat S^{\rho}
 =(1/2)\epsilon^{\rho\mu\nu}\hat J_{\mu\nu}$,
it is possible to rewrite the commutation relations (\ref{komm0}) in the
next form:
\be
[\hat{p}_{\mu },\hat{p}_{\nu }]=0,\quad [\hat{p}^{\mu },\hat{J}^{\nu}]
=-i\epsilon ^{\mu \nu \eta }\hat{p}_{\eta }, \quad [\hat{J}^{\mu },
\hat{J}^{\nu }]=-i\epsilon ^{\mu \nu \eta }\hat{J}_{\eta }\;.
\label{M21comm}
\ee
The invariant measure on the group is given by the formulas \cite{Vilen68t}
\ba
d\mu(x,\uz) = d\mu(z)d^3x = C d^3x \delta (|z_1|^2 -|z_2|^2-1) d^2z_1 d^2z_2=
\frac 1{8\pi^2}d^3x \sinh \theta d\theta d\phi d\psi.
\label{21mera}
\\ \nonumber
-\infty <x<+\infty ,\quad 0<\theta <\infty ,\quad 0<\phi <2\pi ,\quad
-2\pi <\psi <2\pi ,
\ea
where $z_1=\cosh \frac\theta 2 e^{i(\psi-\phi)/2}$,
$z_2=\sinh \frac\theta 2 e^{i(\psi+\phi)/2}$ are the elements of the
first column of matrix $Z$ (\ref{Z3}), and $\theta ,\phi ,\psi $ are the
analogs of Euler angles, $z^2=-z_1$, $z^1=z_2$.
The spin projection operators acting in the space of the functions on the
group $f(x,\bz)$ have the form
\ba
&&\hat S^\mu=\frac 12(z\gamma^\mu\partial_z -
 \cc z \cc\gamma^\mu \partial_{\ccc z}\,),  \quad z=(z^1\; z^2),\;
 \partial_z=(\partial/\partial{z^1}\;\partial/\partial{z^2})^T,
 \nonumber
\\
&&\hat S_R^\mu = -\frac 12 (\chi \cc\gamma^\mu\partial_\chi
 -\cc \chi \gamma^\mu\partial_{\ccc \chi}\,), \quad
 \chi =(z^1\;\cc z^{\dot 2}), \;
 \partial_\chi=(\partial/\partial{z^1}\; \partial/\partial{\cc z^{\dot 2}})^T,
 \label{M21S}
\ea
where $\gamma^\mu$ are three-dimensional $\gamma$-matrices,
\be \label{3gamma}
\gamma^\mu =(\sigma_3,i\sigma_2,-i\sigma_1), \qquad
\gamma^\mu \gamma^\nu=\eta^{\mu\nu}-i\varepsilon^{\mu\nu\rho}\gamma_{\rho}.
\ee
Notice that nonequivalent set of $\gamma$-matrices,
$\gamma^\mu \gamma^\nu=\eta^{\mu\nu}+i\varepsilon^{\mu\nu\rho}\gamma_{\rho}$,
is used in some papers.
In terms of the Euler angles one may obtain
$\hat S^0=-i\partial /\partial \phi$, $\hat S^0_R=i\partial /\partial \psi$.
The sets of commuting operators are the same as in Euclidean case.

In consequence of the identity $\sigma_1\cc U\sigma_1=U$ one may show that
matrix $\sigma_1$ is the invariant symmetric tensor converting
dotted and undotted indices,
\be  \label{21sig}
\cc z_\alpha =(\sigma_1)_\alpha^{\;\;\da} \cc z_\da.
\ee

According to (\ref{transf21}), the invariant tensor $\sigma_{\mu\alpha\da }$
connects vector index and two spinor indices of different types.
On the other hand, using the identity mentioned above one may rewrite
(\ref{transf21}) in the form
$x^{\prime\nu}(\sigma_\mu \sigma_1)=x^\mu U(\sigma_\mu \sigma_1)U^T$.
Thus the invariant tensor, which we denote as
\be  \label{3sig}
\check\sigma_{\mu\alpha\beta}=(\sigma_\mu \sigma_1)_{\alpha\beta}, \quad
\check\sigma_{\mu\alpha\beta}=\check\sigma_{\mu\beta\alpha},
\ee
connects vector index and two spinor indices of one type.
Thus, one can write the generators $\hat S^\mu$ in the form
$\hat S^\mu=\qq \check\sigma^\mu_{\;\;\alpha\beta}
(z^\alpha\partial^\beta+\cc z^\alpha \cc\partial^\beta)$.
An analog of $\sigma^{\mu\nu}$-matrices in 2+1 dimensions is
$(\sigma^{\mu\nu})_{\alpha\beta}=
\varepsilon^{\mu\nu\lambda} \check\sigma_{\lambda\alpha\beta}$.
Raising one of the spinor indices of $\check\sigma_{\mu\alpha\beta}$,
one may obtain two sets of three-dimensional $\gamma$-matrices
differed only by signs of $\gamma^0$ and $\gamma^2$.

Similarly to the Euclidean case, there are two standard realizations of
the representation spaces. These realizations correspond to eigenvalues
$n=\pm 2S$ and $n=0$ of the operator $\hat S^R_3$.

The first realization is the space of analytic ($n=-2S$) functions $f(x,z)$
or antianalytic ($n=2S$) functions $f(x,\cc z)$ of two complex variables
$z^1, z^2$, $|z^2|^2-|z^1|^2=1$, i.e. the space of functions of
two-component spinors. The eigenfunctions of $\hat S_\mu\hat S^\mu$ are
homogeneous functions of degree $2S$ in $z$.
According to (\ref{M3S}), we have
$\hat S_R^0=-(z^1\partial/\partial z^1 + z^2\partial/\partial z^2)$
for the space of analytic functions and
$\hat S_R^0=\cc z^{\dot 1}\partial/\partial \cc z^{\dot 1}
  +\cc z^{\dot 2}\partial/\partial \cc z^{\dot 2}$
for the space of antianalytic functions.
The eigenfunctions of $\hat S_\mu\hat S^\mu$ in these spaces are also
eigenfunctions of $\hat S_R^0$ with eigenvalues $n=\mp 2S$ respectively.

The second realization is the space of eigenfunctions of $\hat S_R^0$ with
zero eigenvalue. It is the space of functions of five real parameters on
the manifold
$$
\RS^3\times \CS D^1, \qquad d\mu = (2\pi)^{-1}d^3x\sinh\theta d\theta d\phi ,
$$
where $\CS D^1\sim SU(1,1)/U(1)$ is a complex disk.
This functions do not depend on the angle $\psi$.

Remember some facts from the representation theory of $SU(1,1)$.
For finite-dimensional nonunitary irreps $T^0_S$ of 2+1 Lorentz group
$SU(1,1)\sim SO(2,1)$ spin projection $s$ (the eigenvalue of $\hat S^0$)
can be only integer or half-integer, $s=-S,\dots,S$, where $S\ge 0$.
However, in 2+1 dimensions Lorentz group has not compact non-Abelian subgroup.
Therefore there are infinite-dimensional unitary representations
corresponding to fractional $S$. These representations are multi-valued
representations of $SU(1,1)$.
For single-valued representations of $SU(1,1)$ the spin projection $s$ can
be only integer or half-integer (for $SO(2,1)$ only integer).

The representations of discrete seria correspond to $S<-1/2$.
Irreps of the positive discrete series $T^+_S$ are bounded by lowest weight
$s=-S$, irreps of the negative discrete series $T^-_S$ are bounded by highest
weight $s=S$, irreps of the principal series correspond to $S=-1/2+i\lambda$
and can be bounded by highest (lowest) weight only for $S=-1/2$.
For other irreps of principal series the spectrum of $s$ is not bounded.
Supplementary series correspond to $-1/2<S<0$ and are characterized by
nonlocal scalar product.

A visual picture for weight diagrams of all seria on the plain
$S,s$ one can find in \cite{GitSh97,Wybou74}.

Thus there are only two possibilities for description of a particle with
fractional spin by means of unitary irreps of $SU(1,1)$ with local scalar
product. The first corresponds to the discrete or principal seria irreps
bounded by lowest (highest) weight, $|s|\ge|S|\ge 1/2$.
The second corresponds to the principal series irreps which is not bounded.

Unitary irreps of discrete seria are used for the description of anyons
\cite{JacNa91,Plyus92,CorPl94,GitSh97}.
In \cite{JacNa91,Plyus92,CorPl94} corresponding unitary infinite-component
representations of $M(2,1)$ were constructed in the space of functions of
$x^\mu$ and complex variable $z=z^1/z^2$, i.e. on the coset space
$M(2,1)/U(1)$.
It was shown that RWE associated with irreps of the discrete seria have
solutions only with definite sign of energy. Thus mentioned RWE are analogs
of Majorana equations in 3+1 dimensions; this aspect was considered more
detail in \cite{CorPl94}.
Various formulations of the higher spin theory based on
finite-component representations were considered, in particular, in
\cite{DesKa83,Deser84,AraDe84,GitTy97,VasTy97}.

\subsection {Relativistic wave equations in 2+1 dimensions}

Let us fix eigenvalues of the Casimir operators of the Poincar\'e group and
of spin Lorentz subgroup:
\ba
&&\hat p^2 f(x,\bz) = m^2 f(x,\bz), \label{12m}
\\
&&\hat p_\mu\hat S^\mu  f(x,\bz) = K f(x,\bz), \label{12s}
\\
&&\hat S_\mu\hat S^\mu f(x,\bz) = S(S+1)f(x,\bz).  \label{12S}
\ea
Below we will call the operator $\hat S_\mu\hat S^\mu$ as operator of the
Lorentz spin square.

Equations (\ref{12m}),(\ref{12s}) define some sub-representation of the
left GRR of $M(2,1)$, which is characterized by mass $m$, Lorentz spin $S$,
and by the eigenvalue $K$ of Lubanski-Pauli operator. At $m=0$ we suppose
$K=0$, that is true for irreps with finite number of spinning degrees of
freedom. The general cases for $m=0$ and for $m$ imaginary were discussed in
\cite{Bin82,GitSh97}.

Possible values of $K$ can be easily described in the massive case. Here we
can use a rest frame, where
$\hat p_{\mu} \hat S^{\mu}=\hat S^0 m \mathop{\rm sign}p_0$. Thus,
$K=sm=s^0m$ for $p_0>0$ and $K=sm=-s^0m$ for $p_0<0$, where $s^0$ is the
eigenvalue of $\hat S^0$. The spectrum of $\hat S^0$ depends on the
representation of the Lorentz group.

Variable $s$ labels irreps of the group $M(2,1)$ and can take both positive
and negative values. Thus there exist the analogy with characterized by
helicity massless particles in 3+1 dimensions. In both cases $SO(2)$ is the
little group, and single-valued irreps of $SO(2)$ are labelled by integer
number $2s$. (It is a particular case of the connection between the massive
fields in $d$ dimensions and massless fields in $d+1$ dimensions,
see \cite{AraDeY87,VasTy97}).
Therefore we will call $s$ as helicity and $|s|$ as spin.

Corresponding to (\ref{P}), space reflection reduces to the rotation at $\pi$
around axis $x^0$ and converts $Z$ to
$(Z^\dagger)^{-1}=\sigma_3 Z\sigma_3$, or $z_1\to z_1$, $z_2\to -z_2$.
Operators $\hat p^0$, $\hat S^0$ do not change.
Thus, distinct from 3+1-dimensional case, space reflection leaves helicity
unaltered.

Fixing $S$ in (\ref{12S}), we pass to the space of homogeneous functions
of degree $2S$ in $z_1,z_2$. According to the sign of $S$, below we will
consider two possible choices of $SU(1,1)$ irreps bounded either with both
sides or with one side respectively.

Finite-dimensional nonunitary irreps $T^0_S$ of $SU(1,1)$ are labelled by
positive integer or half-integer $S$. The basis in the representation space
is formed by the polynomials of power $2S$ in $z$, see (\ref{d18}).
We denote corresponding representations of $M(2,1)$ as $T^0_{m,s}$.

Infinite-dimensional unitary irreps $T^-_S$ ($T^+_S$) of $SU(1,1)$ are
labelled by negative $S<-1/2$ and are bounded by highest (lowest) weight.
The basis in the representation space is formed by the quasipolynomials of
power $2S$ in $z$, see (\ref{d22}). We denote corresponding representations
of $M(2,1)$ as $T^-_{m,s}$ ($T^+_{m,s}$).

One may present a function $f(x,z)$ in the form
\be    \label{21zx}
f(x,z)=\phi(z)\psi(x),
\ee
where $\phi(z)$ is a line composed of the basis elements $\phi_n(z)$ of
the corresponding $SU(1,1)$ representation, and $\psi(x)$ is a column
composed of the coefficients in the decomposition over this basis.
The action of the differential operators $\hat S^\mu$ on a function $f(x,z)$
may be presented in terms of matrices
\be
\hat S^\mu f(x,z)=\phi^n(z)(S^\mu)_n^{\;\;n'}\psi_{n'}(x),
\ee
where $S^\mu$ are $SU(1,1)$ generators in the representation $T_S$
(see Appendix and also \cite{GitSh97}). They obey the commutation
relations of the $SU(1,1)$ group
$[{S}^{\mu },{S}^{\nu }]=-i\epsilon ^{\mu \nu \eta }{S}_{\eta }$.

For fixed $S$ in the matrix representation equations (\ref{12m}),(\ref{12s})
have the form
\begin{eqnarray}
&&(\hat{p}^2-m^2)\psi (x)=0,  \label{f4} \\
&&(\hat p_{\mu} S^{\mu}-sm)\psi (x) =0,  \label{12Lub}
\end{eqnarray}
According to (\ref{12Lub}),
\[
\psi ^{\dagger} (x)(S^{\dagger \mu} \overleftarrow {{\hat p}_\mu} + sm)=0.
\]
It follows from the explicit expressions (\ref{d21}) that for $T^0_{m,s}$
the relation $S^{\dagger \mu}=\Gamma S^\mu \Gamma$, where relations
$(\Gamma )_{nn'} = (-1)^{n}\delta _{nn'}$, $\Gamma^2=1$ take place.
For $T^+_{m,s}$ and $T^-_{m,s}$ matrices $S^\mu$ are Hermitian,
$S^{\dagger \mu}=S^\mu$, according to (\ref{d23}).
Let us introduce the notation
\ba  \nonumber
&&\overline \psi = \psi^{\dagger} \Gamma  \quad {\rm for }\quad T^0_{m,s},
\\   \nonumber
&&\overline \psi = \psi^{\dagger} \quad {\rm for }\quad T^+_{m,s},T^-_{m,s}.
\ea
The function $\overline \psi(x)$ obeys the equation
\begin{equation}  \label{f6}
\overline \psi (x)(S^{\mu} \overleftarrow{{\hat p}_\mu} + sm)=0.
\end{equation}

As a consequence of the relations $S^{\dagger \mu}=\Gamma S^\mu \Gamma$ and
$(S^\mu)^\dagger=-(-1)^{\delta_0\mu}S^\mu$ we obtain that for irrep
$T^0_{m,s}$ finite transformation matrices obey the equation
$\Gamma T^\dagger(g)\Gamma=T^{-1}(g)$.
Therefore $\overline\psi(x)\psi(x)$ is a scalar density, and one may define
a scalar product in the space of columns
\be
(\psi'(x),\psi(x))=\int\overline\psi'(x)\psi(x)d^3x.
\ee
The scalar density is positive definite for $T^+_{m,s}$ and $T^-_{m,s}$ in
contrast to the case of $T^0_{m,s}$.

As a consequence of (\ref{12Lub}) and (\ref{f6}), the continuity equation holds
\begin{equation}  \label{f7}
\partial_\mu j^\mu =0, \quad j^\mu =\overline \psi S^\mu \psi.
\end{equation}

Together with the current vector $j^\mu$, by analogy with four-dimensional
case \cite{GelMiS63}, one can associate with the linear equation (\ref{12Lub})
the energy-momentum tensor $T^{\mu\nu}$ and the energy density $W=-T^{00}$:
\be \label{Ep}
T^{\mu\nu}={\rm Im}\left(S^\mu\frac{\partial\psi}{\partial x^\nu},\psi\right),
\qquad
W=-T^{00}=-{\rm Im}\left(S^0\frac{\partial\psi}{\partial x^0},\psi\right).
\ee
If matrix $S^0$ is diagonable, then the positiveness of $W(x)$ is equivalent
to the requirement that
\be
(S^0\psi,S^0\psi)\ge 0
\ee
for all vectors $\psi$ \cite{GelMiS63}.
In particular, for $T^+_{m,s}$ and $T^-_{m,s}$ the relation
$(S^0\psi,S^0\psi)=\psi^{\dagger}S^0S^0\psi\ge 0$ takes place,
and energy density is positive definite.

There are two cases when equation (\ref{f4}) is the consequence of
(\ref{12Lub}). Indeed, multiplying equation (\ref{12s}) by
$\hat p_{\mu} S^{\mu} +ms$, one gets
\be  \label{12kv}
(\hat p_{\mu} S^{\mu} +ms)(\hat p_{\nu} S^{\nu} -ms)\psi(x)= \left(
{\textstyle \qq}\hat p_\mu \hat p_\nu [S^\mu,\; S^\nu]_+ -m^2 s^2\right)\psi(x)=0.
\ee
In the particular case $S=1/2$ we have $s=\pm 1/2$, $S^\mu = \gamma^ \mu /2$,
and (\ref{12kv}) is merely the Klein-Gordon equation (\ref{f4}). In general
case the matrices $S^\mu$ are not $\gamma$-matrices in higher dimensions,
and the squared equation (\ref{12kv}) does not coincide with the Klein-Gordon
equation (\ref{f4}). Using the rest frame, one may show that the
equation (\ref{f4}) follows from (\ref{12Lub}) also in the case of vector
representation $S=1$, $s=\pm 1$.
In the other cases for the identification of the irrep of $M(2,1)$ it is
necessary to use both equations of the system (\ref{f4}),(\ref{12Lub}).
Notice that another approach to the description of fields with fixed spin
and mass was suggested in \cite{Plyus97}; this approach is based on the
system of spinor linear equations.

It is naturally to connect spin value with the highest (lowest) weight
of the irrep of Lorentz group, $s=\pm S$. This means that up to a sign
($+$ for $p_0>0$, $-$ for $p_0<0$) $s$ is equal to maximal or
minimal eigenvalue of the operator $\hat S^0$ in the representation
$T_S$ of the Lorentz group. According to (\ref{12m})-(\ref{12S}),
in this case functions $f(x,z)$ obey the equations
\ba
&&\hat p^2 f(x,\bz) = m^2   f(x,\bz), \label{12m2}
\\
&&\hat p_\mu\hat S^\mu f(x,\bz) = ms f(x,\bz), \quad s=\pm S, \label{12s2}
\\
&&\hat S_\mu\hat S^\mu f(x,\bz) = S(S+1)f(x,\bz). \label{12S2}
\ea

In the framework of the system (\ref{12m2})-(\ref{12S2}) {\it there are two
possibilities to describe one and the same spin}:

1. Equations for $f(x,\bz)=\phi(\bz)\psi(x)$, where $\phi(\bz)$ transform
under finite-dimensional nonunitary irrep of the Lorentz group.

2. Equations for $f(x,\bz)=\phi(\bz)\psi(x)$, where $\phi(\bz)$ transform
under infinite-dimensional unitary irrep of the Lorentz group. These
equations allow us to describe also particles with fractional spin (anyons).

(1) At first, consider the Poincar{\'e} group representations $T^0_{m,s}$
associated with {\it finite-dimensional non-unitary irreps} of $SU(1,1)$.
In this case $S$ has to be positive, integer or half-integer.
In the rest frame the solutions of the system (\ref{12m2})-(\ref{12S2})
in the space of analytic functions (polynomials of power $2S$ in $z^1$,
$z^2$) are
\ba  \label{splus}
&& s=S>0: \quad  f(x,z)=C_1(z^1)^Se^{imx^0} + C_2(z^2)^Se^{-imx^0},
\\   \label{sminus}
&&s=-S<0: \quad  f(x,z)=C_3(z^1)^Se^{-imx^0} + C_4(z^2)^Se^{imx^0}.
\ea
For unique mass and spin there exist four independent components differed by
signs of $p_0$ and $s$, which correspond to four irreps of $M(2,1)$.
The separation by the sign of helicity $s$ has absolute character
since these states are solutions of different equations.
But the states with different sign of $p_0$ are solutions of one and the
same equation. Hence, the energy spectrum of solutions is not bounded below
or above.

In the space of antianalytic functions (polynomials of power $2S$ in
$\cc z^{\dot 1}$, $\cc z^{\dot 2}$) solutions of the system
(\ref{12m2})-(\ref{12S2}) are
\ba \nonumber
&& s=S>0: \quad
f(x,\cc z)=C_1(\cc z^{\dot 1})^Se^{-imx^0} + C_2(\cc z^{\dot 2})^Se^{imx^0},
\\  \nonumber
&& s=-S<0: \quad
f(x,\cc z)=C_3(\cc z^{\dot 1})^Se^{imx^0} + C_4(\cc z^{\dot 2})^Se^{-imx^0}.
\ea
These solutions are connected with previous case (\ref{splus}),(\ref{sminus})
by charge conjugation (\ref{C}) and therefore may be treated as the solutions
describing antiparticles.

The wave function (\ref{splus}) corresponding to the helicity $s=-S$ has the
form $C (z^2)^{2S}e^{ip_0x^0}$, $p_0=m$, in the rest frame. Acting
on it by finite transformations, we get a solution in the form of the plane
wave, which is characterized by the momentum $p$:
\begin{eqnarray}
&&P=U P_0 U^\dagger ,\quad P_0=mI, \quad Z=UZ_0, \quad Z_0=(z_1\, z_2)^T,
\nonumber \\  \label{f19}
&&f(x,z)=(2\pi)^{-3/2}\left(z^2 u_1 - z^1 u_2 \right)^{2S}e^{ipx}.
\end{eqnarray}
The state with $P_0=mI$ has the stationary subgroup $U(1)$, and we can take
elements $u_1=\cosh{\theta/2}$ and $u_2=\sinh{\theta/2}\,e^{i\omega}$
of the first line of matrix $U$,  
that depends on two parameters only.
Thus $p_0=E=m\cosh\theta ,\quad -p_1+ip_2 = m\sinh \theta e^{i\omega }$, and
one can express the parameters $u_1$ and $u_2$ via the momentum $p$:
\be \label{f20}
{{u_1}\choose{u_2}}=\frac{1}{\sqrt{2m(E+m)}}{{E+m}\choose{-p_1+ip_2}}.
\ee
$2S+1$ components $\psi_n(x)$ are the coefficients in the
decomposition of the function (\ref{f19}) over the basis $\phi^n(z)$,
$f(x,z)=\phi^n(z)\psi_n(x), \quad n=0,1,\dots,2S$:
\ba  \nonumber
\psi_n(x)&=&(2\pi)^{-3/2}\left( C_{2S}^{n}\right)^{1/2}(u_1)^{2S-n}
 (-u_2)^{n}e^{ipx}
\\ \label{f22}
 &=&(2\pi)^{-3/2}\left( C_{2S}^{n}\right)^{1/2}\frac{\left( E+m\right)^{2S-n}
 \left(p_1-ip_2\right)^n}{\left( 2m(E+m)\right) ^{S}}e^{ipx}.
\ea
In the particular case $S=1/2$ we get
$$
\psi (x)=\frac 1{\sqrt{2m(E-m)}}{{p_2-ip_1}\choose{E+m}}e^{ipx}.
$$

Considering the system (\ref{12s2})-(\ref{12S2}) without the condition
of the mass irreducibility (\ref{12m2}), it is easy to see
that the charge density $j^0=\psi^\dagger \Gamma S^0 \psi$ is positive
definite only for $S=1/2$, and the energy density $-T^{00}$ is positive
definite only for $S=1$.
The scalar density $\overline \psi \psi =\psi^\dagger \Gamma \psi $ is not
positive definite.

Let us show that for the particles with half-integer spin described by
the system (\ref{12m2})-(\ref{12S2}) the charge density $j^0$ (\ref{f7})
is positive definite. In the rest frame solutions of the system
(\ref{12m2})-(\ref{12S2}) have only two components (labelled by $s_0=\pm S$),
which we denote as $\psi_{S}(x)$ and $\psi_{-S}(x)$. For half-integer spin
an inequality
$j^0=\psi^\dagger \Gamma S^0 \psi = S(|\psi_{S}|^2+|\psi_{-S}|^2) >0$ holds.
At $S\ge 3/2$ from the explicit form of matrices $S^1$ and $S^2$ (\ref{d21})
one can obtain that in the rest frame $j^1=j^2=0$, therefore the square of
the current vector $(j^0)^2-(j^1)^2-(j^2)^2$ is positive.
Therefore $j^0>0$ for any plane wave.

Thus the charge density $j^0$ is positive definite for half-integer spin
particles described by representations $T^0_{m,s}$ of $M(2,1)$.
The scalar density and the energy density are proportional to
$\psi^\dagger\Gamma\psi=|\psi_{S}|^2-|\psi_{-S}|^2$ in the rest frame and
therefore are indefinite.

Let us consider now particles with integer spin.
In the rest frame solutions of the system also have only two components:
$\psi_{S}(x)$ and $\psi_{-S}(x)$,
$(S^0\psi,S^0\psi)=
 \psi^\dagger \Gamma S^0 S^0\psi = S^2 (|\psi_{S}|^2+|\psi_{-S}|^2) >0$.
Thus the energy density is positive definite for integer spin particles
described by representations $T^0_{m,s}$ of $M(2,1)$.
The charge density is proportional to $|\psi_{S}|^2-|\psi_{-S}|^2$ in the
rest frame and therefore is indefinite.

Consider two particular cases explicitly.
For $S=1/2$ the decomposition (\ref{21zx}) has the following form
\be
f(x,z)=z^1\psi_1(x)+z^2\psi_2(x), \quad \psi'(x')=U^{-1}\psi (x), \quad
\psi (x) = (\psi_1(x) \; \psi_2(x))^T.
\label{12S12}
\ee
Taking into account the relation $U^{-1}=\sigma_3U^{\dagger }\sigma_3$,
which is valued for the $SU(1,1)$ matrices, we get the transformation
law for the line $\bar\psi=\psi^\dagger \sigma_3$,
$\bar \psi'(x')=\bar \psi (x)U$.
The product $\bar \psi (x)\psi (x)=|\psi_1(x)|^2-|\psi_2(x)|^2$ is the
scalar density.

Thus, in the case under consideration, we have two equivalent descriptions.
One in terms of functions (\ref{12S12}) and another one in terms of lines
$\bar\psi(x)$ or columns $\psi (x)$. One can find the action of the
operators $\hat S^\mu $ in the latter representation, and equation
(\ref{12Lub}) can be rewritten in the form of $2+1$ Dirac equation
\be
\hat S^\mu \psi (x) = \frac 12\gamma^{\mu} \psi (x), \quad
(\hat p_{\mu} \gamma^{\mu}\mp m)\psi (x) =0,         \label{dir21}
\ee
where minus corresponds to $s=1/2$, plus corresponds to $s=-1/2$, and
$\gamma^\mu$ are $2\times 2$ $\gamma $-matrices (\ref{3gamma}) in $2+1$
dimensions. The functions $\psi =(\psi^1\;0)^T$ and $\psi =(0\;\psi^2)^T$
are eigenvectors of the operator $\hat S^0$ with the eigenvalues $\pm 1/2$.

Sometimes two equations for $s=\pm 1/2$ are written as one equation for
the four-component reducible representation \cite{GitTy97,VshMaZ98},
$(\hat p_{\mu} \Gamma^{\mu} - m)\Psi (x) =0$, where
$\Gamma^\mu = \diag (\gamma^{\mu}, \, -\gamma^{\mu})$,
that corresponds to the simultaneous consideration of particles with opposite
helicities.

For $S=1$ the decomposition (\ref{21zx}) has the following form
\begin{equation}
f(x,z)=\psi_{11}(x)(z^1)^2 + \psi_{12}(x) z^1 z^2
+ \psi_{22}(x)(z^2)^2,  \label{f9}
\end{equation}
where $\psi (x)=(\psi_{11}(x)\;\psi_{12}(x)/\sqrt {2}\;\psi_{22}(x))^T$
is subjected to equation (\ref{12Lub}) with the matrices
\be \nonumber
S^0=\left(\begin{array}{rrr}
1\; & 0 & 0 \\ 0\; & 0 & 0 \\ 0\; & 0 & -1
\end{array} \right), \quad S^1=-\frac 1{\sqrt 2}\left( \begin{array}{rrr}
0 & -1 & 0  \\ 1 & 0 & -1  \\ 0 & 1 & 0
\end{array} \right), \quad S^2=-\frac i{\sqrt 2}\left( \begin{array}{rrr}
0 & 1 & 0   \\ 1 & 0 & 1   \\ 0 & 1 & 0
\end{array} \right).
\ee
If instead of the cyclic components $\psi_{\alpha\beta}(x)$ one introduces
new (Cartesian) components
${\cal F}_{\mu}=\check\sigma_{\mu}^{\;\;\alpha\beta}\psi_{\alpha\beta}(x)$,
where $\check\sigma_{\mu\alpha\beta}$ is defined in (\ref{3sig}),
${\cal F}_0=-2\psi^{12}$, ${\cal F}_1=\psi^{11}+\psi^{22}$,
${\cal F}_2=i(\psi^{22}-\psi^{11})$,
then equation (\ref{12Lub}) takes the form \cite{GitSh97}
\begin{equation}
\partial _{\mu} \varepsilon^{\mu\nu\eta}{\cal F}_{\eta} - sm{\cal F}^{\nu}=0.
\label{f11}
\end{equation}
A transversality condition follows from (\ref{f11}),
$\partial _{\mu}{\cal F}^{\mu}=0$. One can see now that equations
(\ref{f11}) are in fact field equations of the so called ''self-dual'' free
massive field theory \cite{TowPiN84}. As remarked in \cite{DesJa84}, this
theory is equivalent to the topologically massive gauge theory with the
Chern-Simons term \cite{DesJaT82}. Indeed, the transversality condition
allows introducing gauge potentials $A_\mu $, namely a transverse vector can
be written as a curl
${\cal F}^\mu =\varepsilon ^{\mu \nu \lambda }\partial _\nu A_\lambda
=\varepsilon ^{\mu \nu \lambda }F_{\nu \lambda }/2$,
where $F_{\nu \lambda }=\partial _\nu A_\lambda -\partial _\lambda A_\nu $
is the field strength. Thus, ${\cal F}^\mu $ appears to be dual field
strength, which is a tree-component vector in 2+1 dimensions. Then
(\ref{f11}) implies the following equations for $F_{\mu \nu }$
$$
\partial _\mu F^{\mu \nu } - \frac{sm}2\varepsilon ^{\nu \alpha \beta}
F_{\alpha \beta }=0,
$$
which are the field equations of the topologically massive gauge theory.

To describe neutral spin 1 particle coinciding with its antiparticle
one may consider a function
\be \label{f9n}
f(x,\bz)=\psi_{11}(x)z^1\cc z^1 +\psi_{12}(x)(z^1\cc z^2 +\cc z^1 z^2)/2+
\psi_{22}(x)z^2\cc z^2,
\ee
where we have used (\ref{21sig}) for the convertation to undotted indices.
The spin part of the function (\ref{f9n}) depends not on three angles as in
the case (\ref{f9}), but on two angles only.
This function is an eigenfunction of operator $\hat S_R^3$ with zero
eigenvalue. Substituting (\ref{f9n}) into (\ref{12s2}), we again obtain
equation (\ref{f11}).

(2) Consider now Poincar\'e group representations $T^+_{m,s}$ and
$T^-_{m,s}$ associated with {\it unitary infinite-dimensional irreps}
of $SU(1,1)$ with highest (lowest) weight. In this case $S$ can be
non-integer, $S<-1/2$ (discrete series) or $S=-1/2$ (principal series).
Eigenvalues of $\hat S^0$ can take only positive values for discrete positive
series, $s^0=-S+n$, and only negative valuses for negative one, $s^0=S-n$,
where $n=0,1,2,...\,$.

Let us consider the energy spectrum of the system (\ref{12m2})-(\ref{12S2})
at $m\ne 0$. According to the first equation $p_0=\pm m$.
The second equation ensures the relation between spectra of the operators
$\hat p_0$ and $\hat S^0$,
\be  \label{Espektr}
p_0s^0=ms.
\ee
For representations $T^0_{m,s}$, which correspond to finite-dimensional
irreps $T^0_S$ of the Lorentz group, the value of $s^0$ can be both positive
and negative. Therefore for any $s$ there exist both positive-frequency and
negative-frequency solutions, and the representations $T^0_{m,s}$ splits
into two irreps, characterized by $\sign p_0=\pm 1$.

For unitary $SU(1,1)$ irreps with highest (lowest) weight the spectrum of
$\hat S^0$ has definite sign. For $T^+_S$, which act in the space of
analytic functions, the spectrum of operator $\hat S^0$ is positive,
and for $T^-_S$, which act in the space of antianalytic functions, is
negative. Therefore the sign of energy $p_0$ coincides with the sign of $s$
for $T^+_S$, and the signs of $p_0$ and $s$ are opposite for $T^-_S$.
Thus $T^+_{m,s}$ and $T^-_{m,s}$ are irreps of $M(2,1)$.

As well as in the case of representations $T^0_{m,s}$, for unique mass and
spin there are four states distinguished in signs of $p_0$ and $s$. In the
rest frame there are two solutions of the system in the space of analytic
functions:
\ba  \label{21pp}
&&p_0>0, s>0: \quad  f(x,z)=(2\pi)^{-3/2}(z^2)^S e^{imx^0},
\\   \label{21mp}
&&p_0<0, s<0: \quad  f(x,z)=(2\pi)^{-3/2}(z^2)^S  e^{-imx^0}.
\ea
The solutions are connected by time reflection $T'$ (\ref{T}). In the space
of antianalytic functions there are also two solutions:
\ba   \label{21pm}
&&p_0>0, s<0: \quad  f(x,\cc z)=(2\pi)^{-3/2}(\cc z^{\dot 2})^S e^{imx^0}
\\   \label{21mm}
&&p_0<0, s>0: \quad  f(x,\cc z)=(2\pi)^{-3/2}(\cc z^{\dot 2})^S e^{-imx^0} .
\ea
These solutions are connected respectively with (\ref{21pp}),(\ref{21mp})
by Schwinger time reversal $T_{sch}=CT'$, which turns particles into
antiparticles.
Thus, there exist four equations distinguished in sign of $s$ and by the used
functional space (irrep $T^+_S$ or $T^-_S$ of the Lorentz group), and
any equation has the solutions only with definite sign of $p_0$.

In contrast to the case of $T^0_{m,s}$, where the energy spectrum
$p_0$ is not bounded both above and below, the energy spectrum has definite
sign. In any inertial frame the spectrum is bounded below by $p_0=m$ for
the solutions (\ref{21pp}), (\ref{21pm}) and above by $p_0=-m$ for the
solutions (\ref{21mp}), (\ref{21mm}).

For the unitary irreps of $M(2,1)$ under consideration, which correspond to
the irreps of the discrete seria of the Lorentz group, integration of the
functions (\ref{d22}) in the invariant measure (\ref{21mera}) gives
\ba  \label{M21norm}
&&\int \cc f_{S_1}(x,\bz) f'_{S_2}(x,\bz) d\mu (x,\bz)
 = \delta _{S_1S_2} \int \sum_{n=0}^{\infty} \psi^n(x)\psi^{\prime n}(x)d^3x,
\\ \nonumber
&&\int \cc f_{S_1}(x,\bz) f'_{S_2}(x,\bz) d\mu (\bz)
 = \delta _{S_1S_2} \psi^{\dagger}(x)\psi' (x),
\ea
In particular, the states (\ref{21pp})-(\ref{21mm}) have the norm
$\delta_{SS'}\delta(p-p')$. For the principal series $j=-1/2+i\lambda$, and
$\delta _{j_1j_2}$ in (\ref{M21norm}) is changed by
$\delta (\lambda_1-\lambda_2)$.
At the same time, the integral over the spin space diverges for the
representations $T^0_{m,s}$, which correspond to finite-dimensional irreps of
the Lorentz group.

Arbitrary plain wave solutions can be obtained by analogy with considered
above case of $T^0_{m,s}$. For example, for the states (\ref{21pp}) one can
get the formula (\ref{f22}), where now $C_{2S}^n$ are the coefficients from
(\ref{d22}) and $n=0,1,2,...\,$. The power $2S$ is negative, and
the decomposition $f(x,z)=\phi_n(z)\psi^n(x)$ contains infinite number of
terms.

Let us summarize some properties of the unitary irreps under consideration.
Irreps $T^+_{m,s}$ and $T^-_{m,s}$ of the Poincar\'e group describe
particles and antiparticles respectively. Charge density
$j^0=\psi ^{\dagger}S^0 \psi$ is positive definite for particles and negative
definite for antiparticles. The energy density is positive
definite in both cases since $(S^0\psi,S^0\psi)=\psi^\dagger S^0S^0\psi >0$.
Besides, for the unitary irreps the scalar density $\psi ^{\dagger}\psi$ is
also positive definite in contrast to the finite-dimensional case.
The existence of positive definite scalar density ensures the possibility
of probability amplitude interpretation of $\psi(x)$.

Thus in 2+1 dimensions the problem of the construction of
{\it positive-energy RWEs} is solved by the system (\ref{12m2})-(\ref{12S2})
for the infinite-dimensional unitary irreps $T^+_{m,s}$ (signs of
$p_0$ and $s$ are the same) or $T^-_{m,s}$ (signs of $p_0$ and $s$ are
opposite) characterized by the mass $m$ and the helicity $s$. These irreps
of the Poincar\'e group are associated with irreps $T^{+}_S$ and $T^{-}_S$
of the Lorentz group with lowest (highest) weight. Charge conjugation,
changing signs of $p_0$ and $s^0$, leaves the helicity $s$ invariant and
turns $T^+_{m,s}$ into $T^-_{m,s}$.

An interesting problem is to find an explicit form of the intertwining
operator $A$ for the unitary irreps $T^+_{m,s}$, $T^-_{m,s}$ and the
representation $T^0_{m,s}$ labelled by the same mass $m$ and helicity $s$
but assotiated with finite-dimensional nonunitary irreps of the Lorentz group,
$AT^0_{m,s}=(T^{+}_{m,s}\oplus T^{-}_{m,s})A$.
The intertwining operator is nonunitary and must be a function of the
generators of right translations, since other generators commute with
Lorentz spin square operator $\hat S_\mu\hat S^\mu$ and can't change the
representation of spin Lorentz subgroup.

Notice that the 2+1 Dirac equation arises also in the case of unitary
infinite-dimensional irreps $T^{+}_S$ and $T^{-}_S$ of the Lorentz group not
as an equation for a true wave function, but as an equation for spin coherent
states evolution. In this case the equation includes effective mass
$m_s=|\frac sS| m$, $s=-S,-S+1,\dots$ \cite{GitSh97}.

Among the above considered RWE there exist ones which describe particles with
fractional real spin. These equations are associated with unitary
multi-valued irreps of the Lorentz group and can be used to describe anyons.

In spite of the fact that the number of independent polarization states for
massive 2+1 particle is one, the vectors of the corresponding representation
space of irreps $T^+_{m,s}$, $T^-_{m,s}$ have infinite number of components
in matrix representation. Thus, $z$-representation is more convenient in this
case.

\section {Four-dimensional case}
\subsection {Field on the group $M(3,1)$}

The generators and the action of the left GRR on the functions $f(x,\bz)$ are
given by formulas (\ref{komm0}), (\ref{scal_xz}). For spin projection
operators it is convenient to use three-dimensional vector notation
$\hat S_k =\frac 12\epsilon_{ijk} \hat S^{ij}$, $\hat B_k =\hat S_{0k}$.
The explicit calculation gives
\ba
&&\hat S_k=\frac 12 (z\sigma_k\partial _z -\cc z\cc\sigma_k\partial _{\ccc z}\,)+... \; ,
\nonumber \\
&&\hat B_k=
\frac i2 (z\sigma_k\partial _z + \cc z\cc\sigma_k\partial _{\ccc z}\,)+... \; ,
\quad z=(z^1\; z^2), \quad
\partial_z=(\partial/\partial{z^1}\; \partial/\partial{z^2})^T ;
\\
&&\hat S_k^R=-\frac 12 (\chi\cc\sigma_k\partial_\chi
-\cc\chi\sigma_k\partial _{\ccc \chi}\,)+... \; ,
\nonumber \\
&&\hat B_k^R=
-\frac i2 (\chi\cc\sigma_k\partial_\chi + \cc\chi\sigma_k\partial _{\ccc \chi}\,)+... \; ,
\quad \chi=(z^1\;\uz^1), \quad
\partial_\chi=(\partial/\partial{z^1}\; \partial/\partial{\uz^1})^T ;
\ea
Dots in the formulas replace analogous expressions obtaining by the
substitutions
$z\to z'=(\uz^1\;\uz^2)$, $\chi\to \chi'=(z^2\; \uz^2)$.

Since $\det Z=1$, then any of $z_\alpha$, $\uz_\alpha$ can be expressed in
terms of other three, for example $\uz_2=(1-z_2\uz_1)/z_1$.
Invariant measure on $\RS^4\times SL(2,C)$ has the form \cite{GelGrV66}
\be
d\mu(x,\bz)=(i/2)^3 d^4x d^2z_1d^2z_2d^2\uz_1|z_1|^{-2}.
\ee

The functions on the Poincar\'e group depend on 10 parameters, and
correspondingly there are 10 commuting operators (two Casimir operators, four
left and four right generators).

Both the Poincar\'e group and the spin Lorentz subgroup have two Casimir
operators:
\ba
&&\hat p^2 =\hat p_\mu \hat p^\mu, \quad
 \hat W^2 =\hat W_\mu \hat W^\mu, \quad  {\rm where} \quad
 \hat W^\mu=\frac 12\epsilon^{\mu\nu\rho\sigma}\hat p_\nu \hat J_{\rho\sigma}
 = \qq\epsilon^{\mu\nu\rho\sigma}\hat p_\nu \hat S_{\rho\sigma},
\label{Pcas}
\\
&&\frac 12 \hat S_{\mu\nu}\hat S^{\mu\nu}=
  \frac 12 \hat S^R_{\mu\nu}\hat S_R^{\mu\nu}=
  \hat {\bf S}^2 -\hat {\bf B}^2, \quad
 \frac 1{16}\epsilon^{\mu\nu\rho\sigma}\hat S_{\mu\nu}\hat S_{\rho\sigma}
 =\frac 1{16}\epsilon^{\mu\nu\rho\sigma}\hat S^R_{\mu\nu}\hat S^R_{\rho\sigma}
 =\hat{\bf S}\hat{\bf B}.
\label{Lcas}
\ea

Let us consider a set of ten commuting operators
\be \label{31set}
\hat p_\mu,\; \hat { W}^2,\; \hat {\bf p}\hat {\bf S}, \;
\hat {\bf S}^2 -\hat {\bf B}^2, \; \hat{\bf S}\hat{\bf B}, \;
\; \hat S_3^R,\; \hat B_3^R.
\ee
This set consists of operators of momenta, the Lubanski-Pauli operator
$\hat { W}^2$, the proportional to helicity operator
$\hat{\bf p}\hat{\bf J}=\hat {\bf p}\hat {\bf S}$, and four operators,
which are the functions of the right generators.
This four operators commute with the left rotations and translations
and allow one to distinguish equivalent irreps in the decomposition of GRR.
In the rest frame $\hat {\bf p}\hat {\bf S}=0$, and the complete set of
commuting operators can be obtained from (\ref{31set}) with the help of
the replacement of $\hat {\bf p}\hat {\bf S}$ by $\hat S_3$.

Functions $f(x,\bz)$ on the group $M(3,1)$ are the functions of four real
variables $x^\mu$ and four complex variables $z_\alpha$, $\uz_\alpha$
with the constraint $z_1\uz_2-z_2 \uz_1=1$.

The space of functions on the Poincar\'e group contains the subspace of
analytic functions $f(x,z,\cc\uz)$ of the elements of the Dirac $z$-spinor
\be  \label{dirz}
Z_D=(z^\alpha,\cc\uz_{\dot\alpha}).
\ee
Charge conjugation means the transition to subspace of antianalytic
functions (i.e. analytic functions of $\uz^\alpha,\cc z_{\dot\alpha}$).

According to (\ref{P}), for the space inversion we have
$Z\stackrel{P}{\to}(Z^{-1})^\dagger$ or
\be \label{PZ}
\left(\begin{array}{cc} z^1 & \uz^1 \\ z^2 & \uz^2
\end{array}\right)\;  \stackrel{P}{\to} \;
\left(\begin{array}{rr} -\cc\uz_{\dot 1} & \cc z_{\dot 1}\\ -
\cc\uz_{\dot 2} & \cc z_{\dot 2} \end{array}\right),
\ee
This transformation reverses the sign of the boost operators $\hat B_k$.
It is easy to see that, in contrast to charge conjugation, space inversion
conserves the analyticity (or antianalyticity) of functions of $Z_D$.

Similarly to three-dimensional case (see (\ref{sr_mix})),
eigenfunctions of $\hat S_3^R$ and $\hat B_3^R$ differ only by a phase factor.
Fixing eigenvalues of operators $\hat S_3^R$ and $\hat B_3^R$, one may pass
to the space of functions of $x^\mu$ and elements of Majorana $z$-spinor
\be  \label{majz}
Z_M=(z^\alpha,\cc z_{\dot\alpha}),
\ee
i.e. the space of functions of 8 real independent variables on the manifold
\be
\RS^4\times \CS^2, \qquad d\mu=d^4x d^2z_1d^2z_2.
\ee
Thus in this space we have 8 commuting operators (2 Casimir operators,
4 operators distinguish states inside the irrep, 2 operators distinguish
equivalent irreps). Notice that physical argumentation of necessity to make
use at least 8 variables in order to describe spinning particles contains in
\cite{Kihlb64}.
The space reflection takes the functions of $Z_M$ to the functions of
$\underline Z_M=(\uz^\alpha,\cc\uz_{\dot\alpha})$; as was mentioned above,
$\uz^\alpha$ and $z^\alpha$ have the same transformation rule. The charge
conjugation leave the space of functions of $Z_M$ invariant.

Below we will consider the massive case characterized by the symmetry with
respect to space reflection and therefore the space of the analytic
functions of Dirac $z$-spinor $Z_D$, unless otherwise stipulated.
In this space the action of $M(3,1)$ is given by a formula
\ba \nonumber
&&T(g)f(x,z,\uz)=f(g^{-1}x,g^{-1}z,g^{-1}\uz),
\\  \label{act31a}
&&(g^{-1}x)^\mu=(\Lambda^{-1})^\mu_{\;\;\nu} x^\nu, \quad
  (g^{-1}z)^\alpha=U^\alpha_{\;\;\beta} z^\beta,    \quad
  (g^{-1}\uz)_\da=(\cc U^{-1})_{\da}^{\;\;\db} \cc\uz_{\db}.
\ea
Spin projection operators have the form
\be
\hat S_k=\frac 12 (z\sigma_k\partial _z -\cc\uz\cc\sigma_k\partial _{\ccc\uz}\,), \quad
\hat B_k=\frac i2 (z\sigma_k\partial _z + \cc\uz\cc\sigma_k\partial _{\ccc\uz}\,).
\label{SBgen}
\ee
It is known that one can compose the combinations $\hat M_k$, $\barM_k$à:
\ba
&&\hat M_k=\frac 12(\hat S_k-i\hat B_k)=z\sigma_k\partial _z, \quad
\hat M_+=z^1\partial /\partial{z^2}, \quad
\hat M_-=z^2\partial /\partial{z^1},
\nonumber \\
&&\barM_k=-\frac 12(\hat S_k+i\hat B_k)=\cc\uz\cc\sigma_k\partial _{\ccc\uz}\,, \quad
\barM_+=\cc\uz^{\dot 1}\partial /\partial{\cc\uz^{\dot 2}}, \quad
\barM_-=\cc\uz^{\dot 2}\partial /\partial{\cc\uz^{\dot 1}},
\label{MNgen}
\ea
such that $[\hat M_i,\barM_k]=0$. For unitary representations of the Lorentz
group $\hat S_k^\dagger =\hat S_k$, $\hat B_k^\dagger =\hat B_k$,
and these operators obey the relation $\hat M_k^\dagger =\barM_k$
(for finite-dimensional nonunitary irreps $\hat S_k^\dagger =\hat S_k$,
$\hat B_k^\dagger =-\hat B_k$ and $\hat M_k^\dagger =-\barM_k$).
Introducing spin operators with spinor indices
\be  \label{Mdef}
\hat M_{\alpha\beta}= (\sigma_{\mu\nu})_{\alpha\beta}\hat S^{\mu\nu},
\quad \hat {\bar M}_{\da\db}= (\bar\sigma_{\mu\nu})_{\da\db}\hat S^{\mu\nu},
\ee
where $\sigma_{\mu\nu}$ and $\bar\sigma_{\mu\nu}$ are defined in (\ref{sigmn}),
we obtain
\ba  \label{SM}
&&\hat S^{\mu\nu}=-\qq \left((\sigma^{\mu\nu})^{\alpha\beta}
  \hat M_{\alpha\beta} + (\bar\sigma^{\mu\nu})^{\da\db} \barM_{\da\db}\right),
\\
&&\hat M_{\alpha\beta}\hat M^{\alpha\beta}=2\hat {\bf M}^2, \quad
  \barM_{\da\db} \barM^{\da\db}=2\barMM.
\ea  \label{Msqr}
In the space of analytic functions of $z,\cc\uz$ we have:
\be  \label{Mab}
\hat M_{\alpha\beta}=\qq (z_\alpha\partial_\beta+z_\beta\partial_\alpha), \quad
\hat {\bar M}_{\da\db}=\qq(\cc\uz_\da\upartial_\db+\cc\uz_\db\upartial_\da).
\ee
Taking into account that operators $\hat M_k$, $\barM_k$ are subjected
to commutation relations of $su(2)$ algebra, we obtain spectra of the Casimir
operators of the Lorentz subgroup:
\ba
&&\hat {\bf S}^2 -\hat {\bf B}^2 = 2(\hat {\bf M}^2+\barMM)
  =2j_1(j_1+1)+2j_2(j_2+1)=-\frac 12 (k^2-\rho ^2 -4), \quad
\nonumber \\
&&\hat {\bf S}\hat {\bf B} = -i(\hat {\bf M}^2-\barMM)
  =-i\left( j_1(j_1+1)-j_2(j_2+1)\right) =k\rho,\qquad
\nonumber \\
&&{\rm where} \quad \rho=-i(j_1+j_2+1), \quad  k=j_1-j_2.
\label{Lcas1}
\ea
Thus irreps of the Lorentz group $SL(2,C)$ are labelled by the pair
$(j_1,j_2)$. It is convenient to label unitary irreps by $[k,\rho]$, where
irreps $[k,\rho]$ and $[-k,-\rho]$ are equivalent \cite{BarRa77,GelGrV66}.

Notice that the formulas (\ref{act31a})-(\ref{Lcas1}) are also valid if,
using substitution $\cc\uz_\da\to\cc z_\da$, we consider the
functions of elements of Majorana $z$-spinor $Z_M$ instead of $Z_D$.

The formulas of reduction on the compact $SU(2)$-subgroup have the form
\be  \label{reduc1}
T_{(j_1,j_2)}=\sum_{j=|j_1-j_2|}^{j_1+j_2}T_j, \qquad
T_{[k,\rho]}=\sum_{j=k}^{\infty}T_j
\ee
for finite-dimensional nonunitary irreps and infinite-dimensional unitary
irreps of $SL(2,C)$ respectively \cite{GelGrV66}.
Analogously with $2+1$ case, there are two types of the Poincar\'e group
representations describing the same spin $s$. These types correspond to
finite-dimensional and infinite-dimensional unitary representations of
the Lorentz group. In particular, one may choose:
(i) $s=j_{\max}=j_1+j_2$ for {\it nonunitary finite-dimensional} irreps
$(j_1,j_2)$;
(ii)~$s=j_{\min}=j_0=|j_1-j_2|$ for {\it unitary infinite-dimensional}
irreps $[j_0,\rho]$,
where $j_{\max}$ and $j_{\min}$ are maximal and minimal $j$ in the
decomposition (\ref{reduc1}) of an irrep of the Lorentz group over irreps
$T_j$ of compact $SU(2)$ subgroup. Below we will study only the case of
finite-dimensional representations of the Lorentz group.

Consider monomial basis
$$
(z^1)^a(z^2)^b (\cc\uz_1)^c(\cc\uz_2) ^d
$$
in the space of functions $\phi(z,\cc\uz)$.
The values $j_1=(a+b)/2$ and $j_2=(c+d)/2$ are conserved under the action
of generators (\ref{MNgen}). Therefore the space of irrep $(j_1,j_2)$ is
the space of homogeneous analytic functions depending on two pairs of
complex variables of power $(2j_1,2j_2)$. We denote these functions as
$\phi_{j_1j_2}(z,\cc\uz)$.

For finite-dimensional nonunitary irreps of $SL(2,C)$ $a,b,c,d$ are integer
nonnegative, therefore $j_1,j_2$ are integer or half-integer nonnegative
numbers. One can write functions $f_s(x,z,\cc\uz)$, which are polynomial of
the power $2s=2j_1+2j_2$ in $z,\cc\uz$, in the form
\be \label{31decomp}
 f_s(x,z,\cc\uz)=\sum_{j_1+j_2=s}\sum_{m_1,m_2}
 \psi^{m_1m_2}_{j_1j_2}(x)\varphi^{m_1m_2}_{j_1j_2}(z,\cc\uz),
\ee
where functions
\ba
&&\varphi^{m_1m_2}_{j_1j_2}(z,\cc\uz) = N^\qq (z^1)^{j_1+m_1}(z^2)^{j_1-m_1}
 (\cc\uz_{\dot 1})^{j_2+m_2} (\cc\uz_{\dot 2})^{j_2-m_2},
\\  \label {31zbas}
&&N=(2s)![(j_1+m_1)!(j_1-m_1)!(j_2+m_2)!(j_2-m_2)!]^{-1},
\ea
form basis of the irrep of the Lorentz group. This basis corresponds to
chiral representation (see Appendix B). On the other hand, one can write the
decomposition of the same function in terms of symmetric multispinors
$\psi_{\alpha_1\dots\alpha_{2j_1}}^{\db_1\dots\db_{2j_2}}(x)=
\psi_{\alpha_{(1}\dots\alpha_{{2j_1})}}^{\db_{(1}\dots\db_{{2j_2})}}(x)$:
\be  \label{31ten}
f_s(x,z,\cc\uz)=\sum_{j_1+j_2=s}f_{j_1j_2}(x,z,\cc\uz), \quad
f_{j_1j_2}(x,z,\cc\uz)=\psi_{\alpha_1\dots\alpha_{2j_1}}^{\db_1\dots\db_{2j_2}}(x)
z^{\alpha_1} \dots z^{\alpha_{2j_1}}\cc\uz_{\db_1} \dots \cc\uz_{\db_{2j_2}}.
\ee
Notice that similar generating functions summed over all $s$ have been used
in \cite{Vasil92,Vasil96} to describe massless fields.
Comparing decompositions (\ref{31decomp}) and (\ref{31ten}), we obtain the
relation
\be
N^\qq\psi^{m_1m_2}_{j_1j_2}(x)=
  \psi_{\underbrace{\scriptstyle{1\,\dots\,1}}_{j_1+m_1}\,
        \underbrace{\scriptstyle{2\,\dots\,2}}_{j_1-m_1}}
      ^{\overbrace{\scriptstyle{\dot 1\,\dots\,\dot 1}}^{j_2+m_2}\,
        \overbrace{\scriptstyle{\dot 2\,\dots\,\dot 2}}^{j_2-m_2}}(x).
\ee

Using invariant tensor $\sigma^\mu_{\da\alpha}$ and spinors $z^{\alpha}$,
$\cc\uz_{\dot\alpha}$, $\partial_{\alpha}=\partial /\partial z^{\alpha}$,
$\upartial^{\dot\alpha}=\partial /\partial \cc\uz_{\dot\alpha}$,
it is possible to construct just four vectors:
\ba \label{Vik}
&&\hat V^\mu_{12}=\frac 12
  \bar\sigma^{\mu\da\alpha}\cc\uz_{\da}\partial_{\alpha}, \quad
  \hat V^\mu_{21}=\frac 12
  {\sigma^\mu}_{\alpha\da}z^{\alpha}\upartial^{\da},
\\  \label{Vkk}
&&\hat V^\mu_{11}=\frac 12  \sigma^\mu_{\;\;\alpha\da}
   z^{\alpha}\cc\uz^{\dot\alpha}, \quad
  \hat V^\mu_{22}=\frac 12  \sigma^\mu_{\;\;\alpha\da}
  \partial^{\alpha}\upartial^{\da}.
\ea
These operators are not functions of generators of $M(3,1)$ and
interlock irreps with different $(j_1,j_2)$, however, as we will see
below, play an impotent role in the theory of RWE.

\subsection {Relativistic wave equations, invariant under proper
             Poincar\'e group}

Let us fix eigenvalues of the Casimir operators of the Poincar\'e group
and of the Lorentz subgroup:
\ba
&&\hat { p}^2 f(x,\bz)=m^2 f(x,\bz),       \label{31m}
\\
&&\hat { W}^2 f(x,\bz)=-s(s+1)m^2f(x,\bz),  \label{31s}
\\
&&\hat {\bf M}^2 f(x,\bz)=j_1(j_1+1)f(x,\bz),  \label{31j0}
\\
&&\barMM f(x,\bz)=j_2(j_2+1)f(x,\bz),  \label{31j}
\ea
Spectrum (\ref{31s}) of the operator $\hat W^2$ corresponds to the
consideration of massive spin $s$ particles and massless particles with
discrete spin.
(For tachyons and massless particles with continuous spin spectrum differ
from (\ref{31s}), see \cite{BarRa77,Macke68}.)
As a consequence of two last equations
(remind that we consider the space of analytic functions of $z,\cc\uz$)
we obtain that eigenvalues of the belonging to the complete set (\ref{31set})
operators $\hat S_3^R$ and $\hat B_3^R$ are also fixed,
\be  \label{addR}
\hat S_3^R f(x,\zz)=-(j_1+j_2)f(x,\zz), \quad
i\hat B_3^R f(x,\zz)=(j_1-j_2)f(x,\zz).
\ee
Equations (\ref{31m})-(\ref{31j}) define reducible representation of
the proper Poincar\'e group $M(3,1)$. This representation splits into two
representations labelled by the sign of $p_0$, which are irreducible for
$m\ne 0$.

Nonequivalent representations are distinguished by eigenvalues of the Casimir
operators $\hat p^2$, $\hat W^2$ and by the sign of $p_0$
(see also \cite{BarRa77,Macke68,KimNo86}).
The case of zero eigenvalues of the operators $\hat p^2$ and $\hat W^2$ is
an exception. This case corresponds to massless particles with discrete spin,
and nonequivalent irreps are labelled by the helicity and by the sign of
$p_0$. On the other hand, one can introduce a chirality as $\lambda=j_1-j_2$
(or as the difference of numbers of dotted and undotted indices).
The explicit form of the chirality operator in the space of analytic
functions of $z,\cc\uz$ is given by the formula (see (\ref{31G5}))
\be  \label{chir}
\hat \Gamma^5 = {\textstyle \qq} \left( z^{\alpha} \partial_{\alpha} -
 \cc\uz_\da \upartial^{\da} \right).
\ee
In the massless case helicity is equal to chirality up to sign \cite{BarRa77}.
In the massive case irreps of the proper Poincar\'e group, which are labelled
by the same $m,s,\sign p_0$ but by different chiralities, are equivalent.
Thus, for fixed mass $m$ and spin $s=j_1+j_2$ the system
(\ref{31m})-(\ref{31j}) has $2s+1$ solutions differed by $\lambda=j_1-j_2$.

Using (\ref{SM}), we rewrite the Lubanski-Pauli vector (\ref{Pcas})
and the Casimir operator $\hat W^2$ in the form
\ba
&&\hat W^\mu=\qq\epsilon^{\mu\nu\rho\sigma}\hat p_\nu\hat S_{\rho\sigma}=
 \qq i\hat p_\nu\left( (\sigma^{\mu\nu})_{\alpha\beta}\hat M^{\alpha\beta}-
 (\bar\sigma^{\mu\nu})_{\da\db}\barM^{\da\db}\right),
\\
&&\hat W^2=-\hat p^2(\hat{\bf M}^2+\barMM) - \qq\hat p_\mu \hat p_\nu
 (\sigma^{\mu\rho})_{\alpha\beta}(\bar\sigma_\rho^{\;\;\nu})_{\da\db}
 \hat M^{\alpha\beta}\barM^{\da\db}.
\ea
Taking into account the explicit form of spin operators (\ref{Mab})
and symmetry of $(\sigma^{\mu\nu})_{\alpha\beta}$ and
$(\bar\sigma^{\mu\nu})_{\da\db}$ with respect to permutation of spinor
indices, we rewrite the last relation as
$$
\hat W^2=-\hat p^2(\hat{\bf M}^2+\barMM) - 2\hat p_\mu \hat p_\nu
(\sigma^{\mu\rho})_{\alpha\beta}(\bar\sigma_\rho^{\;\;\nu})_{\da\db}
z^\alpha\partial^\beta \cc\uz^\da\upartial^\db.
$$
Finally, using the identity
$$
4(\sigma^{\mu\rho})_{\alpha\beta}(\bar\sigma_\rho^{\;\;\nu})_{\da\db}=
-\eta^{\mu\nu}\epsilon_{\alpha\beta}\epsilon_{\da\db}+
{\sigma_{\alpha\da}}^\mu {\sigma_{\beta\db}}^\nu +
{\sigma_{\alpha\da}}^\nu {\sigma_{\beta\db}}^\mu
$$
and the condition of mass irreducibility (\ref{31m}), we obtain
\be  \label{31W}
\hat W^2=-m^2(j_1+j_2)(j_1+j_2+1) +
4\hat p_\mu \hat V_{11}^\mu \hat p_\nu \hat V_{22}^\nu,
\ee
where operators $\hat V_{11}^\mu$ and $\hat V_{22}^\mu$ are defined in
(\ref{Vkk}). Therefore for $s=j_1+j_2$ the necessary and sufficient
condition of spin irreducibility is
\be  \label{sub0}
\hat p_\mu\hat V_{11}^\mu \hat p_\nu \hat V_{22}^\nu  f(x,\zz)=0.
\ee
For the representations $(s\,0)$ and $(0\,s)$ we have
$\hat V_{22}^\mu f(x,\zz)=0$ and the condition (\ref{sub0}) is fulfilled
identically. In general case, observing that in momentum representation an
action of operator $\hat V_{11}^\mu \hat p_\mu$ reduces to multiplication by
the number $p_\mu {\sigma^\mu_{\;\;\alpha\da}} z^\alpha z^\da$, we come to
the {\it alternative conditions}:
\ba
&&\hat p_\mu\hat V_{11}^\mu =0,
\\  \label{V2p}
&&\hat p_\nu \hat V_{22}^\nu f(x,\zz)=0.
\ea

The first condition connects the components of momentum $p_\mu$ and complex
spin variables $q_\mu=\sigma_{\mu\alpha\da} z^\alpha \cc\uz^\da/2$,
$q_\mu q^\mu=0$. Thus we have the space of functions of two 4-vectors $p_\mu$,
$q_\mu$, which are subject to the invariant constraints
\be  \label{pqconstr}
p^2=m^2,\quad p_\mu q^\mu=0, \quad q^2=0.
\ee
According to (\ref{pqconstr}), in the rest frame we get
$z^1\cc\uz^1+z^2\cc\uz^2=0$. Similar approach to the constructing of
wave functions describing the elementary particles was suggested by E.Wigner
in \cite{Wig63}, where discussion was restricted to particles of integer spin
and real $q_\mu$ with constraints $p^2=m^2$, $p_\mu q^\mu=0$, $q^2=-1$.
Different generalizations of the approach \cite{Wig63} were considered later
in \cite{KimWi87,BieBrT88,HasSi92,KuzLy95,LyaSeS96}.

The second condition (\ref{V2p}) does not affect spin variables and can be
written in terms of $\psi(x)$. Really, for fixed $j_1,j_2$, using the
decomposition (\ref{31ten}) of $f(x,\zz)$ in terms of multispinors
and also the relation
$\partial_\alpha\psi_{\alpha_1\alpha_2\dots\alpha_k}
z^{\alpha_{(1}} z^{\alpha_2} \dots z^{\alpha_{k)}} ={\delta_\alpha}^{\alpha_1}
\psi_{\alpha_1\alpha_2\dots\alpha_k}z^{\alpha_{(2}} \dots z^{\alpha_{k)}}$,
one can rewrite the system
\be
(\hat p^2-m^2) f_{j_1j_2}(x,z)=0, \quad
\hat p_\mu{\sigma^\mu}_{\alpha\da}\partial^\alpha\upartial^\da
f_{j_1j_2}(x,z)=0
\ee
in the form
\ba  \label{31m2}
&&(\hat p^2-m^2)\psi_{\alpha_1\dots\alpha_k\da_1\dots\da_l}(x) = 0,
\\   \label{aasupp}
&&\partial^{\da\alpha}
\psi_{\alpha\alpha_1\dots\alpha_{k-1}\da\da_1\dots\da_{l-1}}(x) = 0,
\ea
where $\partial^{\da\alpha}=\partial_\mu \bar\sigma^{\mu\da\alpha}$, $k=2j_1$,
$l=2j_2$. These equations describe a particle with unique mass $m$ and spin
$s=j_1+j_2$. Subsidiary condition (\ref{aasupp}) is necessary to exclude
components corresponding to other possible spins $s$,
$|j_1-j_2|\le s < j_1+j_2$, see (\ref{reduc1}).

On the other hand, in order to describe spin $s$ one may use representations
$(j_1\,j_2)$, $j_1+j_2\ne s$. In this case, according to (\ref{31W}), the
condition (\ref{aasupp}) should be changed by new subsidiary condition
\be  \label{gensub}
\partial_{\beta\db}\partial^{\da\alpha}
 \psi_{\alpha\alpha_1\dots\alpha_{k-1}\da\da_1\dots\da_{l-1}}(x)
=-m^2[(j_1+j_2)(j_1+j_2+1)-s(s+1)]
 \psi_{\beta\alpha_1\dots\alpha_{k-1}\db\da_1\dots\da_{l-1}}(x).
\ee
Notice that an approach using this general subsidiary conditions was not
considered earlier.

Passing on to vector indices, one can see that for integer spins and irreps
$(\frac s2 \, \frac s2)$ equations (\ref{31m2}), (\ref{aasupp})
take the form
\be  \label{31teneq}
(\hat p^2-m^2)\Phi_{\mu_1\mu_2\dots\mu_s}(x)=0, \quad
\partial^\mu \Phi_{\mu\mu_2\dots\mu_s}(x)=0, \quad
{\Phi^\mu}_{\mu\dots\mu_s}(x)=0,
\ee
where
$$
\Phi_{\mu_1\mu_2\dots\mu_s}(x)=(-1)^s2^{-s}
\bar\sigma_{\mu_1}^{\;\;\da_1\alpha_1}
\dots\bar\sigma_{\mu_s}^{\;\;\da_s\alpha_s}
\psi_{\alpha_1\dots\alpha_s\da_1\dots\da_s}(x).
$$
Just equations (\ref{31teneq}) known also as massive tensor field
equations or Fierz--Pauli equations are used most often to describe integer
spins.

For half-integer spins and irreps $(\frac{2s\pm 1}4 \; \frac{2s\mp 1}4)$
after passage to vector indices subsidiary conditions (\ref{aasupp}) take
the form
\ba \nonumber
&& 
\partial^\mu \Psi_{\mu\mu_2\dots\mu_n\alpha}(x)=0, \;
\bar\sigma^{\mu\da\alpha}\Psi_{\mu\mu_2\dots\mu_n\alpha}(x)=0, \;
{\Psi^\mu}_{\mu\mu_2\dots\mu_n\alpha}(x)=0,
\\
&& 
\partial^\mu \Psi_{\mu\mu_2\dots\mu_n\da}(x)=0, \;
{\sigma^\mu}_{\alpha\da}{\Psi_{\mu\mu_2\dots\mu_n}}^\da(x)=0, \;
{\Psi^\mu}_{\mu\mu_2\dots\mu_n\da}(x)=0, \label{31subs2}
\ea
where $n=(2s-1)/2$.

\subsection {Relativistic wave equations, invariant under improper
             Poincar\'e group }

Improper Poincar\'e group includes continuous transformations of the proper
group and space reflection operator (parity operator) $\hat I_P$.
According to (\ref{P}),(\ref{PZ}) this operator obeys the condition
$\hat I_P^2=\hat 1$ and commutation relations
\ba
&&[\hat I_P,\hat p_0]=[\hat I_P,\hat p^2]=[\hat I_P,\hat W^2]
  =[\hat I_P,\hat S_k]=[\hat I_P,\hat S_k^R]=0,
\\    \label{PAcomm}
&&[\hat I_P,\hat p_k]_+=[\hat I_P,\hat B_k]_+=[\hat I_P,\hat B_k^R]_+=0.
\ea
States with definite total parity are defined as eigenfunctions of operator
$\hat I_P$:
\be
\hat I_P f(x,\bz)=\pm f(x,\bz).
\ee

For $m>0$ irreps of the improper Poincar\'e group are labelled by an orbit
defining the mass $m$ and the sign of $p_0$, and by irrep of the little group
$O(3)$ defining spin $s$ and intrinsic parity \cite{BarRa77,Macke68,Tung85}.
In the rest frame the intrinsic parity coincides with the total.

The Casimir operators of the Lorentz group do not commute with parity operator,
$[\hat I_P,\hat {\bf M}^2]=\barMM, \quad [\hat I_P,\barMM]=\hat {\bf M}^2$,
and parity transformation combines two labelled by Lorentz indices
$(j_1,j_2)$ and $(j_2,j_1)$ (by chiralities $\pm \lambda$)
equivalent irreps of the proper Poincar\'e group into one representation of
the improper group. The latter representation is reductible and splits into
two irreps differed by intrinsic parity $\eta=\pm 1$.
Thus we can't make use the operators $\hat {\bf M}^2$, $\barMM$ to
select invariant subspaces, and instead of the set of eight commuting
operators
\be  \label{Mset}
\hat p_\mu, \hat W^2,\; \hat {\bf p}\hat {\bf S}, \;\hat {\bf M}^2,\; \barMM
\ee
used above in order to construct the system (\ref{31m})-(\ref{31j}) we
should consider an another set.
Notice that parity operator $\hat I_P$ can't be used directly for
identification of invariant subspaces since according to (\ref{PAcomm}) it
does not commute with translations and boosts.

The simplest possibility is to consider a system
\ba
&&\hat { p}^2 f(x,\zz) = m^2 f(x,\zz),        \label{31m1}
\\
&&\hat { W}^2 f(x,\zz) = -s(s+1)m^2f(x,\zz),  \label{31s1}
\\
&&\hat S^R_3 f(x,\zz) = -sf(x,\zz).           \label{31j1}
\ea
The last equation fixes the power $2s=2(j_1+j_2)$ of the polynomial in $\zz$,
see (\ref{addR}). The first two equations are the conditions of mass and spin
irreducibility. Therefore the system describes fixed mass and spin, but
the Poincar\'e group representation defined by this system is reducible.
This representation decomposes into $2(2s+1)$ irreps differed by the chirality
$\lambda=-s,\dots,s$ and sign of $p_0$.

Supplementing the system (\ref{31m1})-(\ref{31j1}) by the equation
\be
i\hat B^R_3 f(x,\zz) = \pm(j_1-j_2)f(x,\zz),            \label{31l1}
\ee
which change the sign under space reflection, it is possible to extract
components corresponding to the representation $(j_1,j_2)\oplus(j_2,j_1)$.
If we consider only the components labelled by $(j_1,j_2)$ and $(j_2,j_1)$,
then for $j_1\ne j_2$ mass and spin irreducibility conditions
(\ref{31m1}),(\ref{31s1}) leave $4(2s+1)$ independent components
corresponding to the direct sum of four improper Poincar\'e group irreps
differed by signs of energy $p_0$ and intrinsic parity $\eta$.
But states with definite intrinsic parity arise in such an approach only as
linear combinations of the solutions of {\it two} systems
(\ref{31m1})-(\ref{31l1}) with different sign in the last equation (i.e.
solutions with fixed chirality).

Let us investigate the possibility to construct the system of equations,
which remains invariant under space reflection and has solutions with
definite intrinsic parity. For this purpose it is
necessary to consider equations, which combine labelled by different
chiralities $\lambda=j_1-j_2$ equivalent irreps of the proper Poincar\'e
group. In the other words, {\it it is necessary to consider supplementary
operators, which define some extension of the Lorentz group}.
These operators, replacing $\hat {\bf M}^2$ and $\barMM$
in the set (\ref{Mset}), must commute with all the left generators of the
proper Poincar\'e group and with parity operator $\hat I_P$.
{\it We suppose that one of these commuting operators is linear in $\hat p$.}

A general form of the invariant equations linear in $\hat p$ is
\be  \label{V0}
\hat p_\mu \hat V^\mu f(x,\bz) = \varkappa f(x,\bz),
\ee
where $\hat V^\mu$ is a transforming as four-vector function of $\bz$ and
$\partial /\partial \bz$.

The introduced above vector operators $V^\mu_{ik}$ (\ref{Vik}), (\ref{Vkk})
interlock irreps with different $(j_1,j_2)$.
Operators $\hat V^\mu_{12}$, $\hat V^\mu_{21}$ conserve $j_1+j_2$, and
operators $\hat V^\mu_{11}$, $\hat V^\mu_{22}$ conserve $j_1-j_2$.
Any of four connecting two scalar functions relations
\ba \label{V12}
&&\hat p_\mu \hat V^\mu_{12} f_{j_1,j_2}(x,\zz)
 \!=\!\varkappa_{12} f_{j_1-\qq,j_2+\qq}(x,\zz), \;
 \hat p_\mu \hat V^\mu_{21} f_{j_1,j_2}(x,\zz)
 \!=\!\varkappa_{21} f_{j_1+\qq,j_2-\qq}(x,\zz), \;
\\  \label{V11}
&&\hat p_\mu \hat V^\mu_{11} f_{j_1,j_2}(x,\zz)
 \!=\!\varkappa_{11} f_{j_1+\qq,j_2+\qq}(x,\zz), \;
 \hat p_\mu \hat V^\mu_{22} f_{j_1,j_2}(x,\zz)
 \!=\!\varkappa_{22} f_{j_1-\qq,j_2-\qq}(x,\zz),
\ea
one may consider as a RWE. Thus the operator
$\hat V^\mu$ in (\ref{V0}) is a linear combination of $\hat V^\mu_{ik}$.

Let us consider finite-component equations invariant with respect to space
reflection. This means:
\par 1. The operator $\hat p_\mu\hat V^\mu$ is invariant under space reflection.
\par 2. The equation has solutions $f(x,\zz)=\sum \psi_n(x)\phi_n(\zz)$,
where functions $\phi_n(z)$ carry a representation containing finite number
of irreps $(j_1,j_2)$.

It is easy to see that at $\varkappa_{11} \ne 0$ operator $\hat V^\mu_{11}$
can't be contained in $\hat V^\mu$. In this case at $\varkappa_{22} \ne 0$
one can separate from the system of equations for functions
$f_{j_1,j_2}(x,\zz)$, $f(x,\zz)=\sum f_{j_1,j_2}(x,\zz)$
the independent equation for the characterized by maximal $j_1+j_2$ function,
which does not contain $\hat V^\mu_{22}$.
(Besides, it is not necessary to use operators $\hat V^\mu_{11}$ and
$\hat V^\mu_{22}$ since these operators leave $j_1-j_2$ invariable
and can't connect irreps with different $\lambda$.)

Relating to operators $\hat V^\mu_{12}$ and $\hat V^\mu_{21}$,
one can see that only the combination $\hat p_\mu \hat\Gamma^\mu$,
\be
\hat\Gamma^\mu= \hat V^\mu_{12}+\hat V^\mu_{21}=
\frac 12 \left(
 \bar\sigma^{\mu\da\alpha}\cc\uz_{\dot\alpha}\partial_{\alpha}+
 {\sigma^\mu}_{\alpha\da}z^{\alpha}\upartial^{\dot\alpha} \right),
\ee
is invariant under space reflections. Operators $\hat\Gamma^\mu$ connect
representation $(j_1\,j_2)$ with $(j_1\!+\!1\;j_2\!-\!1)$ and
$(j_1\!-\!1\;j_2\!+\!1)$ and conserve $j_1+j_2$. These operators obey the
commutation relations
\ba
&&[\hat S^{\lambda\mu},\hat\Gamma^\nu]
=i(\eta^{\mu\nu}\hat \Gamma^\lambda - \eta^{\lambda\nu}\hat \Gamma^\mu ),
\label{comg1} \\
&&[\hat\Gamma^\mu,\hat\Gamma^\nu]=-i\hat S^{\mu\nu}, \label{comg2}
\ea
which coincide with the commutation relations of matrices $\gamma^\mu /2$.
The explicit calculation shows that $\hat\Gamma_\mu \hat\Gamma^\mu$
depends on irrep of the Lorentz subgroup,
\be  \label{gg}
\hat\Gamma_\mu \hat\Gamma^\mu = 2j_1+2j_2+4j_1j_2.
\ee

Supplementing generators of the Lorentz group by four operators
\be
\hat S^{4\mu}=\hat\Gamma^\mu, \quad \hat S^{ab}=-\hat S^{ba},
\ee
we obtain
\be  \label{deSitt}
[\hat S^{ab},\hat S^{cd}] = i(\eta^{bc}\hat S^{ad}-\eta^{ac}\hat S^{bd}
-\eta^{bd}\hat S^{ac}+\eta^{ad}\hat S^{bc}),   \quad \eta^{44}=\eta^{00}=1.
\ee
Thus operators $\hat S^{ab}$, $a,b=0,1,2,3,4$, obey the commutation
relations of the generators of the 3+2 de Sitter group $SO(3,2)\sim Sp(4,R)$.
This group has two fundamental irreps, namely four-dimensional spinor irrep
$T_{[10]}$ (by matrices $Sp(4,R)$) and five-dimensional vector irrep
$T_{[01]}$ (by matrices $SO(3,2)$).

Using (\ref{Lcas}), (\ref{Lcas1}) and (\ref{gg}), we obtain for the second
order Casimir operator of the group $Sp(4,R)$
$$
\hat S_{ab} \hat S^{ab}f(x,\zz) = 4S(S+2)f(x,\zz), \quad S=j_1+j_2.
$$
Thus we deal with symmetric representations of $Sp(4,R)$, which we denote
as $T_{[2S\, 0]}$ (see Appendix). These irreps can be obtained as a symmetric
term in the decomposition of the direct product $(\otimes T_{[1\, 0]})^{2S}$.
Irreps $T_{[2S\, 0]}$ characterized by dimensionality $(2S+3)!/(6(2S)!)$
combines all finite-dimensional irreps of the Lorentz group with $j_1+j_2=S$.

However, it is obvious that the equation
\be
\hat p_\mu \hat \Gamma^\mu f(x,\zz) = \varkappa f(x,\zz)
\ee
by itself does not fix spin $s$ and mass $m$, defined by (\ref{31m}) and
(\ref{31s}), or the power $j_1+j_2$ of the $f(x,\zz)$ in $\zz$. In the rest
frame it is easy to see that even for fixed $S=j_1+j_2$ this equation fix
only product $ms=\varkappa$, $s\le S$.

Let us consider the set of eight commuting operators
\be  \label{Gset}
\hat p_\mu,\; \hat W^2,\; \hat {\bf p}\hat {\bf S}\;({\rm or}\;\hat S_3 \;
\hbox{in the rest frame}), \;
\hat p_\mu\hat\Gamma^\mu,\; \hat S_{ab}\hat S^{ab}
\ee
acting in the space of functions of eight variables
$x^\mu, z^\alpha,\cc\uz_\da$.
In compare with the set (\ref{Mset}) we have replaced two right operators
${\bf M}^2$, $\barMM$ by invariant under parity transformation operators
$\hat p_\mu\hat\Gamma^\mu$, $\hat S_{ab}\hat S^{ab}$.
Notice that instead of $\hat S_{ab}\hat S^{ab}$ one can use operator
$\hat S_3^R$ with eigenvalues equal to the minus power of polynomial in
$z,\cc\uz$, see (\ref{31j1}).

Invariant subspaces are labelled by eigenvalues of operators
\ba
&&\hat p^2 f(x,\zz) = m^2 f(x,\zz),           \label{ffmeq}
\\
&&\hat W^2 f(x,\zz) = -m^2s(s+1)f(x,\zz),           \label{ffweq}
\\
&&\hat p_\mu \hat\Gamma^\mu f(x,\zz) = m\tilde s f(x,\zz), \label{ffseq}
\\
&&\hat S_{ab}\hat S^{ab}f(x,\zz) = 4S(S+2)f(x,\zz). \label{ffjeq}
\ea
Unlike equations (\ref{31j0}),(\ref{31j}),
which fix $j_1$ and $j_2$ separately, the last equation of the system fixes
irrep $T_{[2S\, 0]}$ of the 3+2 de Sitter group and therefore the power
$2S=2j_1+2j_2$ of the polynomial in $z,\cc\uz$.
Irreps of the Poincar\'e group characterized by spin $s\le S$ can be realized
in the space of these polynomials.

In the rest frame
\ba \nonumber
&&\hat p_0^2 f(x,\zz)= m^2 f(x,\zz), \\
&&\hat p_0 \hat \Gamma^0 f(x,\zz)= m\tilde s f(x,\zz),\quad  \hat\Gamma^0 = \qq
 (\sigma^{0\da\alpha} \cc\uz_{\da}\partial_{\alpha} +
 \sigma^0_{\;\;\alpha\da} z^{\alpha}\upartial^{\da}).
 \label{g0p0}
\ea
According to the first equation, $p_0=\pm m$ and correspondingly $\tilde s$
is a product of eigenvalue of operator $\hat \Gamma^0$ and
$\mathop{\rm sign} p_0$. For $p_0=m$, any characterized by
$n_1-n_2=2s$ function is the solution of equation (\ref{g0p0}),
where $n_1$ is the power of homogeneity in the variables
$(z^1+\cc\uz_{\dot 1})$, $(z^2+\cc\uz_{\dot 2})$, and $n_2$ is the power
of homogeneity in the variables $(z^1-\cc\uz_{\dot 1})$,
$(z^2-\cc\uz_{\dot 2})$. Therefore, for $p_0=-m$ any
characterized by $n_1-n_2=-2s$ function is the solution of equation
(\ref{g0p0}).

Let us show that the relation
\be
|\tilde s|\le s \le S
\ee
takes place. Variables $z^\alpha$ and $z_\da$ have the same transformation
rule under space rotations. Thus, the pairs
$(z^1+\cc\uz_{\dot 1})$, $(z^2+\cc\uz_{\dot 2})$
and $(z^1-\cc\uz_{\dot 1})$, $(z^2-\cc\uz_{\dot 2})$ are spinors of rotation
group, but are characterized by opposite parity.
The polynomials of power $2j'$ in the first pair of variables or $2j''$ in
the second pair of variables transform under $T_{j'}$ or $T_{j''}$ of
the rotation group.
At fixed $j'$ and $j''$ the relation $\tilde s=(j'-j'')\mathop{\rm sign} p_0$
takes place, and the space of polynomials of the power $2S=2j'+2j''$
corresponds to direct product of the representations $T_{j'}$ and $T_{j''}$.
This direct product decomposes into sum of irreps, labelled by
$s=|j'-j''|,\dots,j'+j''$, and therefore spin $s$ runs from $|\tilde s|$
up to $S$.

In particular, for $|\tilde s|=S$ the spin irreducibility condition
(\ref{ffweq}) is a consequence of other equations of the system, and spin
is equal to one half of the polynomial power.
Below we restrict our consideration by this case, which
allows to describe spin $s$ by means of the irrep of the 3+2 de Sitter group
with minimal possible dimensionality. Correspondingly, for $\tilde s=S$ we
will consider the system
\ba
&&\hat p^2 f(x,\bz) = m^2 f(x,\bz),           \label{fmeq}
\\
&&\hat p_\mu \hat\Gamma^\mu f(x,\bz) = ms f(x,\bz), \label{fseq}
\\
&&\hat S_{ab}\hat S^{ab}f(x,\bz) = 4s(s+2)f(x,\bz). \label{fjeq}
\ea
In the rest frame the general solution in the set of polynomial of the
power $2s$ in $z,\cc\uz$ has the form
\be  \label{sol1}
f_{m,s}(x,\bz) = \sum_{s_3=-s}^s
 C_{s_3}e^{imx^0} (z^1+\cc\uz_{\dot 1})^{s+s_3}(z^2+\cc\uz_{\dot 2})^{s-s_3}+
 C'_{s_3}e^{-imx^0}(z^1-\cc\uz_{\dot 1})^{s+s_3}(z^2-\cc\uz_{\dot 2})^{s-s_3},
\ee
where $s_3$ is the spin projection,
\be
\hat S_3 f(x,\bz)=s_3 f(x,\bz), \quad \hat S_3=\qq
(z^1\partial_1 + \cc\uz_{\dot 1}\upartial^{\dot 1} -
 z^2\partial_2 - \cc\uz_{\dot 2}\upartial^{\dot 2}).
\ee
Thus for unique mass $m$ and spin $s$ there are $2s+1$ independent
positive-frequency solutions and $2s+1$ independent negative-frequency
solutions belonging to two irreps of improper Poincar\'e group.
In the case $\tilde s=-S$, which corresponds to the change of sign in the
equation (\ref{fseq}), general solution is obtained from (\ref{sol1}) by the
substitution $(z^\alpha+\cc\uz_\da)\leftrightarrow (z^\alpha-\cc\uz_\da)$.
It follows from (\ref{sol1}) that for half-integer spins the sign of
$\tilde s$ is the product of $\sign p_0$ and intrinsic parity. \footnote{
 According to (\ref{sol1}), in the rest frame for half-integer spin
 positive-frequency and negative-frequency states are characterized by
 opposite parity. One can show (see \cite{Ryder88,AhlJoG93,GaiAl95}) that for
 fixed mass $m$ and representation $(\qq\,0)\oplus(0\,\qq)$ of the Lorentz
 group this condition is sufficient to derive the Dirac equation.}

Only corresponding to spin 1/2 four-dimensional irrep of the 3+2 de
Sitter group remains irreducible under the reduction on the
improper Lorentz group. For spin one 10-dimensional irrep splits into 6+4
(antisymmetric tensor and four-vector), for spin 3/2 20-dimensional irrep
splits into 8+12, and so on.

Consider plain wave solutions corresponding to a moving along $x^3$ particle.
They can be obtained from the solutions in the rest
frame (\ref{sol1}) by means of the Lorentz transformation
$$
P=UP_0U^\dagger, \quad {\rm where}\;\; P_0=\pm\diag\{m,m\},
     \quad U=\diag\{e^{-a},e^a\}\in SL(2,C),
$$
where the sign corresponds to the sign of $p_0$,
\be  \label{sol2a}
p_\mu=k_\mu{\rm sign}\ p_0, \quad k_0= m \cosh 2a, \quad k_3= m \sinh 2a,
\quad  e^{\pm a}=\sqrt{(k_0\pm k_3)/m}.
\ee
Thus it follows that
\ba \nonumber
f'_{m,s,s_3}(x,\bz)=& C_1 e^{ik_0x^0+k_3x^3}
 (z^1e^a+\cc\uz_{\dot 1}e^{-a})^{s+s_3}(z^2e^{-a}+\cc\uz_{\dot 2}e^a)^{s-s_3}+
\\ \label{sol2}
& C_2 e^{-ik_0x^0-k_3x^3}
  (z^1e^a-\cc\uz_{\dot 1}e^{-a})^{s+s_3}(z^2e^a-\cc\uz_{\dot 2}e^{-a})^{s-s_3}.
\ea
In the ultrarelativistic case for positive $a$ it is convenient to rewrite
(\ref{sol2}) in the form
\ba \nonumber
&&f_{m,s,s_3}(x,\bz) = {\textstyle\left( \frac {k_0+k_3}m \right)^s} \times
\\ \label{sol3}
&&  \left( \left(
  C_1 e^{ik_0x^0+k_0x^3} + C_2 (-1)^{s-s_3} e^{-ik_0x^0-k_0x^3} \right)
  (z^1)^{s+s_3}(\cc\uz_{\dot 2})^{s-s_3}+
  {\textstyle O\left( \frac {k_0-k_3}{k_0+k_3} \right)^{\qq}} \right)
\ea
The main term in (\ref{sol3}) corresponds to functions carrying
irrep $(\frac{s+\lambda}2\,\frac{s-\lambda}2)$, $\lambda=s_3$, of the Lorentz
group. The contribution of other irreps
$(\frac{s+\lambda'}2\,\frac{s-\lambda'}2)$ are
damped by a factor $(\frac {k_0-k_3}{k_0+k_3})^{|\lambda-\lambda'|}$.
Passing to the limit $a\to +\infty$ (or $m\to 0$), we obtain the states with
certain chirality $\lambda=j_1-j_2=s_3$ (for $a\to -\infty$
with chirality $\lambda=j_1-j_2=-s_3$ respectively).
In particular, in the limit the states characterized by helicity $s_3=\pm s$
correspond to the representation $(s\,0)\oplus(0\,s)$ of the Lorentz group.

Taking into account that operators $\hat V^\mu_{21}$ ($\hat V^\mu_{12}$)
lower (raise) chirality $\lambda$ by 1 and the decomposition
\be  \label{decjj}
f_{s}(x,\zz)=\sum_{\lambda=-s}^s f_{j_1j_2}(x,\zz), \;\;
{\rm where }\; s=j_1+j_2, \;\; \lambda=j_1-j_2,
\ee
one can write equation (\ref{fseq}) in chiral representation in the form
\be  \label{fseq1}
\left( \begin{array}{c} \hat p_\mu \hat V_{21}^\mu f_{s-\qq,\qq}
\\ \hat p_\mu \hat V_{12}^\mu f_{s,0}+ \hat p_\mu \hat V_{21}^{\mu}f_{s-1,1}
\\ \dots
\\ \hat p_\mu \hat V_{12}^\mu f_{\qq,s-\qq} \end{array}\right) =
ms\left( \begin{array}{c} f_{s,0} \\ f_{s-\qq,\qq} \\ \dots \\ f_{0,s}
\end{array} \right).
\ee
For $m\ne 0$ this equation binds $1+[s]$ irreps of the improper Lorentz group
and allows one to express components corresponding to irrep $(s \; 0)$ in
terms of components corresponding to irrep $(s-\qq\;\qq)$ and so on.
This, in turn, for $s=1,\,3/2,\,2$ allows one to pass from the first order
equations for the reducible representation to second order equations for
irrep of improper Poincar\'e group.
For example, for $s=1$, excluding $f_{1,0}$ and $f_{0,1}$, we obtain
\be
m^2 f_{\qq\qq}(x,\bz)=
[\hat p_\mu \hat V^\mu_{12},\hat p_\nu \hat V^\nu_{21} ]_+ f_{\qq\qq}(x,\bz).
\ee
In general case one also can to pass from the system of first order
equations (\ref{fseq1}) on the reducible representation to higher order
equations for irrep, for example, to the equations of $1+[s]$ order on the
components transforming under irreps $(\frac s2 \;\frac s2)$ or
$(\frac {2s+1}4 \;\frac {2s-1}4)\oplus (\frac {2s-1}4 \;\frac {2s+1}4)$
for the cases of integer or half-integer spin respectively.

Let us consider some particular cases.

For $s=j_1+j_2=1/2$ we have
\be
f_{\qq}(x,\bz)
=\chi_\alpha (x)z^\alpha + \cc\psi^{\da} (x)\cc\uz_{\da}=Z_D\Psi_D(x), \quad
\Psi_D(x)={{\chi_\alpha (x)} \choose {\cc \psi^{\da}(x)} },
\label{31fdir}
\ee
where $Z_D$ is given by the formula (\ref{dirz}).
If we substitute (\ref{31fdir}) into equation (\ref{fseq}) and
compare the coefficients at $z^\alpha$ and at $\cc\uz_{\dot\alpha}$
in the left and right side, then we obtain the Dirac equation
\be   \label{psiD}
\hat p_\mu \gamma^\mu \Psi_D(x)=m\Psi_D(x), \quad
\gamma^\mu=\left( \begin{array}{cc} 0 & \sigma^\mu
                    \\ \bar\sigma^\mu & 0 \end{array} \right).
\ee
According to (\ref{PZ}) for space inversion we obtain
$Z_D\Psi_D(x)\stackrel{P}{\to}Z_D\Psi^P_D(\bar x)=Z_D \gamma^0\Psi_D(\bar x)$,
where $\bar x =(x^0,-x^k)$.
The matrix $\gamma^5=\diag \{\sigma^0,-\sigma^0\}$ corresponds to the
chirality operator (\ref{chir}).

A complex conjugate function corresponds to charge conjugate state,
$$
\cc f_{1/2}(x,\bz) = - \psi_{\alpha}(x)\uz^\alpha
                   - \cc\chi^{\dot\alpha}(x)\cc z_{\dot\alpha},
$$
(the minus sign is from anticommutation of spinors,
$\psi_\alpha z^\alpha= -z_\alpha \psi^\alpha$) or in the matrix form
\be
Z_D\Psi_D(x)\stackrel{C}{\to}\cc Z_D\cc\Psi_D(x)=\underline Z_D \Psi^c_D(x),
\quad
\Psi^c_D(x)=-{ {      \psi_{\alpha}(x)} \choose
            { \cc\chi^{\dot\alpha} (x)} }
=i\sigma^2{ {        \psi^{\alpha}(x)} \choose
            {-\cc\chi_{\dot\alpha} (x)} },
\ee
where $\underline Z_D=(\uz^\alpha, \cc z_\da)$ and $Z_D$ obey the same
transformation law.
Thus we get the different scalar functions to describe particles and
antiparticles and hence two Dirac equations for both signs of charge
respectively. That matches completely with the results of the article
\cite{GavGi00}. It was shown there that in the course of a consistent
quantization of a classical model of spinning particle namely such
(charge symmetric) quantum mechanics appears. At the same time it
is completely equivalent to the one-particle sector of the corresponding
quantum field theory.

Real functions $f_{1/2}(x,\bz)=\cc f_{1/2}(x,\bz)$ describing Majorana
particle depend on the elements of $Z_M$ (\ref{majz}), and correspondingly
$\psi^\alpha (x)=-\chi^\alpha (x)= i\sigma^2 \chi_\alpha (x)$.
Space reflection maps this functions into the functions of $\underline Z_M$.

For $s=j_1+j_2=1$ we have
\be \label{31fDK}
f_1(x,\bz)=\chi_{\alpha\beta}(x) z^\alpha z^\beta
   +\phi_{\alpha}^{\;\;\db}(x) z^{\alpha} \cc\uz_{\dot\beta}
   +\psi^{\dot\alpha\db}(x) \cc\uz_{\dot\alpha} \cc\uz_{\dot\beta}
   =\Phi_\mu (x) q^\mu + \qq F_{\mu\nu}(x) q^{\mu\nu} ,
\ee
where
\ba \label{31qq}
q^\mu=\qq{\sigma^{\mu}}_{\alpha\db}z^\alpha \cc\uz^{\db}, \quad
q_\mu q^\mu=0, \qquad
q_{\mu\nu}=-q_{\nu\mu}=\qq \left((\sigma_{\mu\nu})_{\alpha\beta}z^\alpha z^\beta+
(\bar\sigma_{\mu\nu})_{\da\db} \cc\uz^{\da}\cc\uz^{\db}\right),
\\
\Phi_\mu(x)= -\bar\sigma_\mu^{\;\;\db\alpha} \phi_{\alpha\db}(x), \quad
F_{\mu\nu}(x)=-2\left( (\sigma_{\mu\nu})_{\alpha\beta}\chi^{\alpha\beta}(x)+
(\bar\sigma_{\mu\nu})_{\da\db}\psi^{\da\db}(x)\right).
\ea
Substituting (\ref{31fDK}) into equation (\ref{fseq}), we obtain
\ba \nonumber
&&m\psi^{\da\db}(x)=\qq\hat p_\mu \bar\sigma^{\mu\da\gamma}
 \phi_{\gamma}^{\;\;\db}(x), \quad
 m\chi_{\alpha\beta}(x)=\qq\hat p_\mu {\sigma^\mu}_{\dot\gamma\alpha}
 \phi_{\beta}^{\;\;\dot\gamma}(x),
\\ \label{DKs}
&&m\phi_{\alpha}^{\;\;\db}(x)=\hat p_\mu (\bar \sigma^{\mu\db\gamma}
 \chi_{\alpha\gamma}(x) + \sigma^\mu_{\;\;\alpha\da} \psi^{\da\db}(x)),
\\ \label{DK}
&&mF_{\mu\nu}(x)=\partial_\mu\Phi_\nu(x) - \partial_\nu \Phi_\mu(x), \qquad
  m\Phi_\mu(x) = \partial^\nu F_{\mu\nu}(x).
\ea
The Duffin--Kemmer equation in the form (\ref{DKs}) or (\ref{DK}) is the
equation for irrep $T_{[20]}$ of the 3+2 de Sitter group and thus for the
reducible representation $(1\,0)\oplus (\qq\,\qq)\oplus(0\,1)$ of the
Lorentz group.
This representation contains both four-vector $\Phi_\mu(x)$ and
antisymmetric tensor $F_{\mu\nu}(x)$, which correspond to chiralities
$\lambda =0$ and $\lambda =\pm 1$. Excluding components $F_{\mu\nu}(x)$,
we obtain second order system only for the components $\Phi_\mu(x)$
transforming under irrep $(\qq \qq)$ of the Lorentz group:
\be  \label{Proca}
(\hat { p}^2 - m^2)\Phi_\mu(x) = 0, \qquad \hat p^\mu\Phi_\mu(x) = 0.
\ee
One can rewrite operator $\hat p_\mu\hat \Gamma^\mu$ in terms of
complex variables $q^\mu$ and $q^{\mu\nu}$,
\be  \label{31gq}
\hat p_\mu\hat \Gamma^\mu = -i\hat p_\mu (q^{\mu\nu}\partial/\partial q^\nu
- q_\nu \partial/ \partial q_{\mu\nu}).
\ee
Such conversion to vector indices is possible under considering of any
integer spin. Notice that two sets of real spin variables with vector indices
can be obtained by the substitution of elements of $Z_M$ and $\underline Z_M$
instead of $Z_D$ into (\ref{31qq}).

One may describe neutral spin 1 field, in particular, by real function
of the elements of Majorana $z$-spinor, $f_1(x,\bz)=\cc f_1(x,\bz)$.
However, the spaces of quadratic functions of Dirac $z$-spinor $Z_D$
and Majorana $z$-spinor $Z_M$ are noninvariant with respect to charge
conjugation and space reflection respectively.
To describe spin 1 neutral particle coinciding with its antiparticle,
one may use bilinear functions of $Z_D$ and $\underline Z_D$.

For the cases $s=1/2$ and $s=1$ first equation of the system
(\ref{fmeq})-(\ref{fjeq}) (Klein-Gordon equation) is the consequence of
other equations. For $s>1$ the solutions of (\ref{fseq}) are characterized by
spin and mass spectrum,
$s_i=\{s,\, s-1,\dots 1\}$ or $s_i=\{s,\, s-1,\dots 1/2\}$, $m_i=ms/s_i$.
Thus for higher spin fields the Klein-Gordon equation is independent
condition, allowing to exclude from spin spectrum all spins, except maximal
$s=j_1+j_2$.

The cases $s=1/2$ and $s=1$ are also the exceptions in sense of simplicity
of labelling the components by spinor or vector indices.
The number of indices of symmetric spin-tensors, necessary for labelling
higher spin components, increases in spite of the fact that it is sufficient
to use only three operators and therefore only three numbers for
labelling of the states belonging to symmetric irreps of $SO(3,2)$.

In particular, for spin 3/2 particle there exist four kinds of components,
namely $\psi_{\alpha\beta\gamma}$, $\psi_{\da\beta\gamma}$,
$\psi_{\da\db\gamma}$, $\psi_{\da\db\dot\gamma}$, corresponding to four
possible values of the chirality. For the spin 2 particle the representation
in terms of $q^\mu$ and $q^{\mu\nu}$ is also cumbersome,
\be
f_2(x,q)=\Phi_{\mu\nu}(x) q^\mu q^\nu + \qq F_{\mu\nu,\rho}(x)
q^{\mu\nu}q^\rho + \frac 14 F_{\mu\nu,\rho\sigma}(x) q^{\mu\nu}q^{\rho\sigma},
\ee
with the necessity to fix independent components by means of relations
$q_\mu q^\mu=0$, $q_{\mu\nu} q^\mu + q_{\mu\nu} q^\nu=0$ and so on.

Thus, beginning from the spin 3/2, it is convenient to use the universal
notation $\psi^{m_1m_2}_{j_1j_2}(x)$ associated with the decomposition
(\ref{31decomp}) over monomial chiral basis (\ref{31zbas}) (see also
(\ref{31bas5})-(\ref{31dec})). Two indices $j_1,j_2$ label spin $s=j_1+j_2$
and chirality $\lambda=j_1-j_2$ and two indices $m_1,m_2$ label independent
components inside the irrep of the Lorentz group. This notation are also
suitable for infinite-dimensional representations.

By analogy with 2+1 case, one can find plain wave solutions of the system
(\ref{fmeq})-(\ref{fseq}) for the any spin $s$ in general form without
using matrix representation. Corresponding to the particle moving along $x^3$
states are eigenstates of the operator $\hat p_i\hat S^i$
with eigenvalues $|p|\sigma$, where $\sigma=s_3{\rm sign}\,p_3$ is the
helicity. These states have the form
\ba \nonumber
f_{m,s,\sigma}(x,\bz)=& \sum_{\sigma=-s}^s C_{\sigma} e^{ik_0x^0+k_3x^3}
 (z^1e^a+\cc\uz_{\dot 1}e^{-a})^{s+\sigma}(z^2e^{-a}+\cc\uz_{\dot 2}e^a)^{s-\sigma}+
\\ \label{sol4}
 & \sum_{\sigma'=-s}^s C_{\sigma'} e^{-ik_0x^0-k_3x^3}
 (z^1e^a-\cc\uz_{\dot 1}e^{-a})^{s-\sigma}(z^2e^a-\cc\uz_{\dot 2}e^{-a})^{s+\sigma},
\ea
where $e^a$ is given by (\ref{sol2a}). For the rest particle one can obtain
the general solution characterized by the spin projection $s'$ on the
direction ${\bf n}$ from (\ref{sol1}) by the rotation
$z'_\beta =U^{\;\;\alpha}_\beta z_\alpha$, $U\subset SU(2)$.
For particle characterized by momentum direction ${\bf n}$ and
helicity $\sigma$, starting from the state (\ref{sol4}), one can obtain
the solution by the analogous rotation.

Improper Poincar\'e group includes space reflection, which interchanges
representations $(j_1\,j_2)$ and $(j_2\,j_1)$. Therefore we consider
equations binding these representations (and corresponding components
of the solutions of the system (\ref{fmeq})-(\ref{fjeq})) more detail.

In the case $j_1=j_2$ solutions of the system (\ref{fmeq})-(\ref{fjeq})
are characterized by fixed spin $s=j_1+j_2$ and mass $m$. Thus the relations
(\ref{31m})-(\ref{31j}) are valid, and corresponding $2(2s+1)$
components obey the equations for massive tensor field (\ref{31teneq}).

In general case equations connecting the components
transforming under irreps $(j_1,j_2)$ and $(j_2,j_1)$ of the Lorents
subgroup have the form
\ba   \nonumber
&&(2j_2)!(\hat p_\mu\hat V_{12}^\mu)^{2|\lambda|}f_{j_1\,j_2}(x,\zz) =
  (2j_1)!m^{2|\lambda|} f_{j_2\,j_1}(x,\zz), \quad
\\
&&(2j_1)!(\hat p_\mu\hat V_{21}^\mu)^{2|\lambda|}f_{j_2\,j_1}(x,\zz) =
  (2j_2)!m^{2|\lambda|} f_{j_1\,j_2}(x,\zz),     \label{chet}
\ea
where $j_1>j_2$, $|\lambda|=j_1-j_2$. This set of equations are invariant
under space reflection. Using the decomposition (\ref{decjj}) and explicite
form of the general solution (\ref{sol1}) of the system
(\ref{fmeq})-(\ref{fjeq}) in the rest frame, one can prove the validity of
(\ref{chet}) by direct calculation.
Going over to spin-tensor notation, we get
\ba  \nonumber
&&p_\mu\bar\sigma^{\mu\da_{2j_2+1}\alpha_{2j_2+1}} \dots
  p_\nu\bar\sigma^{\nu\da_{2j_1}\alpha_{2j_1}}
 \psi_{\alpha_1\dots\alpha_{2j_1}}^{\da_1\dots\da_{2j_2}}(x)
=m^{2|\lambda|} \psi_{\alpha_1\dots\alpha_{2j_2}}^{\da_1\dots\da_{2j_1}}(x),
\\   \label{chet1}
&&p_\mu\sigma^\mu_{\;\;\alpha_{2j_2+1}\da_{2j_2+1}} \dots
  p_\nu\sigma^\nu_{\;\;\alpha_{2j_1}\da_{2j_1}}
 \psi_{\alpha_1\dots\alpha_{2j_2}}^{\da_1\dots\da_{2j_1}}(x)
=m^{2|\lambda|} \psi_{\alpha_1\dots\alpha_{2j_1}}^{\da_1\dots\da_{2j_2}}(x).
\ea
Equations (\ref{chet1}) are consequences of the system
(\ref{fmeq})-(\ref{fjeq}), but unlike this system in general case require
some supplementary conditions to fix mass and spin.

Equations (\ref{chet1}) are the first order equations only in the case
$|\lambda|=1/2$, which corresponds to representations
$(\frac{2j\pm 1}4 \, \frac{2j\mp 1}4)$, $j=j_1+j_2$, describing half-integer
spins. In this case, going over to vector indices and supplementing the
equations by subsidiary conditions (\ref{31subs2}) (which also are the
consequences of the system (\ref{fmeq})-(\ref{fjeq}) and exclude components
with $s<j_1+j_2$), we obtain the Rarita--Schwinger equations \cite{RarSc41}
\be   \label{rars1}
(\hat p_\mu \gamma^\mu-m)\Psi_{\mu_1\mu_2\dots\mu_n}(x)=0, \quad
\gamma^\mu \Psi_{\mu\mu_2\dots\mu_n}(x)=0,
\ee
where $n=2s-1$ and $\Psi_{\mu_1\dots\mu_n}(x)$ is a four-component column
composed of $\Psi_{\mu_1\dots\mu_n\alpha}(x)$ and
${\Psi_{\mu_1\dots\mu_n}}^{\dot\alpha}(x)$.
The conditions $\partial^\mu \Psi_{\mu\mu_2\dots\mu_n}(x)=0$ and
${\Psi^\mu}_{\mu\dots\mu_n}(x)=0$ appear as consequences of these two
equations \cite{Ohnuk88}.

A case $|\lambda|=s$ corresponding to representations $(s\,0)$ and $(0\,s)$
is preferred because of minimal number of components. In this case the
equations (\ref{chet1}) are $2s$ order Joos-Weinberg equations
\cite{Joos62,Weinb64,Weinb69} of so called $2(2s+1)$-component theory,
\ba  \nonumber
&&p_\mu\bar\sigma^{\mu\da_1\alpha_1} \dots
  p_\nu\bar\sigma^{\nu\da_{2s}\alpha_{2s}}
 \psi_{\alpha_1\dots\alpha_{2s}}(x) = m^{2s} \psi^{\da_1\dots\da_{2s}}(x),
\\   \label{chet2}
&&p_\mu\sigma^\mu_{\;\;\alpha_1\da_1} \dots
  p_\nu\sigma^\nu_{\;\;\alpha_{2s}\da_{2s}}
 \psi^{\da_1\dots\da_{2s}}(x) = m^{2s} \psi_{\alpha_1\dots\alpha_{2s}}(x).
\ea
In the rest frame as a consequence we obtain $p_0^{4s}=m^{4s}$ and for
$s\ge 1$ Joos-Weinberg equations have solutions with complex energy $p_0$,
$|p_0|=m$. The existence of such solutions was pointed out also in
\cite{AhlEr92}.

\subsection{Relativistic wave equations, invariant under improper
            Poincar\'e group. \\ Equations for several scalar functions}

Above we have considered the linear equations for {\it one} scalar function on
the group. The condition of invariance under space reflection leads us to the
system (\ref{fmeq})-(\ref{fjeq}) for particle with spin $s=j_1+j_2$ and mass
$m$.

For the construction of invariant wave equations one may also use the
operators $\hat p_\mu \hat V^\mu_{ik}$, which are not invariant under space
reflections. Using several scalar functions $f(x,\bz)$, it is possible to
restore the invariance under space reflections.

In particular, equations (\ref{V12}) containing operators $\hat V^\mu_{12(k)}$
and $\hat V^\mu_{21(k)}$ interlock two scalar functions.
Using the decomposition (\ref{31ten}) in terms of spin-tensors, we obtain
Dirac--Fierz--Pauli equations \cite{Dir36,FiePa39},
\ba \nonumber
&&\hat p_\mu\bar\sigma^{\mu\da\beta}
 \psi^{\da_1\,\dots\,\da_l}_{\beta\,\beta_1\,\dots\,\beta_n}(x)=
 \varkappa
 \psi^{\da\,\da_1\,\dots\,\da_l}_{\beta_1\,\dots\,\beta_n}(x)
\\  \label{DFPeq}
&&\hat p_\mu {\sigma^\mu}_{\beta\da}
 \psi^{\da\,\da_1\,\dots\,\da_l}_{\beta_1\,\dots\,\beta_n}(x)=
 \varkappa
 \psi^{\da_1\,\dots\,\da_l}_{\beta\,\beta_1\,\dots\,\beta_n}(x).
\ea
These equations connect two functions transforming under irreps
$(\frac n2 +\qq\,\frac l2)$ and $(\frac n2\,\frac l2 +\qq)$ of the Lorentz
group and at $n=l$ mapping to one another under parity transformation.

Let us consider a system of equations of the form (\ref{V12}),(\ref{V11}),
which connect several scalar functions with different $j_1,j_2$. The equations
of this system interlock the representation $(j_1,j_2)$ with at least one of
the representations $(j_1\pm 1,j_2\mp 1)$, $(j_1\pm 1,j_2\pm 1)$.
This allows one to identify this system with general Gel'fand--Yaglom
equations \cite{GelYa48,GelMiS63}
\be \label{Geleq}
(\alpha^\mu \hat p_\mu - \varkappa )\psi=0, \quad
[S^{\lambda\mu},\alpha^\nu]
=i(\eta^{\mu\nu}\alpha^\lambda - \eta^{\lambda\nu}\alpha^\mu ) .
\ee
In the present approach the latter relation is a consequence of the
commutation relations $[\hat S^{\lambda\mu},\hat V^\nu_{ik}]
=i(\eta^{\mu\nu}\hat V^\lambda_{ik} - \eta^{\lambda\nu}\hat V^\mu_{ik} )$.
This relation is necessary for Poincar\'e invariance of the equation
\cite{BarRa77,GelMiS63}.

Supplemented by commutation relations $[\alpha^\mu ,\alpha^\nu]= S^{\mu\nu}$
finite-component equations of the form (\ref{Geleq})
are known as Bhabha equations \cite{Bhabh45}, although for the first time
they were systematically considered by Lubanski \cite{Luban42}. These
equations are classified according to the finite-dimensional irreps of the
3+2 de Sitter group $SO(3,2)$. Other possible commutation relations of
matrices $\alpha^\mu$ are discussed in \cite{Caste67}.

Considered above equation (\ref{fseq}) on a scalar function is the particular
case of Bhabha equations. This case correspond to symmetric irreps
$T_{[2s\,0]}$ of the 3+2 de Sitter group. Generally speaking, the Bhabha
equations are characterized by finite number of different $m$ and $s$.
Therefore, this equations connect the fields transforming under
nonequivalent irreps of the Poincar\'e group.

If equations include the operators $\hat p_\mu \hat V^\mu_{11}$ and
$\hat p_\mu \hat V^\mu_{22}$, then either the equations describe at least
two different spins $s$ or the condition $s=j_1+j_2$ connecting spin $s$ with
a highest weight of irrep of Lorentz group is not valid.

Cite as an example the system interlocking irreps (00) and
$(\frac 12 \frac 12)$ of the Lorentz group:
\be  \label{sp01}
\hat p_\mu\hat V_{11}^\mu f_{00}(x,\bz)=\varkappa_1 f_{\qq \qq}(x,\bz), \quad
\hat p_\mu\hat V_{22}^\mu  f_{\qq \qq}(x,\bz)=\varkappa_2 f_{00}(x,\bz)
\ee
where $f_{00}(x,\bz)=\psi (x)$,
$f_{\qq \qq}(x,\bz)=\psi_{\alpha}^{\;\;\db}(x) z^{\alpha}\cc\uz_\db$;
in component-wise form we have $\hat p_\mu\psi= 2\varkappa_1\psi_\mu$,
$\hat p_\mu\psi^\mu= \varkappa_2\psi$.
In the rest frame one may obtain $\varkappa_2=2\varkappa_1=m$.
Thus the system (\ref{sp01}) is equivalent to Duffin equation for scalar
particles, which correspond to five-dimensional vector irrep $T_{[01]}$ of
$SO(3,2)$ group.

\subsection{Relativistic wave equations, invariant under improper
            Poincar\'e group. \\ Equations for particles with composite spin}

Many-particle systems are described by the functions of the sets of variables
$x_{(i)},z_{(i)},\cc\uz_{(i)}$. But here we will consider not many-particle
systems in usual sense, but some objects corresponding to functions
$f(x,z_{(1)},\cc\uz_{(1)},\dots, z_{(n)},\cc\uz_{(n)})$, (or, briefly,
$f(x,\{\bz_{(i)} \})$), i.e. to functions of one set of $x$ and several sets
of $\bz$. These objects one may interpret as particles with composite spin.

As an example we will consider the Ivanenko-Landau-K\"{a}hler (or
Dirac-K\"{a}hler) equation \cite{IvaLa28,Kahle62}. Let us write some linear
in $\bz_{(1)}$ and $\bz_{(2)}$ scalar function $f(x,\bz_{(1)},\bz_{(2)})$ in
the form
\be
f(x,\bz_{(1)},\bz_{(2)})=Z^{(1)}_D \Psi (x) (Z^{(2)}_D)^\dagger=
\sum_{i,j=1}^4 (Z_D^{(1)})_i \Psi_{ij}(x) (\cc Z_D^{(2)})_j,
\ee
where $Z_D=(z^1\,z^2\,\cc\uz_{\dot 1}\,\cc\uz_{\dot 2})$, and $\Psi (x)$
is a $4\times 4$ matrix with a transformation rule
$$
\Psi'(x')={\check U}\Psi(x) ({\check U})^\dagger, \quad
\check U=\diag\{ U,(U^{-1})^\dagger \},
$$
in contrast to the transformation rule $\Psi_D'(x')={\check U}\Psi_D(x)$
of Dirac spinor (\ref{psiD}).
Let us impose the equation on the first ("left") spin subsystem,
\be  \label{ILK0}
(\hat p_\mu\hat\Gamma^\mu_{(1)}-m/2)f(x,\bz_{(1)},\bz_{(2)})=0,
\ee
and do not impose any conditions on the second ("right") spin subsystem.
Writing (\ref{ILK0}) in component-wise form, one can obtain
Ivanenko-Landau-K\"{a}hler equation in spinor matrix representation
\be  \label{ILK}
(\hat p_\mu \gamma^\mu -m)\Psi(x)=0.
\ee
According to (\ref{ILK}), 16 components $\Psi_{ij}(x)$ obey Klein-Gordon
equation, therefore mass is equal to $m$.
Spin of both subsystems is equal two 1/2. The spin of system is indefinite,
and there are both spin 0 and spin 1 components.

The consideration of this equation is associated mainly with the attempts to
describe fermions by the antisymmetric tensor fields (see, for example,
\cite{BenTu88,BenTu83,Bulli86,ObuSo93} and also \cite{IvaObS85} as a good
introduction). The spin subsystems ("left-spin" and "right-spin")
were considered in \cite{BenTu83,ObuSo93}).

Let us consider now linear symmetric functions of $\bz_{(1)},\dots,\bz_{(n+l)}$:
\be  \label{fsym1}
f_{\frac n2 \frac l2}(x,\{\bz_{(i)} \}) =
\psi^{\dot\alpha_1,\dots,\dot\alpha_l}_{\beta_1,\dots,\beta_n}(x)
\sum z_{(1)}^{\beta_1}\dots z_{(n)}^{\beta_n}
\cc\uz_{(n+1)\dot\alpha_1}\dots \cc\uz_{(n+l)\dot\alpha_l},
\ee
where symmetric spinors
$\psi^{\dot\alpha_1,\dots,\dot\alpha_l}_{\beta_1,\dots,\beta_n}(x)$
transform under irreps $(n/2,l/2)$, and all permutations of $1,\dots,n+l$
are summed over. As a consequence of symmetry of the multispinors with
respect to index permutations, spin subsystems are indistinguishable, and
this allow us to use functions of several sets of spin variables for
describing usual particles.

So, one may obtain Dirac--Fierz--Pauli equations (\ref{DFPeq}), acting
by the operators $\hat V^\mu_{12(k)}$ and $\hat V^\mu_{21(k)}$ on the
functions (\ref{fsym1}) corresponding to irreps
$(\frac n2 +\qq,\frac l2)$ and $(\frac n2, \frac l2 +\qq)$
of the Lorentz group:
\ba \nonumber
&&\hat V^\mu_{12(k)} f_{\frac n2+\qq,\frac l2}(x,\{\bz_{(i)}\})
 =\varkappa f_{\frac n2,\, \frac l2+\qq}(x,\{\bz_{(i)}\}), \quad   \\
&&\hat V^\mu_{21(k)} f_{\frac n2,\frac l2+\qq}(x,\{\bz_{(i)}\})
 =\varkappa f_{\frac n2+\qq,\, \frac l2}(x,\{\bz_{(i)}\}). \quad
\ea

In general case a linear symmetric function of $\bz_{(k)}$, $k=1,\dots,2j$
has the form
\be  \label{fsym2}
f_j(x,\{ \bz_{(i)} \}) = \sum_{n,l;n+l=2j}
\psi^{\da_1\,\dots\,\da_l}_{\beta_1\,\dots\,\beta_n}(x)
z_{(1)}^{\beta_1}\dots z_{(n)}^{\beta_n}
\cc\uz_{(n+1)\da_1}\dots \cc\uz_{(n+l)\da_l} .
\ee
Functions (\ref{fsym2}) correspond to symmetric part of the representation
$\left( (\qq\,0)\oplus(0\,\qq) \right)^{2j}$. This symmetric part expand
into direct sum of irreps $(j_1\, j_2)$, $j_1+j_2=j$.
Impose on each spin subsystem the condition
\be  \label{BarWig}
(\hat p_\mu\hat\Gamma^\mu_{(k)}-m/2)f(x,\bz_{(1)},\dots,\bz_{(2j)})=0, \quad
k=1,\dots,2j.
\ee
Rewriting this equations in four-component form, we obtain Bargmann--Wigner
equations \cite{BarWi48,Ohnuk88,Grein97}
\be
(\hat p_\mu \gamma^\mu_{(k)}-m)_{\alpha_k\beta_k}
\psi_{\beta_1\dots \beta_k \dots \beta_{2j}}(x)=0.
\ee
As a consequence of (\ref{BarWig}), one may obtain equations for system
as a whole:
\ba   \nonumber
&&(\hat p^2 -m^2) f(x,\bz_{(1)},\dots,\bz_{(2j)}) = 0,
\\    \label{fseqbw}
&&(\hat p_\mu \hat\Gamma^\mu - ms)f(x,\bz_{(1)},\dots,\bz_{(2j)})=0, \quad
\hat\Gamma^\mu=\sum \hat\Gamma^\mu_{(k)},
\ea
which are analogous with equations (\ref{fmeq})-(\ref{fseq}) at the case
$s=j_1+j_2$. Both Bargmann--Wigner equations and system (\ref{fseqbw}) have
$2(2s+1)$ independent solutions $\psi(x)$, and therefore
this systems are equivalent.

\subsection{Relativistic wave equations: comparative consideration}

In the framework of the group-theoretical classification of the scalar fields
$f(x,\bz)$ on the Poincar\'e group we have obtained two types of equations
describing unique spin and mass, namely equations for the eigenfunctions
of the Casimir operator of the Lorentz spin subgroup ($j_1$ and $j_2$ are
fixed, see (\ref{31j})) and equations for the eigenfunctions of the Casimir
operator of the $SO(3,2)$ group (the sum $j_1+j_2$ is fixed). Below we will
consider comparative characteristics of these equations and also the case
$(j_1\,j_2)\oplus(j_2\,j_1)$ corresponding to irreps of the improper
Poincar\'e group but requiring two scalar functions for its formulation.

1. Equations for the functions corresponding to the fixed irrep $(j_1\, j_2)$
of the Lorentz group. Mass and spin irreducibility conditions
leave $2(2s+1)$ independent components corresponding to two improper
Poincar\'e group irreps differed by sign of $p_0$. For $s=j_1+j_2$
the equations in spin-tensor form constitute the system of the Klein-Gordon
equation and the subsidiary condition (\ref{aasupp}), which
eliminates components with other possible values of spin $s$ for fixed
$j_1,j_2$, $|j_1-j_2|\le s <j_1+j_2$.
For $s\ne j_1+j_2$ one should consider general subsidiary condition
(\ref{gensub}). An alternative to the use of subsidiary condition
is the consideration of functions of momentum and spin variables
with invariant constraints (\ref{pqconstr}).

There exist two preferred cases. The first corresponds to the representations
$(\frac s2\,\frac s2)$ mapping onto themselves under space
reflection and are most often used to describe integer spins.
The second corresponds to the representations $(s\,0)$ and $(0\,s)$. In
this case there is not necessity to impose subsidiary conditions since
they are fulfilled identically.

2. Equations for the functions corresponding to the representations
$(j_1\,j_2)$ and $(j_2\,j_1)$, $j_1\ne j_2$, which are interchanged under
space reflection. Unlike the considered above
equations for fixed $j_1,j_2$, these equations in general case do not assume
formulation as equations for one scalar function $f(x,\bz)$.
The conditions of mass and spin irreducibility leave $4(2s+1)$
independent components corresponding to four improper Poincar\'e group
irreps differed by sign of $p_0$ and intrinsic parity $\eta$.
To choose $2(2s+1)$ components corresponding to fixed sign of $\eta$ or
$p_0\eta$ it is necessary to supplement these conditions by
equations (\ref{chet}) connecting components corresponding to
$(j_1\,j_2)$ and $(j_2\,j_1)$.

Equations (\ref{chet}) are first order equations only for the representations
$(j\!+\!\qq\; j)\oplus(j\; j\!+\!\qq)$.
These representations and associated with them Rarita--Schwinger equations
(\ref{rars1}) are most often used to describe half-integer spins.
However, just as in the case of representations $(j\,j)$,
subsidiary conditions supplement the field equations, and the number of
equations exceeds the number of field components.
Therefore one has an overdetermined set of equations which, although
consistent in the free-field case, for $s>1$ becomes self-contradictory with
minimal electromagnetic coupling \cite{FiePa39}. In order to avoid
inconsistency it is possible to give a Lagrangian formulation, introducing
auxiliary fields \cite{FiePa39,SinHa74,SinHa74a}, but this formulation leads
to acasual propagation with minimal electromagnetic coupling
\cite{Wight78,Zwanz78,Tung67,VelZw69,
Velo72,CapKo80}.

For the case $(s\,0)\oplus(0\,s)$ one can construct 2(2s+1)-component
theory, but corresponding Joos-Weinberg equations of $2s$ order
\cite{Joos62,Weinb64} (see (\ref{chet2})) for $s\ge 1$ have also solutions
with complex energy.

The second order equation for representation $(s\,0)\oplus(0\,s)$,
$(\hat P^2-\frac e{2s}\hat S^{\mu\nu}F_{\mu\nu}-m^2)\psi(x)=0$,
\cite{FeyGe58,Iones67,Hurle71,Hurle74} for free particle possesses $4(2s+1)$
independent components differed by spin projection and by signs of $p_0$ and
$\eta$. On the other hand, this equation describes unique mass and spin and
is characterized by casual solutions. In particular, exact solutions in
external constant uniform electromagnetic field are known \cite{Krugl99}.
One may rewrite the above equation as first order equation with minimal
coupling for representations
$(s\,0)\oplus(s\!-\!\qq\;\qq)\oplus(\qq\;s\!-\!\qq)\oplus(0\,s)$.
As noted in \cite{Tung67}, this is the simplest class of describing
unique mass and spin representations, which led to first order equations
without subsidiary conditions.

3. Equations (\ref{fmeq})-(\ref{fseq}) for eigenfunctions of the Casimir
operator (\ref{fjeq}) of $SO(3,2)$ group with eigenvalues $4s(s+2)$,
$s=j_1+j_2$:
\be  \label{31sys}
(\hat p_\mu \hat \Gamma^\mu - ms)f(x,\zz)=0, \quad (\hat p^2-m^2)f(x,\zz)=0.
\ee
The condition of spin irreducibility (\ref{31s1}) is a consequence of this
system.

The first equation of the system is the Bhabha equation \cite{Luban42,Bhabh45}
corresponding to symmetric irrep $T_{[2s\;0]}$ of the group $Sp(4,R)\sim SO(3,2)$.
This equation represent a straightforward higher spins generalization of
Dirac and spin 1 Duffin--Kemmer equations.
Both Bhabha equations and the problem of minimal coupling for these equations
were most detail considered in papers of Krajcik and Nieto
(see \cite{KraNi77}; it contains references to the six earlier papers).
The theory is casual with minimal electromagnetic coupling \cite{KraNi76},
but in general case Bhabha equations describe multi-mass systems.
Notice that the connection of the Rarita--Schwinger and Bargmann--Wigner
equations with Bhabha equations was considered also in \cite{LoiOtS97}.

The solutions of the system (\ref{31sys}) have the components transforming
under $2s+1$ irreps $(j_1,j_2)$, $j_1+j_2=s$, of the Lorentz group.
But the components corresponding to different chiralities $\lambda=j_1-j_2$
are not independent. In contrast to left generators of the Poincar\'e group
operators $\hat\Gamma_\mu$ do not commute with chirality operator
(which is the right generator of the Poincar\'e group) and combine $2s+1$
representations of the Lorentz group into one irrep of the 3+2 de Sitter
group $SO(3,2)$.

The current component $j^0$ is positive definite for half-integer
spin particles and the energy density is positive definite for integer spin
particles, see Appendix B.

In the rest frame equations (\ref{31sys}) have $2s+1$ positive- and $2s+1$
negative-frequency solutions labelled by different spin projections,
see (\ref{sol1}), and half-integer spins solutions with opposite frequency
are characterized by opposite parity.
In the ultrarelativistic limit two solutions with opposite sign of $p_0$
correspond to any of $2s+1$ of possible values of chirality, see (\ref{sol3}).

Thus the system (\ref{31sys}) describes a particle with unique spin and mass,
is invariant under parity transformation and possesses $2(2s+1)$ independent
components.

Let us briefly consider the problem of equivalence of the different
RWE. In the case of free fields, using the relation
\be  \label{comm0}
[\partial_\mu,\partial_\nu]=0,
\ee
one can establish the equivalence of wide class of RWE.

So, as we have established above, in a free case the system (\ref{31sys})
and the Bargmann--Wigner equations (\ref{BarWig}), which both describe a
particle by means wave functions with components transforming under $2s+1$
irreps $(j_1\,j_2)$, $j_1+j_2=s$, of the Lorentz group, are equivalent.
However, the formulation (\ref{31sys}) is more general since unlike
Bargmann--Wigner equations can be considered also in the case of
infinite-dimensional unitary representations of the Lorentz group, as was
done above with an analogous system in 2+1-dimensional case.

The considered above free equations for representations $(j_1\,j_2)$ or
$(j_1\,j_2)\oplus(j_2\,j_1)$ can be obtained as a consequence of the
Bargmann--Wigner equations \cite{Ohnuk88,Grein97} or the system (\ref{31sys})
by excluding of other components. In general case for $m\ne 0$ one may
express all components in terms of the components corresponding to two
chiralities $\pm\lambda$, where $-s\le\lambda\le s$.

It is obvious that the coupling, which is minimal for one system, is not
minimal for another "equivalent" system if one uses the relation
(\ref{comm0}) to prove this equivalence in a free case. These equations
will differ by the terms proportional to the commutator of covariant
derivatives $[D_\mu,D_\nu]=igF_{\mu\nu}$.

Therefore, when an interaction is introduced, the system of equations can be
found inconsistent if, taking account of (\ref{comm0}), some equations are
the consequences of another.
In particular, spin 1 Bargmann--Wigner equations with minimal electromagnetic
coupling are inconsistent (for the prove see, for example, \cite{BucSh93}),
but equivalent to them in a free case Duffin--Kemmer and Proca equations with
minimal coupling are consistent and characterized by casual solutions
\cite{VelZw69}.

Recently different approaches to introduce interactions for higher spin
massive fields have been considered
(see, in particular, \cite{Krugl99,BucKrP99,BucGiK00,Klish00}).
Keeping in mind the present approach, we hope that new possibilities
to describe interacting higher spin fields will arise.


\section{Equations for fixed spin and mass: general features}

Consider now the general properties of the obtained equations describing a
particle with unique mass $m>0$ and spin $s$ in two dimensions
\ba
&&\hat p^2  f(x,\theta) = m^2 f(x,\theta), \label{11m0}
\\
&&\hat p_\mu\hat\Gamma^\mu f(x,\theta) = ms  f(x,\theta), \label{11s0}
\ea
in three dimensions
\ba
&&\hat p^2 f(x,\bz) = m^2   f(x,\bz), \label{12m0}
\\
&&\hat p_\mu\hat S^\mu f(x,\bz) = ms    f(x,\bz), \label{12s0}
\\
&&\hat S_\mu\hat S^\mu f(x,\bz) = S(S+1)f(x,\bz), \label{12S0}
\ea
in four dimensions
\ba
&& \hat{ p}^2 f(x,\bz) = m^2 f(x,\bz),           \label{fmeq0}
\\
&&\hat p_\mu \hat\Gamma^\mu f(x,\bz) = ms f(x,\bz), \label{fseq0}
\\
&&\hat S_{ab}\hat S^{ab}f(x,\bz) = 4S(S+2)f(x,\bz). \label{fjeq0}
\ea
In the latter case in addition we suppose $s=\pm S$ to avoid nontrivial spin
and mass spectrum.

In all dimensions the first equation (condition of the mass irreducibility)
is the eigenvalue equation for the Casimir operator of the Poincar\'e group.
But the other equations, although seem similar, have different origin in even
and odd dimensions. This is related to the different role of space inversion.

In 2+1 dimensions other equations (\ref{12s0})-(\ref{12S0}) are eigenvalue
equations for the Casimir operator of the Poincar\'e group and the
spin Lorentz subgroup.

In even dimensions the Casimir operators of the Lorentz subgroup do not
commute with the space inversion operator, and space inversion combines two
labelled by chiralities $\pm \lambda$ equivalent representations of the
proper Poincar\'e group into representation of the improper Poincar\'e group.
If one rejects equations, which fix chirality (in 3+1 dimensions this
corresponds to the transition to the system (\ref{31m1})-(\ref{31j1})), then
in the rest frame it is easy to see that there is redundant number of
independent components. Thus it is necessary to construct equation, binding
the states with different chiralities, and correspondingly new set of
commuting operators.
It can be down by using supplementary operators $\hat \Gamma^\mu$, which
extend Lorentz group $SO(D,1)$ up to $SO(D,2)$ group with the maximal compact
subgroup $SO(D)\otimes SO(2)$. Operator $\hat\Gamma^0$ is the generator of
compact $SO(2)$ subgroup.

Third equation of the system fixes the power $2S$ of homogeneity of the
functions $f(x,\bz)$ in $\bz$ and therefore fixes irrep of the Lorentz
group in 2+1 dimensions or of the 3+2 de Sitter group in 3+1 dimensions.
(In 1+1 dimensions there exists analogous equation
$\hat \Gamma_a\hat \Gamma^a f(x,\theta)=s(s+1)f(x,\theta)$, but, in fact,
this equation defines the structure of $\hat\Gamma^\mu$.)

Positive (half-)integer $S=s$ correspond to finite-dimensional nonunitary
irreps of the Lorentz (or de Sitter) group, which realize in the space of
the power $2s$ polynomials in $\bz$.

Negative $S=-s$ correspond to infinite-dimensional unitary irreps. The
unitary property allows one to combine probability amplitude interpretation
and relativistic invariance (the desirability of this combination was
stressed by Dirac in \cite{Dir72}).
Thus the equations under consideration allow two approaches to the
description of the same spin by means of both finite-dimensional
nonunitary and infinite-dimensional unitary irreps.

In 1+1 and 2+1 dimensions there is the possibility of the existence of
particles with fractional spin since the groups $SO(1,1)$ and $SO(2,1)$ do
not contain compact Abelian subgroup. However, the description of massive
particles with fractional spin can be given only in terms of the
infinite-dimensional irreps of the group $SO(2,1)$. This is another reason
to consider infinite-dimensional irreps.

Fixing the irrep of the Lorentz (or de Sitter) group with the help of the
third equation of the system, one can come to usual multicomponent matrix
description by the separation of space and
spin variables: $f(x,\bz)=\sum \phi_n(\bz)\psi_n(x)$, where $\phi_n(\bz)$ form the
basis in the representation space of the Lorentz (or de Sitter) group.
Thus, depending on the choice of the solution of the third equation,
second equation in matrix representation is either finite-component
equation or infinite-component equation of Majorana type.

For fundamental spinor irreps the action of differential operators
$2\hat S^\mu$ in 2+1 dimensions and $2\hat \Gamma^\mu$ in 1+1 and 3+1
dimensions in the space of functions $f(x,\bz)$ on the Poincar\'e group can
be rewritten in terms of action of corresponding $\gamma$-matrices on the
functions $\psi(x)$.

Differential operators $\hat \Gamma^\mu$ and matrices $\gamma^\mu/2$
obey the same commutation relations
$$
[\hat \Gamma^\mu,\hat \Gamma^\nu]=-i\hat S^{\mu\nu}, \quad
[\hat S^\mu,\hat S^\nu]=-i\epsilon^{\mu\nu\rho}\hat S_\rho.
$$
In 3+1 dimensions operators $\hat \Gamma^\mu$ and $\hat S^{\mu\nu}$ obey the
commutation relations of generators of $SO(3,2)$ group, see (\ref{deSitt}).

Anticommutation relations for operators $\hat S^\mu$ in 2+1 and
$\hat \Gamma^\mu$ in 1+1 and 3+1 dimensions are analogous with the relations
for $\gamma$-matrices,
$$
[\hat S^\mu,\hat S^\nu]_+=\qq\eta^{\mu\nu},\quad
[\hat \Gamma^\mu,\hat \Gamma^\nu]_+=\qq\eta^{\mu\nu},
$$
and are valid only for fundamental spinor irreps. This is group-theoretical
property connected with the fact that for these irreps the double action of
lowering or raising operators on any state gives zero as a result.
(Notice that, besides the case of spinor irreps of orthogonal groups,
anticommutation relations also take place for fundamental $N$-dimensional
irreps of $Sp(N)$ and $SU(N)$ groups \cite{GitSh98}.)

For $s=1/2$ and $s=1$ the first equation of the system (condition of mass
irreducibility) is a consequence of (\ref{12s0}) or (\ref{fseq0}).
In general case the second equation of the system describes multi-mass
systems $m_is_i=ms$. Thus for $s>1$ it is necessary to consider both equations.

Consider some characteristics of the equations associated with
finite-dimensional irreps of the Lorentz group.
If we reject the first equation of the system (i.e. the condition of mass
irreducibility), then for the second equation of the system the component
$j^0$ of the current vector is positive definite only for $s=1/2$, and the
energy density $-T^{00}$ (see (\ref{Ep})) is positive definite only for $s=1$.
(The case $s=1$ in 3+1 dimensions has considered in detail in
\cite{GelMiS63,Ghose96}).
However, for the system as a whole the component $j^0$ of the current vector
is positive definite for any half-integer spin, and energy density is positive
definite for any integer spin. Besides, in the rest frame half-integer spin
solutions with opposite sign of $p_0$ are characterized by opposite parity.

For the case of infinite-component equations in 2+1 dimensions,
the energy is positive definite for any spin, and $j^0$ is positive or
negative definite in accordance with the sign of charge.

The consideration of the field on the Poincar\'e group also allows one to
ensure essential progress in the problem of practical computations for
multicomponent equations.
As was noted in \cite{Ginzb56}, the general investigation of Gel'fand--Yaglom
equations "revealed a number of interesting features, but ...
the use of such equations (or more accurately, systems of a large or
infinite number of equations) for any practical computations is not possible".
In the present approach, due to the use of spin differential operators
instead of finite or infinite-dimensional matrices, from the technical
point of view there is no essential
distinction in the consideration of the equations associated with
various finite-dimensional and infinite-dimensional representations of
the Lorentz group. Therefore the present approach is adequate to work with
higher spins and positive energy wave equations.
For example, the use of spin variables $\bz$ has allowed us to obtain
explicit compact form of general plane wave solutions for any spin
(including fractional spin in 2+1 dimensions).

Notice that unlike the equations for particles with unique mass and
spin, in general case RWE with mass and spin spectrum can either interlock
several scalar functions $f(x,\bz)$ (as general Gel'fand--Yaglom equations
and, in particular, Bhabha equations)
or describe objects with composite spin, which correspond to the functions
$f(x,\bz_{(1)},\dots \bz_{(n)})$ of one set of space-time coordinates $x$ and
several sets of spin coordinates $\bz$ (as Ivanenko-Landau-K\"{a}hler or
Dirac-K\"{a}hler equation).

\section{Conclusion}

In this paper we have elaborated a general scheme of analysis of
fields on the Poincar\'e group and have applied it in two, three and
four-dimensional cases.

Considering the left GRR of the Poincar\'e group,
we introduce the scalar field $f(x,\bz)$ on the group, where $x$ are
coordinates in Minkowski space and $\bz$ are coordinates on the Lorentz group.
The connection between the left GRR and the scalar field allows one to use
the powerful mathematical method of harmonic analysis on a group, at the same
time supporting the consideration by physical motivations.

The consideration of the functions $f(x,\bz)$ guarantees the possibility to
describe arbitrary spin particles because any irrep of a group is equivalent
to some sub-representation of GRR. Thus we deal with an unique field
containing all masses and spins. As a consequence, we have:

1. The explicit form of spin projection operators does not depend on the spin
value. These operators are the differential operators with respect to $\bz$.

2. For this scalar field and thus for arbitrary spin discrete transformations
$C,P,T$ are defined as the automorphisms of the Poincar\'e group.

3. RWE arise under the classification of the
functions on the Poincar\'e group by eigenvalues of invariant operators and
have the same form for arbitrary spin.

The switch to the usual multicomponent description by functions
$\psi_n(x)$ corresponds to a separation of the space-time and spin
variables, $f(x,\bz)=\sum \phi_n(\bz)\psi_n(x)$,
where $\phi_n(\bz)$ and $\psi_n(x)$ transform under contragradient
representations of the Lorentz group.
The use of the transformation rules of $x,\bz$ under automorphisms enables
us to deduce the transformation rules of $\psi_n(x)$ under $C,P,T$ without
any consideration of the specific form of equations of motion.

It is shown that in even dimensions the consistent consideration of
invariant with respect to space reflection RWE
requires to use the generators of group $SO(D,2)$, which is an
extension of the corresponding Lorentz group $SO(D,1)$.

The interpretation of the right generators belonging to the complete set of
commuting operators on the Poincar\'e group is given.
This interpretation is similar to Wigner and Casimir interpretation of
right generators of the rotation group in the nonrelativistic theory (see
\cite{Wigne59,BieLo81}).
Like in the nonrelativistic case, right generators define some quantum
numbers, which do not depend on the choice of the laboratory frame.
In particular, in 3+1-dimensional case three right generators of the
Poincar\'e group define Lorentz characteristics $j_1,j_2$ and chirality,
and fourth right generator distinguishes particles and antiparticles.

Using complete sets of the commuting operators on the group, we classify
scalar functions $f(x,\bz)$. As one of the results of this
classification we reproduce essentially all known finite-component RWE.
Moreover, such an approach allows one to consider some alternative
possibilities, which have not been formulated before. In particular, in
3+1-dimensional case we write out general subsidiary conditions
(\ref{gensub}) corresponding to $s\ne j_1+j_2$.
On the other hand, instead of subsidiary conditions one may consider
functions of momentum $p$ and spin variables $\bz$ with invariant constraints
(\ref{pqconstr}).
It is shown that the set of operators related to higher spin equations
in 3+1 dimensions obeys commutation relations of $so(3,3)$ algebra, which
coincide with the algebra of $\gamma$-matrices for spin $1/2$. But unlike
the latter case the set of operators for higher spin equations is not closed
with respect to anticommutation.

In the framework of the classification of scalar functions we get also
positive energy wave equations allowing probability amplitude
interpretation and associated with infinite-dimensional unitary
representations of the Lorentz group.
Along with the alternative description of integer or half-integer spin
fields, just these equations ensure description of fractional
spin fields in 1+1 and 2+1 dimensions.

The consideration of the scalar field on the Poincar\'e group
have allowed us both to obtain new results and to reproduce the main results
of RWE theory, which earlier were obtained by means of different reasons
and methods.
Thus a general approach to the construction of different types of RWE is
established.
Besides, one may consider this approach as an alternative method to
construct a detail theory of the Poincar\'e group representations.

Notice that the approach under consideration can be directly applied to
higher dimensional cases and, as we hope, can be generalized to other
space-time symmetry groups, such as de Sitter and conformal groups.

\section*{Acknowledgments}

This work was partially supported by Brazilian Agencies CNPq (D.M.G.) and
FAPESP (D.M.G., A.L.Sh.).
The authors would like to thank I L Buchbinder, L A Shelepin, S N Solodukhin,
I V Tyutin and A A Deriglazov for useful discussions.


\appendix
\section{Bases of 2+1 Lorentz group representations and $S^\mu$ matrices}

Spin projection operators $\hat S^\mu$ acting in the space of the
functions $f(x,\bz)$ of $x=(x^\mu)$ and two complex variables
$z^1=z_2,\, z^2=-z_1$, $|z_1|^2-|z_2|^2=|z^2|-|z^1|=1$, have the form
\be
\hat S^\mu=\frac 12(z\gamma^\mu\partial_z -
\cc z\cc\gamma^\mu\partial_{\ccc z}\,),  \quad z=(z^1\; z^2),\;
\partial_z=(\partial/\partial{z^1}\;\partial/\partial{z^2})^T,
\ee
where $\gamma^\mu=(\sigma_3, i\sigma_2, -i\sigma_1)$. For $z=(z_1\; z_2)$ the
relation $\hat S^\mu=-\frac 12(z\cc\gamma^\mu\partial_z -
\cc z\gamma^\mu\partial_{\ccc z}\,)$ is valid.

The polynomials of the power $2S$ in $z$, which correspond to
finite-dimensional irreps $T^0_S$ of 2+1 Lorentz group, can be written in
the form
\be \label{d18}
T^0_S:\quad f_S(x,z)=\sum\limits_{n=0}^{2S} \phi^n(z)\psi_n(x), \quad
\phi^n(z)=\left( C_{2S}^n\right)^{1/2}
(z^1)^{2S-n}(z^2)^{n},  \quad  s^0=S-n,
\ee
where $s^0$ is eigenvalue of $\hat S^0$, and $C_{2S}^n$ are binomial
coefficients. The quasipolynomials of the power $2S\le -1$, which correspond
to infinite-dimensional unitary irreps $T^\pm_S$ of 2+1 Lorentz group, can be
written in the form
\begin{eqnarray}
&&T^+_S:\quad f_S(x,z)=\sum\limits_{n=0}^\infty \phi^{n}(z)\psi_n(x), \quad
 \phi^{n}(z)=\left( C_{2S}^n\right) ^{1/2}(z^2)^{2S-n} (z^1)^{n}, \quad
 s^0=-S+n,  \nonumber \\
&&T^-_S:\quad f_S(x,\cc z)=\sum\limits_{n=0}^\infty \phi^{n}(z)\psi_n(x), \quad
 \phi^{n}(z)=\left( C_{2S}^n\right)^{1/2}
 (\cc z^{\dot 2})^{2S-n} (\cc z^{\dot 1})^{n},
 \quad s^0=S-n, \label{d22} \\
&&C_{2S}^n=\left( \frac{(-1)^n\Gamma (n-2S)}{n!\Gamma (-2S)}\right)^{1/2}.
\nonumber
\end{eqnarray}

There is a correspondence between the action of differential operators
$\hat S^{\mu}$ on the functions $f(x,\bz)=\phi(\bz)\psi(x)$ and the
multiplication of matrices $\hat S^{\mu}$ by columns $\psi (x)$ composed of
$\psi_n(x)$, $\hat S^\mu f(x,\bz)=\phi(\bz)S^\mu\psi(x)$. For the
finite-dimensional representations $T^0_S$ we have $(S^0)^\dagger=S^0$,
$(S^k)^\dagger=-S^k$,
\begin{eqnarray}
&&(S^0)_n^{\;\;n'}=\delta _{nn^{\prime }}(S-n), \qquad n=0, 1, \ldots, 2S,
\nonumber \\
&&(S^1)_n^{\;\;n'}=-\frac i2\left (\delta _{n\ n^{\prime}+1}\sqrt{%
(2S-n+1)n} +\delta _{n+1\ n^{\prime}}\sqrt{(2S-n)(n+1)}\right ),
\nonumber \\
&&(S^2)_n^{\;\;n'}=-\frac 12\left (\delta _{n\ n^{\prime}+1}\sqrt{%
(2S-n+1)n} -\delta _{n+1\ n^{\prime}}\sqrt{(2S-n)(n+1)}\right ).
\label{d21}
\end{eqnarray}
Matrices $S^\mu$ satisfy the condition $(S^\mu)^\dagger=\Gamma S^\mu\Gamma$,
where $\Gamma$ is a diagonal matrix, $(\Gamma)_n^{\;\;n'}=(-1)^n\delta_{nn'}$.
The substitution $z\to \cc z\,$ in (\ref{d18}) changes only signs of $S^0$
and $S^2$. For representations $T^+_S$ of discrete positive series is valid
$(S^\mu)^\dagger=S^\mu$,
\begin{eqnarray}
&&(S^0)_n^{\;\;n'}=\delta _{nn^{\prime }}(-S+n), \qquad n=0, 1, 2, \ldots,
\nonumber \\
&&(S^1)_n^{\;\;n'}=-\frac 12\left (\delta _{n\ n^{\prime}+1}\sqrt{%
(n-1-2S)n} +\delta _{n+1\ n^{\prime}}\sqrt{(n-2S)(n+1)} \right ),
\nonumber \\
&&(S^2)_n^{\;\;n'}=\frac i2\left (\delta _{n\ n^{\prime}+1}\sqrt{%
(n-1-2S)n} -\delta _{n+1\ n^{\prime}}\sqrt{(n-2S)(n+1)} \right ).
\label{d23}
\end{eqnarray}
For $T^-_S$ matrices $S^1$ have the same form, whereas $S^0$, $S^2$ change
only their signs.

The case of representations of principal series, which is not bounded
by the highest (lowest) weight, was considered in \cite{GitSh97}.

For the representations, which correspond to finite-dimensional irreps
$T^0_S$, the decomposition (\ref{d18}) can be written in terms of symmetric
spin-tensors:
$\psi_{\alpha_1\dots\alpha_{2S}}(x)=\psi_{\alpha_{(1}\dots\alpha_{2S)}}(x)$,
\be  \label{21ten}
f_S(x,z)=\psi_{\alpha_1\dots\alpha_{2S}}(x)
z^{\alpha_1} \dots z^{\alpha_{2S}}.
\ee
Comparing the decompositions (\ref{d18}) and (\ref{21ten}), we obtain the
relation
\be
(C^n_{2S})^{1/2}\psi_n(x)=
 \psi_{\underbrace{\scriptstyle{1\,\dots\,1}}_{2S-n} \,
       \underbrace{\scriptstyle{2\,\dots\,2}}_{n} }(x).
\ee

\section{Bases of 3+2 de Sitter and 3+1 Lorentz groups representations
and $\Gamma^\mu$ matrices}

Consider polynomials of elements of Dirac $z$-spinor
$Z_D=(z^\alpha,\cc\uz_{\dot\alpha})$.
Any polynomial of power $2S$ can be decomposed in the basis of
$(2S+3)!/(6(2S)!)$ monomials
$$
(z^1)^a (z^2)^b \cc\uz_{\dot 1}^c \cc\uz_{\dot 2}^d, \quad a+b+c+d=2S.
$$
One may write out 16 operators, which conserve the power of polynomial:
\ba  \label{31S}
&&\hat S^{\mu\nu}=
 \qq ( (\sigma^{\mu\nu})^{\;\;\beta}_\alpha z^\alpha \partial_\beta +
      (\bar\sigma^{\mu\nu})^\da_{\;\;\db} \cc\uz_\da \upartial^\db )-c.c.,
\\  \label{31Gmu}
&&\hat\Gamma^\mu= \hat V^\mu_{12}+\hat V^\mu_{21}-c.c.= \frac 12 \left(
 \bar\sigma^{\mu\da\alpha}\cc\uz_{\da}\partial_{\alpha}+
 {\sigma^\mu}_{\alpha\da}z^{\alpha}\upartial^{\da} \right)-c.c.,
\\
&&\uGamma^\mu=
 i(\hat V^\mu_{12}-\hat V^\mu_{21})+c.c.= \frac i2 \left(
 \bar\sigma^{\mu\da\alpha}\cc\uz_{\da}\partial_{\alpha} -
 {\sigma^\mu}_{\alpha\da}z^{\alpha}\upartial^{\da} \right)+c.c.,
\\  \label{31G5}
&&\hat\Gamma^5 = \qq \left( z^{\alpha} \partial_\alpha
 -\cc\uz_{\dot\alpha} \upartial^{\da} \right)+c.c.,
\\  \label{31I}
&&\hat {\cal T} = -\hat S_3^R = \qq \left( z^{\alpha} \partial_{\alpha}
 +\cc\uz_{\dot\alpha} \upartial^{\da} \right)-c.c.,
\ea
where $\partial_{\alpha}=\partial/\partial z^{\alpha}$,
$\upartial^{\da}=\partial/\partial \cc\uz_{\da}$,
\be \label{sigmn}
(\sigma^{\mu\nu})_\alpha^{\;\;\beta} = -\frac i4
(\sigma^\mu\bar\sigma^\nu-\sigma^\nu\bar\sigma^\mu)_\alpha^{\;\;\beta},\quad
(\bar\sigma^{\mu\nu})^\da_{\;\;\db}= -\frac i4
(\bar\sigma^\mu\sigma^\nu-\bar\sigma^\nu\sigma^\mu)^\da_{\;\;\db},
\ee
and $c.c.$ is complex conjugate term corresponding the action in the space of
polynomials of the elements of $\underline Z_D=(\uz^\alpha,\cc z_{\dot\alpha})$.
Operator $\hat {\cal T}$ commutes with other 15 operators and defines $(\pm)$
power of the polynomials, for functions of $Z_D$ and $\underline Z_D$
respectively. Operators (\ref{31S})-(\ref{31G5}) obey
commutation relations of $so(3,3)\sim sl(4,R)$ algebra,
\ba \nonumber
&&[\hat\Gamma^5,\hat S^{\mu\nu}]=0, \quad
 [\hat\Gamma^5,\hat \Gamma^\mu]=i\uGamma^\mu,  \quad
 [\hat\Gamma^5,\uGamma^\mu]=-i\hat \Gamma^\mu,  \\
&&[\uGamma^\mu,\uGamma^\nu]=-i\hat S^{\mu\nu}, \quad
 [\hat S^{\lambda\mu},\uGamma^\nu]
  =i(\eta^{\mu\nu}\uGamma^\lambda-
  \eta^{\lambda\nu}\uGamma^\mu), \quad
 [\hat\Gamma^\mu,\uGamma^\nu]=i\eta^{\mu\nu}\hat\Gamma^5 ,
\ea
see also (\ref{comg1}),(\ref{comg2}). Using the notations
$\hat S^{4\mu}=\hat\Gamma^\mu$, $\hat S^{5\mu}=\uGamma^\mu$,
$\hat S^{54}=\hat\Gamma^5$, one can rewrite commutation relations in the
form (\ref{deSitt}), where $\eta_{55}=\eta_{44}=\eta_{00}=1$,
$\eta_{11}=\eta_{22}=\eta_{33}=-1$.
However, for unitary representations of the Poincar\'e group all the
generators and, in particular $\hat B^R_3=\mp i\hat\Gamma^5$ (for functions
of $Z_D$ and $\underline Z_D$ respectively), are Hermitian.
Thus, setting $\hat S^{5\mu}=i\uGamma^\mu$, $\hat S^{54}=i\hat\Gamma^5$,
for these representations it is natural to consider an algebra
$so(4,2)\sim su(2,2)$ of Hermitian operators.

Supplementing generators $\hat S^{\mu\nu}$ of the Lorentz group by four
operators $\hat \Gamma^\mu$ (or $\uGamma^\mu$), we obtain the algebra of the
3+2 de Sitter group $SO(3,2)$. Generators in finite-dimensional representations
of $SO(3,2)$ obey the relations $\hat\Gamma^{0\dagger} = \hat\Gamma^0$,
$\hat\Gamma^{k\dagger} =-\hat\Gamma^k$.

The linear functions of $z$ $f(x,z)=Z_D\Psi_D(x)$ correspond to
four-dimensional bispinor representation. In the space of columns $\Psi_D(x)$
the operators act as matrices
\be
\hat S^{\mu\nu} \to \sigma^{\mu\nu}/2, \quad \hat \Gamma^\mu \to \gamma^\mu/2,
\quad \hat \Gamma^5 \to \gamma^5/2,
\quad \hat {\underline \Gamma}^\mu \to i\gamma^\mu \gamma^5/2,
\quad \hat {\cal T} \to 1/2.
\ee
In accordance with general theory, Dirac matrices and spin 1 Duffin--Kemmer
matrices obey commutation relations of $so(3,3)$ algebra
\cite{Hepne62,Petra95}.

Using (\ref{P})-(\ref{C}), we get for the action of the discrete
transformations on the operators (\ref{31S})-(\ref{31I}):
\be \label{31diskr}
\begin{array}{ccccrr}
  &\hat S^{\mu\nu} &\hat\Gamma^\mu &\uGamma^\mu &\hat\Gamma^5 &\hat{\cal T}\\
C & -1             & -1            &  1         &  1          & -1    \\
P,T' &(-1)^{\delta_{0\mu}+\delta_{0\nu}} & -(-1)^{\delta_{0\mu}} & (-1)^{\delta_{0\mu}} & -1 &  1 \\
T_{sch} &-(-1)^{\delta_{0\mu}+\delta_{0\nu}} & (-1)^{\delta_{0\mu}}  & (-1)^{\delta_{0\mu}} & -1 & -1 \\
\end{array}
\ee

It is possible to construct two linear in $\hat p^\mu$ equations for the
scalar functions $f(x,\bz)$, which are invariant under proper Poincar\'e
group
\be
(\hat p_\mu \hat\Gamma^\mu -\varkappa)f(x,\bz)=0, \qquad
(\hat p_\mu \uGamma^\mu -\varkappa)f(x,\bz)=0,
\ee
but in accordance with (\ref{31diskr}) only operator
$\hat p_\mu \hat \Gamma^\mu$ is invariant under space reflection,
operator $\hat p_\mu \hat {\underline\Gamma}^\mu$ changes the sign.
Thus only the first equation is invariant under space reflection.

Operators $\hat \Gamma^5$ and $\hat p_\mu\hat \Gamma^\mu$ commute with all
the left generators of the Poincar\'e group but do not commute with each
other, $[\hat\Gamma^5,\hat p_\mu\hat\Gamma^\mu]=i\hat p_\mu \uGamma^\mu$.
Therefore chirality of massive particle describing by the equation
$(p_\mu\hat\Gamma^\mu -ms)f(x,\bz)=0$ is uncertain. Operator $\hat{\cal T}$
commutes both with all left generators of the Poincar\'e group and with
operators $\hat\Gamma^\mu$; therefore one may relate to this operator
some conserved quantum number changing the sign under charge conjugation.

On the polynomials of four complex variables $z^\alpha,\cc\uz_{\dot\alpha}$
one can realize symmetric irreps $T_{[2S\,0\,0]}$ of $SL(4,R)\sim SO(3,3)$.
These irreps are a symmetric part of $2S$-times direct product of fundamental
four-dimensional irreps $T_{[1\,0\,0]}$ and remain irreducible after the
reduction on the subgroup $SO(3,2)$, $T_{[2S\,0\,0]}\to T_{[2S\,0]}$.
Notice that here we use the notation different from \cite{Bhabh45}: $[2S\,j]$
corresponds to $(j+S\,S)$ in the notation of \cite{Bhabh45}.

We will consider two bases of finite-dimensional irrep $T_{[2s\,0]}$ of
$SO(3,2)$, namely, bases consisting of eigenfunctions of operators
$\hat\Gamma^5$ or $\hat\Gamma^0$. The first basis corresponds to chiral
representation,
\be \label{31bas5}
\varphi^{m_1m_2}_{j_1j_2}(\zz) = N^{1/2}
 (z^1)^{j_1+m_1}(z^2)^{j_1-m_1} \cc\uz_{\dot 1}^{j_2+m_2} \cc\uz_{\dot 2}^{j_2+m_2},
\ee
where $s=j_1+j_2$, $\lambda=j_1-j_2$, $m_1$ and $m_2$ are eigenvalues of the
operators $\hat{M_3}$ and $\hat{N_3}$, which are the linear combinations of
$\hat{S_3}$ and $\hat{B_3}$, see (\ref{MNgen}),
$N=(2s)!/((j_1+m_1)!(j_1-m_1)!(j_2+m_2)!(j_2-m_2)!)$.
Consisting of eigenfunctions of $\hat\Gamma^0$ basis
\be  \label{31bas0}
\phi^{n_1n_2}_{k_1k_2}(\zz) = (N')^{1/2}
(z^1+\cc\uz_{\dot 1})^{k_1+n_1}(z^2+\cc\uz_{\dot 2})^{k_1-n_1}
(z^1-\cc\uz_{\dot 1})^{k_2+n_2}(z^2-\cc\uz_{\dot 2})^{k_2-n_2},
\ee
where $s=k_1+k_2$,
$N'=(2s)!/((k_1+n_1)!(k_1-n_1)!(k_2+n_2)!(k_2-n_2)!)$,
for $s=1/2$ corresponds to Dirac representation.
The functions (\ref{31bas0}) are eigenfunctions of operators
$\hat\Gamma^0,\uGamma^3,\hat S_3$ with eigenvalues
$k_1-k_2$, $i(n_1-n_2)/2$, $(n_1+n_2)/2$ respectively. For fixed $s$
we have
\be  \label{31dec}
f_s(x,\zz)=\sum_{j_1+j_2=s}\sum_{m_1,m_2} \psi^{m_1m_2}_{j_1j_2}(x)
\varphi^{m_1m_2}_{j_1j_2}(\zz)= \sum_{k_1+k_2=s}\sum_{n_1,n_2}
\psi^{n_1n_2}_{k_1k_2}(x)\phi^{n_1n_2}_{k_1k_2}(\zz).
\ee

Below we will use the basis (\ref{31bas0}). According to (\ref{sol1}) in the
rest frame for a particle describing by the system (\ref{fmeq})-(\ref{fjeq})
we have
\ba \nonumber
&&f(x,\zz)=\psi^+(x)\phi^+_{s,s^3}(\zz)+\psi^-(x)\phi^-_{s,s^3}(\zz)
 =C_1 e^{imx^0}\phi^+_{s,s^3}(\zz)+C_2 e^{-imx^0}\phi^-_{s,s^3}(\zz),
\\  \label{31rest}
&&\phi^+_{s,s^3}(\zz)=(z^1+\cc\uz_{\dot 1})^{s+s^3}(z^2+\cc\uz_{\dot 2})^{s-s^3}, \quad
 \phi^-_{s,s^3}(\zz)=(z^1-\cc\uz_{\dot 1})^{s+s^3}(z^2-\cc\uz_{\dot 2})^{s-s^3}.
\ea

The equation $(\hat p_\mu \hat\Gamma^\mu -sm)f(x,\zz)=0$ has the matrix form
\be \label{31D}
(\hat p_\mu\Gamma^\mu-sm)\psi(x)=0,
\ee
where $\psi(x)$ is a column. It is convenient to enumerate the basis elements
(\ref{31bas0}) (and the elements of the column $\psi(x)$)
in order of decrease of $k_1-k_2=s,s-1,\dots,-s$.
Matrices $\Gamma^\mu$ obey the relations $\Gamma^{0\dagger}=\Gamma^0$,
$\Gamma^{k\dagger}=-\Gamma^k$. Matrix $\Gamma^0$ is diagonal and has the
elements $k_1-k_2$. Matrices $\Gamma^1$ and $\Gamma^3$ are skew-symmetric
real, and $\Gamma^2$ is symmetric imaginary.
According to (\ref{31Gmu}) matrices $\Gamma^k$ have nonzero elements only in
blocks corresponding to the transitions $(k_1,k_2)\to(k_1\pm 1/2,k_2\mp 1/2)$.
Using this property, it is easy to see that diagonal matrix $\Gamma$ with the
elements $(-1)^{2k_2}$ commutes with $\Gamma^0$ and anticommutes with
$\Gamma^k$, $\Gamma^{\mu\dagger}=\Gamma\Gamma^\mu\Gamma$. This allows one to
rewrite the Hermitian-conjugate equation
$\psi^\dagger(\overleftarrow {{\hat p}_\mu}\Gamma^{\mu\dagger}+sm)=0$
in the form
\be \label{31Dc}
\bar\psi(x)(\overleftarrow {{\hat p}_\mu}\Gamma^\mu+sm)=0, \quad
\bar\psi=\psi^\dagger\Gamma,
\ee
and to define invariant scalar product in the space of columns as
$\int\bar\psi(x)\psi(x)d^3x$. As a consequence of (\ref{31D}) and
(\ref{31Dc}), the continuity equation holds
$$
\partial_\mu j^\mu=0, \quad  j^\mu=\bar\psi\Gamma^\mu\psi.
$$

Now the question concerning the positive definiteness of current vector
component $j^0$ and energy density may be consider similarly to
2+1-dimensional case, see Section 3. For half-integer spin particles
describing by the system (\ref{fmeq})-(\ref{fjeq}) charge density $j^0$ is
positive definite, since in the rest frame (see (\ref{31rest}))
$j^0=\bar\psi\Gamma\Gamma^0\psi=s(|\psi^+(x)|+|\psi^-(x)|)>0$.
Energy density (defined in terms of energy-momentum tensor (\ref{Ep})) and
the scalar product $\bar\psi\psi$ are indefinite since in the rest frame they
are proportional to $|\psi^+(x)|-|\psi^-(x)|$. For integer spin particles
energy density is positive definite, the scalar product and $j^0$ are
indefinite.

Consider discrete transformations in terms of the columns $\psi(x)$.
According to (\ref{PZ}) under space reflection
$\phi^{n_1n_2}_{k_1k_2}(\zz)\to(-1)^{2k_1}\phi^{n_1n_2}_{k_1k_2}(\zz)$.
Whence, taking into account $f(x,\zz)\to f(x',z')=\phi(\zz)\psi'(x')$, we get
\be
\psi(x)\stackrel{P}\to(-1)^{2s}\Gamma\psi(\bar x), \quad\text{where}
\;\; \bar x= (x^0, -x^k).
\ee
According to (\ref{C}) under charge conjugation
$\phi^{n_1n_2}_{k_1k_2}(\zz)\to\phi^{n_1n_2}_{k_1k_2}(\cc z, \uz)$.
Taking into account that $\phi^{n_1n_2}_{k_1k_2}(\cc z,\uz)$ and
$(-1)^{s+n_1-n_2}\phi^{n_2n_1}_{k_2k_1}(\zz)$ have the same transformation
rule, we get
\be
\psi^{n_1n_2}_{k_1k_2}(x)
\stackrel{C}\to (-1)^{s+n_1-n_2}\cc\psi^{n_2\,n_1}_{k_2\;k_1}(x).
\ee
In particular, for $s=1/2$, using the relation $f(x,\zz)=Z_D\Psi(x)$, we get
$\Psi(x)\stackrel{P}\to\gamma^0\Psi(\bar x)$,
$\Psi(x)\stackrel{C}\to\Psi^c(x)=C\bar\Psi^T(x)$, where $C$ is the matrix with elements
$-i\sigma_2$ on secondary diagonal, $C=i\gamma^2\gamma^0$.
The transformation properties of the bilinears $\bar\psi\Gamma^\mu\psi$,
$\bar\psi\Gamma^5\psi$, $\bar\psi{\underline \Gamma}^\mu\psi$ under $C,P,T$
coincide with ones of the corresponding operators, see (\ref{31diskr}).


\end{document}